%% file: Final_accepted.tex
\documentclass[twoside,11pt]{article}

\usepackage{jmlr2e}
\usepackage{multirow}
\usepackage{pifont}
\usepackage{hyperref}
\usepackage{amsmath}%, amsthm}
\usepackage{color}
\usepackage{algpseudocode}
\usepackage{enumerate}
\usepackage{wrapfig}
\usepackage{graphicx} % more modern
\usepackage{caption}
\usepackage{subcaption}
\usepackage{algorithm}
\usepackage{bm}
\usepackage{hyperref}
\usepackage{soul}
\usepackage{algorithm,algcompatible}

\input{revisit/Command.tex}

\def\T{{ \mathrm{\scriptscriptstyle T} }}
\newtheorem{property}{Property}

\newcolumntype{P}[1]{>{\centering\arraybackslash}p{#1}}

\algblock{ParFor}{EndParFor}
\algnewcommand\algorithmicparfor{\textbf{parfor}}
\algnewcommand\algorithmicpardo{\textbf{do}}
\algnewcommand\algorithmicendparfor{\textbf{end\ parfor}}
\algrenewtext{ParFor}[1]{\algorithmicparfor\ #1\ \algorithmicpardo}
\algrenewtext{EndParFor}{\algorithmicendparfor}
\algnewcommand\INPUT{\item[\textbf{Input:}]}%
\algnewcommand\OUTPUT{\item[\textbf{Output:}]}
\newcommand{\cmark}{\ding{51}}%
\newcommand{\xmark}{\ding{55}}%

\usepackage{lastpage}
\jmlrheading{24}{2023}{1-\pageref{LastPage}}{7/22; Revised
2/23}{8/23}{22-0833}{Quan Zhang, Yanxun Xu, Mei-Cheng Wang, Mingyuan Zhou}
\ShortHeadings{Weibull Racing Survival Analysis}{Zhang, Xu, Wang, and Zhou}

\firstpageno{1}

\begin{document}

\title{Weibull Racing Survival Analysis with Competing Events, Left Truncation, and Time-Varying Covariates}

\author{\name Quan Zhang \email quan.zhang@broad.msu.edu \\
       \addr Department of Accounting and Information Systems\\
       %Broad College of Business\\
       Michigan State University\\
       East Lansing, MI 48824, U.S.A.
       \AND
       \name Yanxun Xu \email yanxun.xu@jhu.edu \\
       \addr Department of Applied Mathematics and Statistics\\
    Johns Hopkins University\\
       Baltimore, MD 21218, U.S.A.
       \AND
       \name Mei-Cheng Wang \email mcwang@jhu.edu\\
       \addr Department of Biostatistics\\
Johns Hopkins University\\
Baltimore, MD 21205, U.S.A.
       \AND
       \name Mingyuan Zhou \email mingyuan.zhou@mccombs.utexas.edu \\
       \addr Department of Information, Risk, and Operations Management\\
       Department of Statistics and Data Sciences\\
       The University of Texas at Austin\\
       Austin, TX 78712, U.S.A.
       }

\editor{Sanmi Koyejo}
\maketitle

\begin{abstract}%   <- trailing '%' for backward compatibility of .sty file
We propose Bayesian nonparametric Weibull delegate racing (WDR) to fill the gap in interpretable nonlinear survival analysis with competing events, left truncation, and time-varying covariates. 
We set a two-phase race among a potentially infinite number of sub-events to model nonlinear covariate effects, which does not rely on transformations or complex functions of the covariates. Using gamma processes, the nonlinear capacity of WDR is parsimonious and data-adaptive.
In prediction accuracy, WDR dominates cause-specific Cox and Fine-Gray models and is comparable to random survival forests in the presence of time-invariant covariates. More importantly, WDR can cope with different types of censoring, missing outcomes, left truncation, and time-varying covariates, on which other nonlinear models, such as the random survival forests, Gaussian processes, and deep learning approaches, are largely silent. We develop an efficient MCMC algorithm based on Gibbs sampling. 
%and provide an \texttt{R} package.
%We use WDR to study time to death of three types of lymphoma, illustrate how the hazards are nonlinearly affected by gene expression, and show the potential of WDR in discovering and diagnosing new diseases using its interpretability. Moreover, we investigate the age at onset of mild cognitive impairment and interpret the accelerating or decelerating effects of biomarkers on the progression of Alzheimer's disease.
We analyze biomedical data, interpret disease progression affected by covariates, and show the potential of WDR in discovering and diagnosing new diseases.
\end{abstract}


\section{Introduction} 
In survival analysis, it is common to consider competing events (also known as competing risks) that are mutually exclusive. In other words, the occurrence of one event precludes the occurrence of another. %other events. 
For example, when studying the time to death of a patient, all possible causes of death are competing events since a patient who died of one cause would never die of others. In the presence of competing events, people model both the event time and the event type or which one of the competing events occurs first. One may argue that every survival model 
can handle competing events if each competing event is analyzed separately and meanwhile, subjects having suffered from other events are treated as right-censored\footnote{Right-censored data, or right censoring, means that we do not observe the exact event time of a subject but know it is larger than some value, and this value is defined as the right censoring time.}. However, this approach can be problematic because it violates the assumption of independent or non-informative censoring \citep{kalbfleisch2011statistical}, which means the censoring time is stochastically independent of the length of survival time,
 %; \yx [the assumption requires %\red{that the censoring mechanism be non-informative of the length of survival time.}
%subjects who are exposed to the risk to have the same future risk for the occurrence of the event as the censored ones do as if censoring is random and thus non-informative] \yx this sentence is a little bit hard to read. Also random censoring and non-informative censoring are not the same thing. I would suggest to delete this sentence. \zz
and thus often leads to biased estimation of the cumulative incidence function \citep{austin2016introduction}.

Left truncation and time-varying covariates often exist in biomedical studies and should be accommodated in survival analysis. 
%Censoring is a form of missing event time whose true value is not observed but known to fall into some interval \citep{kleinbaum2010survival}. %We have right censoring if this interval is only lower-bounded, left censoring if upper-bounded, and interval censoring if both the lower and upper bounds are known. 
Left truncation occurs when subjects who have already experienced a failure event or who have passed some milestone are not eligible to be recruited. Restricting the analysis to the recruited subjects without accounting for left truncation results in an immortal time bias \citep{levesque2010problem,mcgough2021penalized} because these subjects cannot have the failure event prior to entering the study. For example, in Alzheimer's disease studies, the failure time can be the age at the onset of symptoms of mild cognitive impairment (MCI), and only subjects who are disease-free at baseline are recruited. In this setting, a patient having MCI  at the time of recruitment is excluded from the study, leading to an underestimated risk at her age if the analysis does not account for left truncation. 
Time-varying covariates occur when a covariate value changes over time during the follow-up period. In Alzheimer's disease studies, biomarkers such as cerebrospinal fluid variables and cognitive measures can be collected longitudinally and are informative of the disease progression. In this paper, we consider covariate measurements at discrete points in time, which is often the case in practice.
 
When studying competing events in biomedical research, interpreting model inference is often of interest. 
Two of the most popular models are the cause-specific Cox model \citep{prentice1978analysis,lunn1995applying,putter2007tutorial} and the Fine-Gray proportional subdistribution hazards model \citep{fine1999proportional}, which are both semi-parametric. The former models the cause-specific hazard function and is often applied to studying the etiology of diseases, while the latter models the subdistribution function and is favorable when developing prediction models and risk-censoring systems \citep{austin2016introduction}. Both models assume proportional hazards and use a linear function of covariates to interpret how much a unit increase in a covariate is multiplicative to the hazard functions. We refer the readers to \cite{austin2020review} for a detailed review on interpreting the Fine-Gray model in various applications.  

%One often faces a trade-off between model interpretability and flexibility in competing event analysis. 
However, semi-parametric and parametric models for competing events often achieve interpretability by sacrificing flexibility. Specifically, 
%though the Cox and Fine-Gray models and other semi-parametric approaches based on partial likelihoods provide a framework for incorporating censoring, left truncation, and time-varying covariates \citep{sueyoshi1992semiparametric,fine1999proportional,mcgough2021penalized}, their interpretation depends
the Cox and Fine-Gray models and other semi-parametric approaches depend 
on the linear function of covariates in partial likelihoods, and thus the model fit can be undermined in the presence of non-monotonic covariate effects. Data transformation and stratification can alleviate such a problem but often require expert opinions and/or an excessive number of parameters. One can also replace the linear function of covariates with complex functions, such as splines \citep{danieli2019competing} and neural networks \citep{kvamme2019time}, but this practice leads to difficulty in model interpretation and may overburden practitioners who try to explain a covariate effect.
% st{by the existence of non-monotonic covariate effects. }
Other (semi-)parametric methods for competing event analysis have been categorized by \cite{haller2013applying} into mixture models \citep{ng2003based,lau2011parametric}, vertical modeling \citep{nicolaie2010vertical,nicolaie2019vertical}, and the pseudo-observation approach \citep{klein2005regression,rahman2021deeppseudo}; they are also incapable of nonlinear modeling unless making data transformations, and/or they cannot account for left truncation or time-varying covariates.   

Nonparametric or deep learning approaches can enhance model flexibility, but their interpretation is not straightforward, and how to deal with left truncation or time-varying covariates by these approaches has not been %is insufficiently 
sufficiently investigated. For example, the random survival forests for competing events \citep{ishwaran2014random} has simplified the modeling of nonlinear covariate effects, and Bayesian models have been developed, such as Gaussian processes \citep{barrett2013gaussian,alaa2009deep} and Lomax racing \citep{zhang2018nonparametric}. However, no extensions of these models have been made to accommodate left truncation or time-varying covariates. 
%However, the former may lack interpretation if the dimension of covariates is high and the latter requires decreasing hazards as ground truths. 
%Besides, random survival forests have been widely used for competing event analysis \citep{ishwaran2014random} and compared as a benchmark. To the best of our knowledge, however, no extensions of these approaches have been made to accommodate left truncation or time-varying covariates.
\cite{shi2021dependent} propose a dependent Dirichlet process mixture model for competing events in the presence of a binary time-varying covariate for treatment switching, but it is limited to the scenario where only one binary time-varying covariate exists. 
In addition to replacing the linear function of covariates in a (semi-)parametric model by neural networks \citep{kvamme2019time,nagpal2021deep,rahman2021deeppseudo}, another stream of deep learning methods is to discretize the time and transform competing-event survival analysis to a classification problem \citep{lee2018deephit,lee2019dynamic,tjandra2021hierarchical}. % with the continuous survival and cumulative incidence functions becoming probability mass functions. 
These classification-based methods cannot easily handle left truncation and are sensitive to
%This practice results in loss of information of continuity and 
time discretization. Specifically,   a fine-grained discretization causes imbalanced labels, while a coarse-grained one leads to inaccurate survival estimation. Moreover, a potential
overfitting %over fit
on smaller data and a general lack of interpretation in neural networks can concern practitioners.

To address the aforementioned challenges, we construct Weibull delegate racing (WDR), a gamma process-based nonparametric Bayesian hierarchical model, for survival analysis with competing events. It achieves both interpretability and flexibility without transformations or complex functions of covariates. 
WDR assumes non-informative censoring and uses the race among Weibull random variables to jointly model event times, event types, and potential subtypes. It enables data-adaptive nonlinear modeling and has the interpretation as a race among latent sub-events. 
%\yx [delete]Consequently, WDR can potentially be  used for discovery of new diseases   where covariates linearly accelerate or decelerate the disease progression. 
We propose an efficient MCMC algorithm based on Gibbs sampling and slice sampling for posterior inference.  The MCMC can handle different types of censoring and impute missing event types. 
%An \texttt{R} package is provided for general dissemination. 
%We also propose a maximum a posteriori estimation for big data application.
To the best of our knowledge, WDR is the first approach for
%WDR distinguishes itself from literature as a one-stop solution to 
interpretable nonlinear modeling of competing events in the presence of left truncation and time-varying covariates. Moreover, it delivers outstanding performance in prediction accuracy.%, various types of censoring, and missing event times or types.

The paper proceeds as follows.
In Section~\ref{sec_survival}, we propose Weibull racing and Weibull delegate racing. Section~\ref{sec:inference} shows the Bayesian inference and how WDR deals with time-varying covariates, censoring, and missing event times or types, inducing the connection between WDR and discrete choice models for classification. In Section~\ref{sec:experiments}, we use synthetic data to showcase WDR's parsimonious nonlinear modeling capacity and outstanding performance. In Section~\ref{sec:real}, we analyze real data of lymphoma and Alzheimer's disease to understand how hazards of competing events are influenced by covariates and show the potential of WDR in discovering and diagnosing new diseases. Section~\ref{sec_conclusion} concludes the paper. We defer to the Appendices
the proofs, some technical details of WDR including the inference algorithms, 
some experiment settings, 
and supplementary results.

\section{Weibull Racing and Weibull Delegate Racing}\label{sec_survival}
We first define the left-truncated Weibull distribution and introduce the Weibull racing property that can be directly used for competing event analysis with left truncation. Then we propose the Weibull racing survival model assuming monotonically accelerating or decelerating covariate effects on event times. Finally, we extend it to the Weibull delegate racing model that allows monotonic or non-monotonic covariate effects. %We also shed light on Weibull delegate racing classification as a discrete choice model. Weibull (delegate) racing will be reduced to Lomax (delegate) racing when the Weibull shape parameter is equal to $1$. Compared to Lomax (delegate) racing that can only model decreasing hazard functions, Weibull (delegate) racing is more flexible when the shape parameter varies.

\subsection{Left-Truncated Weibull Distribution and Weibull Racing}
Let $\textstyle{T\sim\mbox{Weibull}(a, \lambda)}$ denote a Weibull random variable $T$ with the density function equal to  
$\textstyle{
f(t\given a, \lambda) = a\lambda t^{a-1} \exp({-\lambda t^a}),~ t\in \mathbb{R}_+
}$, 
and the survival function (that is $\textstyle{\mbox{Pr}(T>t)}$, the probability that $\textstyle{T>t}$)
$\textstyle{S(t\given a, \lambda)=\exp({-\lambda t^a}),~t\in \mathbb{R}_+}$ with $\textstyle{\mathbb{R}_+}$ representing the nonnegative side of the real line.  $\textstyle{a>0}$ is the Weibull shape parameter, and $\textstyle{\lambda>0}$ such that the expectation $\textstyle{\E(T) = \lambda^{-1/a}\Gamma(1+1/a)}$. %Base on this parameterization, w
We introduce the left-truncated Weibull distribution in Definition~\ref{def:tweibull} and show its survival and hazard\footnote{The hazard function is defined as the density function over the survival function.} functions in Corollary~\ref{cor:tweibull}.  
\begin{definition}%[Left-truncated Weibull distribution]
\label{def:tweibull}
A left-truncated Weibull distribution,  $\textstyle{\text{Weibull}_\tau(a,\lambda)}$ with $\textstyle{a>0}$ and $\textstyle{\lambda>0}$ and the left truncation threshold $\textstyle{\tau\geq 0}$,  is defined by the density function $ 
f_\tau(t\given a, \lambda)= 
 f(t\given a, \lambda)/S(\tau\given a,\lambda)= a\lambda t^{a-1}\exp({-\lambda (t^a-\tau^a) })
 $ if  $\textstyle{t\geq \tau}$ and otherwise $0$,
where $\textstyle{f(\cdotv\given a, \lambda)}$ and $\textstyle{S(\cdotv\given a, \lambda)}$ are the density and survival functions of $\textstyle{\text{Weibull}(a,\lambda)}$. 
\end{definition}
\begin{cor}\label{cor:tweibull}
A left-truncated Weibull distribution,  $\textstyle{ \text{Weibull}_\tau(a,\lambda)}$, has the survival function $\textstyle{S_\tau(t\given a, \lambda) %=  S(t\given a, \lambda)/S(\tau\given a,\lambda)
=\exp({-\lambda (t^a-\tau^a)})}$ and the hazard function 
$\textstyle{h_\tau(t\given a, \lambda) = 
%f_\tau(t\given a, \lambda)/S_\tau(t\given a, \lambda)=
a\lambda t^{a-1}}$  for $\textstyle{t\geq \tau}$. If $\textstyle{t<\tau}$, $\textstyle{S_\tau(t\given a, \lambda) =1}$ and $\textstyle{h_\tau(t\given a, \lambda) =0}$.
\end{cor}
Assuming that a subject's event time follows an untruncated $\textstyle{\mbox{Weibull}(a,\lambda)}$ distribution,  $\textstyle{S_\tau(t\given a,\lambda)}$ is essentially the conditional probability of surviving over $t$ given that the subject has already survived $\tau$. Consequently, $\textstyle{f_\tau(t\given a,\lambda)}$ (or $\textstyle{S_\tau(t\given a,\lambda)}$) is the likelihood of the subject with a left truncation time $\tau$ and an event (or right censoring) time  $t$. 
$\textstyle{\text{Weibull}_\tau(a,\lambda)}$ is reduced to the untruncated $\text{Weibull}(a,\lambda)$ if $\tau=0$. 
Property~\ref{prop:truncated_weibull_race} characterizes a race among independent left-truncated Weibull random variables. %We study the distributions of the minimum and the argument of the minimum of independent Weibull random variables that share the shape parameter and the truncation threshold. 
\begin{property}%[Weibull racing]
\label{prop:truncated_weibull_race}
If $\textstyle{t_j\sim{\emph{\mbox{Weibull}}_\tau}(a, \lambda_j)}$, $\textstyle{j=1,\ldots,J}$, are independent to each other,
the minimum $\textstyle{t=\min\{t_1,\ldots,t_J\}}$ 
and the argument of the minimum $\textstyle{y=\mathop{\mathrm{argmin}}\nolimits_{j\in\{1,\ldots,J\}} \{t_j\}}$ are independent and  satisfy
%follows 
\begin{align}
\textstyle{
t\sim \emph{ \mbox{Weibull}}_\tau (a, \sum\nolimits_{j=1}^J \lambda_{j} ), ~y\sim \emph{\mbox{Categorical}} ({\lambda_1} /{\sum\nolimits_{j=1}^J \lambda_{j}},   \cdots, {\lambda_J}/{\sum\nolimits_{j=1}^J \lambda_{j}}).}
\label{eq:Cat}   
\end{align}
\end{property}

Intuitively, suppose there is a race among teams $j=1,\cdots,J$, each of whose completion time $t_j$ follows $\mbox{Weibull}_\tau(a, \lambda_j)$, and the winner is the team that has the minimum completion time. Property \ref{prop:truncated_weibull_race} implies that the winner's completion time $t$ still follows a left-truncated Weibull distribution and is independent of which team wins.
In the context of survival analysis with left truncation, a team $j$ is a competing event with the requirement that the latent survival time $t_j$  exceeds $\tau$, 
%In the context of survival analysis, if we consider a competing event as a team and the latent survival time to this event as the completion time of the team which must be greater than $\tau$,  
$t$ is the observed time to event (or the observed failure time), and $y$ is the event type (or the cause of failure). 
%If $a=1$ and $\tau=0$, a left-truncated Weibull distribution is reduced to an exponential distribution and the Weibull racing is reduced to exponential racing.
We refer to Property~\ref{prop:truncated_weibull_race} as Weibull racing. It not only describes a natural mechanism of surviving under competing events, but also provides an attractive modeling framework amenable to Bayesian inference. %in terms of computation %terms of posterior simulation as 
%because
Conditioning on $a$ and $\lambda_j$'s,  the joint likelihood of the event type $y$  
and the event time $t$ is
\begin{align}
\textstyle{
p(y,t\given a, \{\lambda_{j}\}_{j})%&= \frac{\lambda_y}{\sum_{j=1}^J \lambda_{j}} \mbox{Weibull}\left(t; a, \sum_{j=1}^J \lambda_{j}\right)
=a  \lambda_{y} t^{a-1} \exp ({-(t^a-\tau^a)\sum\nolimits_{j=1}^J\lambda_{j}} ),
}
\label{eq:joint_Weibull}
\end{align}
which is fully factorized and thus facilitates Bayesian inference by MCMC.

\subsection{Linear Weibull Racing Survival Model}\label{sec:weibull_racing}
We model linear covariate effects on event times in the Weibull racing framework by introducing a gamma-mixed Weibull distribution. Let $\textstyle{\lambda\sim \mbox{Gamma}( r,1/b)}$ denote a gamma random variable % parameterized by shape $r$ and scale $1/b$, 
with $\E(\lambda)=r/b$ and $\mbox{var}(\lambda)=r/b^2$. If $\textstyle{t\sim\mbox{Weibull}_\tau(a,\lambda)}$ and $\lambda\sim \mbox{Gamma}( r,1/b)$,  we have the marginal density of $t$ given $a$, $r$, $b$, and $\tau$ as
%$$ f(t\given a,r,b,\tau)= \int_0^\infty \mbox{Weibull}_\tau(t;a,\lambda) \mbox{Gamma}(\lambda; r,1/b) d\lambda =\frac{arb^r t^{a-1}}{(b+t^a)^{r+1}}.$$ 
$$ \textstyle{
f_\tau(t\given a,r,b)= \int_0^\infty \mbox{Weibull}_\tau(t;a,\lambda) \mbox{Gamma}(\lambda; r,1/b) d\lambda =
arb^r t^{a-1}/(b+t^a-\tau^a)^{r+1}.
}
$$ 
If $a\leq 1$, this gamma-mixed Weibull distribution has a decreasing density $f_\tau(t\given a,r,b)$. %or the mode at $(\frac{ab-b}{1+ar})^\frac{1}{a}$ if $a>1$. 
If~$a>1$, the shape of $f_\tau(t\given a,r,b)$ depends on the values of $a$, $r$, $b$, and $\tau$. Specifically, the gamma-mixed Weibull distribution has the mode at $\textstyle{ [(a-1)(b-\tau^a)/({1+ar})]^{1/a}
}$ if this value is greater than $\tau$; otherwise $\textstyle{f_\tau(t\given a,r,b)}$ is monotonically decreasing. 
Leveraging Property~\ref{prop:truncated_weibull_race} and the gamma-mixed Weibull distribution, we define the Weibull racing survival model as follows.
\begin{definition}\label{def:weibull_racing}
Suppose subject $i$ survives competing events $\{j\given j= 1,\ldots,J\}$ with  latent event times $\{t_{ij}\}_j$. Given its left truncation time $\tau_i$ and time-invariant covariates~$\xv_i$, under the Weibull racing survival model, the subject's  observed event time $t_i$ and event type $y_i$ are generated by 
\begin{align*}
\textstyle
{
t_i = t_{iy_i},~y_i=\mathop{\mathrm{argmin}}\nolimits_{j\in\{1,\ldots,J\}} t_{ij}, ~t_{ij}\sim\emph{\mbox{Weibull}}_{\tau_i}(a,\lambda_{ij}), ~\lambda_{ij}\sim \emph{\mbox{Gamma}}(r_j, \exp({\xv_i^\T\betav_j})) .
}
%\label{eq:LomaxRace}
\end{align*}
\end{definition}
We explain the notations using an Alzheimer's disease example where there are two competing events, the onset of mild cognitive impairment (MCI) ($j=1$) and death ($j=2$). Given subject $i$ entering the study at age $\tau_i$ that is her left truncation time, we assume her latent age at onset of MCI is $t_{i1}$ and latent age at death is $t_{i2}$, where $t_{ij}$, $j=1,2$, depends on the linear function $\bm x_i^\T \betav_j$ of the time-invariant covariates $\bm x_i$ and the parameters $a$ and $r_j$. If the onset of MCI happens before death ($t_{i1}<t_{i2}$), we observe the onset of MCI ($y_i=1$) at time $t_i=t_{i1}$. Otherwise, we observe death ($y_i=2$) at time $t_i=t_{i2}$. Note that $t_{ij}$'s, $j=1,\ldots,J$, are independent only if $\{\lambda_{ij}\}_j$ are given. In fact, they are marginally dependent as we only know $\xv_i$ and infer $\{r_j,\betav_j\}_j$.

%An alternative view of Weibull racing is from the perspective of discrete choice models \citep{hanemann1984discrete,greene2003econometric,train2009discrete,zhang2017permuted}. The observed event type $y$ is equal to the event whose latent arrival time is the minimum among all the competing events. Distinct from ordinary discrete choice models where the observed event type corresponds to the one associated with the maximum latent utility and the utility values are not identifiable or of interest, the event type $y$ is determined by Weibull racing to minimize the waiting time for any one of the competing events, and this minimum waiting time $t$ is observed and studied. If the event times are unobserved and $y$ is the only dependent variable of interest, the survival analysis with competing events will be reduced to a classification problem. We show the performance of Weibull racing for classification in Section \ref{app:discrete_choice} of the Appendix.  

Given $\xv_i$, $\tau_i$, $a$, $r_j$, and $\betav_j$ for $j=1,\ldots,J$, the survival function $S(t)$ and hazard function $h(t)$ at $t$, $t>\tau_i$, for the observed event time $t_i$ in the Weibull racing survival model are
\begin{align*}
\textstyle{
S(t)=\mbox{Pr}(t_i>t)=\prod\nolimits_{j=1}^J[\exp({\xv_i^\T\betav_{j}})(t^a-\tau_i^a)+1]^{-r_j},~h(t)=\sum\nolimits_{j=1}^J\frac{ar_{j}t^{a-1}}{t^a-\tau_i^a+\exp({-\xv_i^\T\betav_{j}})}.
}%\label{eq:weibull_surv_hazard}
\end{align*}
We can rewrite the latent event time of competing event $j$ as $\textstyle{t_{ij} \sim \mbox{Weibull}_{\tau_i}(a, \exp({\xv_i^\T\betav_j})\tilde\lambda_{j})}$ with $\tilde\lambda_{j}\sim\mbox{Gamma}(r_j, 1)$. 
Particularly, when $\tau_i=0$, $\log t_{ij} = -{\xv_i^\T\betav_j}/{a}+\log \tilde t_{j}$ with $\tilde t_{j}\sim\mbox{Weibull}(a, \tilde\lambda_{j})$. From this perspective, time to each competing event $j$ in Weibull racing is characterized by an accelerated failure time model \citep{kalbfleisch2011statistical} in that the covariates $\xv_i$ accelerate or decelerate the baseline event time $\tilde t_{j}$ by $|\xv_i^\T\betav_j/a|$. Furthermore, %with $\tilde S_{j}(t)=(t^a+1)^{-r_j}$ and $\tilde h_{j}(t)= ar_jt^{a-1} /(1+t^a)$ and 
given covariates $\xv_i$ and $\tau_i$, we can write the survival and hazard functions of $t_{ij}$  for event~$j$ as 
\begin{align}
\textstyle{
S_j(t) = \mbox{Pr}(t_{ij}>t)= [\exp({\xv_i^\T\betav_{j}})(t^a-\tau_i^a)+1]^{-r_j}, ~h_j(t) = \frac{ar_{j}t^{a-1} }{t^a-\tau_i^a+\exp({-\xv_i^\T\betav_{j}})}   }. \label{eq:sjhjlinear}
\end{align}
% \begin{align*}
% &S_j(t_{ij}) =  (\exp({\xv_i^\T\betav_{j}})t_{ij}^a+1)^{-r_j}=\tilde S_{j}\big(t_{ij}\exp({{\xv_i^\T\betav_j}/{a}})\big), \\
% &h_j(t_{ij}) = \frac{ar_{j}t_{ij}^{a-1}}{t_{ij}^a+\exp({-\xv_i^\T\betav_{j}})}= \exp({{\xv_i^\T\betav_j}/{a}}) \tilde h_{j}\big(t_{ij} \exp({{\xv^\T\betav_j}/{a}})\big),
% \end{align*}
% which are alternative representations of an accelerated failure time model. 
Both $S_j$ and $h_j$ are monotonic in each coordinate of $\xv_i$.

In the Weibull racing survival model, the latent time to a competing event is linearly accelerated by covariates. However, a predefined competing event may be further categorized into subtypes, and each subtype has a different dependence on $\xv_i$. Finding the subtypes is important but can be difficult. In medical research where competing diseases are of interest, the nosology of a disease is often subject to human knowledge, diagnostic techniques, and patient population. Multiple diseases of the same phenotype may have been recognized as one disease (competing event), and their distinct etiology and different impacts on patients' survival can be identified only if the difficulties are overcome. For instance, diabetes is categorized into type I and type II; diffuse large B-cell lymphoma has been known to have three subtypes or arguably more, and each subtype is attributed to a different genotype \citep{rosenwald2002use}. In this regard, the Weibull racing survival model is restrictive in that it requires the competing events to be so well defined that their latent times linearly depend on covariates. To circumvent a fine-grained specification of competing events, which is not always feasible, we further develop Weibull delegate racing, assuming that a competing event consists of a potentially infinite number of sub-events, to each of which the latent time is linearly accelerated by the covariates. Decomposing events into sub-events not only improves the model fit but also helps to explore the underlying mechanisms of disease progression.

\subsection{Weibull Delegate Racing}
Based on the idea of Weibull racing that a subject's observed event time is the minimum of latent times to the predefined competing events, we propose Weibull delegate racing (WDR) survival analysis in the nonparametric Bayesian framework, assuming that the time to a competing event is the minimum of latent times to a number of sub-events appertaining to this competing event. In particular, we denote a gamma process defined on the product space $\mathbb{R}_+\times\Omega$ by $G_j\sim\Gamma\mbox{P}(G_{0j},1/c_{0j})$, where  $G_{0j}$ is a finite and continuous base measure over a complete separable metric space $\Omega$, and $1/c_{0j}$ is a positive scale parameter such that $G_j(A)\sim\mbox{Gamma}(G_{0j}(A), 1/c_{0j})$ for each Borel set $A\subset \Omega$. A draw from the gamma process consists of a countably infinite number of non-negatively weighted atoms, expressed as $G_j=\sum_{k=1}^\infty r_{jk} \delta_{\betav_{jk}}$. We define the WDR model as follows.

\begin{definition}%[Weibull delegate racing]
\label{df:WDR}
Suppose subject $i$ survives competing events $\{j\given j= 1,\ldots,J\}$ with  latent event times $\{t_{ij}\}_j$. Given its left truncation time $\tau_i$ and time-invariant covariates~$\xv_i$ and a random draw of a gamma process $G_j\sim\Gamma\emph{\mbox{P}}(G_{0j},1/c_{0j})$, expressed as $G_j=\sum_{k=1}^\infty r_{jk} \delta_{\betav_{jk}}$ for $j=1,\ldots,J$, Weibull delegate racing models subject $i$'s observed event time $t_i$ and event type $y_i$ as
\begin{align*}
& t_i=t_{iy_i},~y_i=\mathop{\mathrm{argmin}}\nolimits_{j\in\{1,\ldots,J\}} t_{ij}, ~ t_{ij} = t_{ij\kappa_{ij}},~ 
\kappa_{ij}=\mathop{\mathrm{argmin}}\nolimits_{k\in\{1,\ldots,\infty\}} t_{ijk},\nonumber\\
& t_{ijk}\sim\emph{\mbox{Weibull}}_{\tau_i}(a, \lambda_{ijk}), 
~ \lambda_{ijk}\sim \emph{\mbox{Gamma}}(r_{jk},\exp({\xv_i^\T \betav_{jk}})). %\label{eq:argmin_argmin}
\end{align*}
\end{definition}
We assume in WDR an infinite number of latent sub-events under a prespecified competing event~$j$, and each sub-event $k$ has a latent event time $t_{ijk}$ for subject $i$. WDR can be considered as a two-phase race among latent sub-events for each subject. In the first phase, within each competing event $j$ there is a race among its countable sub-events $\{k\given k=1,\ldots,\infty\}$, and the winner, namely sub-event $\kappa_{ij}$ whose time $t_{ij\kappa_{ij}} = \min_k t_{ijk}$, will represent event $j$ in the second phase of the race by letting $t_{ij}$ equal to $t_{ij\kappa_{ij}}$. In the second phase, $J$ competing events associated with the event times $\{t_{ij}\}_j$ compete with each other, and eventually, the winner's event time and type are observed as $t_i=\min_j t_{ij}$ and $y_i=\mathop{\mathrm{argmin}}_{j} t_{ij}$. Although WDR assumes a potentially infinite number of sub-events within each competing event $j$, the total weights of these sub-events $G_j=\sum_{k=1}^\infty r_{jk} \delta_{\betav_{jk}}$ is finite by the gamma process. Consequently, the weights of negligible sub-events are parsimoniously and data-adaptively shrunk towards zero \citep{zhou2016augmentable}. Therefore, the nonlinearity of WDR is fulfilled by the racing of a finite number of significant sub-events, and each sub-event time is linearly accelerated by the covariates $\xv_i$.

Intuitively, the nonlinear modeling capacity of WDR is fulfilled by taking the minimum among $J$ minima, which is a two-step nonlinear operation.  In mathematics, the event time $t_{ij}$ and its survival function $S_j$ and hazard function $h_j$ for competing event $j$ are no longer monotonic in $\xv_i$ as shown by Corollary~\ref{cor:Exp_sum_gamma}.   
\begin{cor}\label{cor:Exp_sum_gamma}
Weibull delegate racing is equivalent to
\begin{align*}
\textstyle{
t_i = t_{y_i},~y_i=\mathop{\mathrm{argmin}}\nolimits_{j%\in\{1,\ldots,J\}
} t_{ij},~t_{ij}\sim\emph{\mbox{Weibull}}_{\tau_i} (a, \sum\nolimits_{k=1}^\infty \exp({\xv_i^\T\betav_{jk}})\tilde\lambda_{jk} ),~ \tilde\lambda_{jk} \sim\emph{\mbox{Gamma}}(r_{jk}, 1).
}
\end{align*}
The survival function and the hazard function of $t_{ij}$ for event $j$ are
\begin{align*}
\textstyle{
S_j(t)=\emph{\mbox{Pr}}(t_{ij}>t)=\prod\nolimits_{k=1}^{\infty} [\exp({\xv_i^\T\betav_{jk}})(t^a-\tau_i^a)+1]^{-r_{jk}}, ~
h_j(t)=
\sum\nolimits_{k=1}^{\infty}\frac{ar_{jk}t^{a-1}}{t^a-\tau^a+\exp({-\xv_i^\T\betav_{jk}})}
}.
\end{align*}
\end{cor}
In stark contrast to \eqref{eq:sjhjlinear} of the linear Weibull racing survival model, $S_j$ and $h_j$ of WDR for event~$j$ are non-monotonic in $\xv_i$. The countable gamma-mixed Weibull survival time $t_{ij}$ has enhanced flexibility, relaxing the parametric restrictions of conventional Weibull survival models \citep{commenges1998modelling,sparling2006parametric}. %To study how the hazard function changes over time by 
%Calculating the derivative of $h_j$ with respect to $t$, 
 One can verify that 
%$\frac{dh_j(t_j)}{dt_j}=\sum_k\frac{a(a-1)r_{jk}e^{-\xv^\T\betav_{jk}}t^{a-2}-ar_{jk}t^{2a-2}}{(t^a+e^{-\xv^\T\betav_{jk}})^2}.$ So 
if~$a\leq 1$, $h_j$ is decreasing in~$t$, and otherwise, $h_j$ can be increasing (for a long enough time) then decreasing or
arbitrarily non-monotonic in $t$. 
% with appropriate values of $r_{jk}$'s and $\betav_{jk}$'s.  
%Corollary \ref{cor:Exp_sum_gamma} formulates WDR as a generalization of Weibull racing, where the truncated Weibull distribution of $t_{ij}$ is parameterized by a weighted summation of a countably infinite number of gamma random variables with covariate-dependent weights.
More importantly, the flexibility of WDR is data-adaptively parsimonious. By Corollary \ref{cor:Exp_sum_gamma}, the truncated Weibull distribution of $t_{ij}$ is parameterized by a weighted sum of an infinite number of covariate-dependent functions $\{\exp(\xv_i^\T \betav_{jk})\}_k$ with gamma-distributed weights $\{\tilde \lambda_{jk}\}_k$. 
With the gamma process regularizing $\{r_{jk}\}_k$, the weight $\tilde\lambda_{jk}$ of a negligible sub-event $k$ approaches to zero. Eventually, a relatively small number of sub-events have considerable weights, and their covariate dependence $\{\betav_{jk}\}_k$ may suggest different mechanisms of disease progression of event $j$.
%{\mm The Weibull (delegate) racing survival model will be reduced to the Lomax (delegate) racing \citep{zhang2018nonparametric} if the Weibull shape parameter $a$ is equal to 1. Learning instead of fixing $a$ helps to introduce much flexibility to hazard estimations (see Section~\ref{sec:synthetic} for details), and this is a significant improvement of WDR over the Lomax delegate racing model which only allows decreasing hazard functions. (Will be moved to introduction or literature review)} \red{(Interpretation and characteristics when $a>1$?)} 
The marginal density of $t_i$ given $a$, $\{r_{jk},\betav_{jk}\}_{j,k}$, and $t_i>\tau_i$ is shown by Theorem \ref{thm:t_marginal} in Appendix~\ref{app:proof}.
%WDR is able to model and predict both the event times and types and potential subtypes. As we see, shortly, in Section~\ref{sec:inference}, WDR can deal with different types of censoring and missing event types or times.

\section{Time-Varying Covariates and Bayesian Inference }\label{sec:inference} 
We formulate survival analysis with time-varying covariates 
in Section~\ref{sec:timevarying} 
by assuming piecewise Weibull hazards and transforming the problem into dealing with left truncation and right censoring. Section~\ref{sec:lh} provides data augmentation schemes for Weibull racing in the presence of censoring and missing outcome. 
The schemes also apply to WDR. We interpret WDR as a discrete choice model for classification if all the event times are missing. Section~\ref{sec:mcmc} shows the hierarchical model of WDR that facilitates an efficient %Gibbs sampling %-based 
MCMC 
algorithm with the weights of unnecessary sub-events shrunk towards zero. The inference accommodates various types of censoring and missing outcome imputation by sampling the event times or types as auxiliary variables.
In addition, we propose for big data analysis a maximum a posteriori estimation that admits optimization by stochastic gradient descent; the details are provided in Appendix~\ref{app:mcmc}.   %\textit{Note that you mentioned the MAP here, also in the abstract, introduction, and discussion. However, there is only one sentence in the main body of the manuscript. I think you should at least include a paragraph in the main body, then refer to Appendix C. } 

\subsection{Weibull Delegate Racing with Time-Varying Covariates}\label{sec:timevarying}
We consider time-varying covariates in WDR. Suppose subject $i$ has $V$-dimensional, time-varying covariates $\textstyle{ \bm X_i(t)=(X_{i1}(t), X_{i2}(t),\ldots, X_{iV}(t))^\T}$. Without loss of generality, some covariates $X_{iv}(t)$ can be time-invariant such that $X_{iv}(t)= x_v$ for all $t$. We consider, as is often the case, intermittent covariate measurements. Concretely, subject $i$ enters the study at time $\tau_i^{(0)}$ with covariates $\bm x_i^{(0)}$ and then is observed at times $\tau_i^{(1)},\ldots, \tau_i^{(L_i)}$ with covariates $\bm x_i^{(1)},\ldots,\bm x_i^{(L_i)}$, respectively, for followup visits before a failure event or censoring. A positive  $\tau_i^{(0)}$ represents a left truncation, and $\tau_i^{(0)}, \tau_i^{(1)},\ldots, \tau_i^{(L_i)}$ and the number of updates $L_i$ are subject-specific. We assume constant covariates between followup visits, that is, $\textstyle{\bm X_i(t)= \bm x_i^{(l)}}$ for $\textstyle{t\in \bm[\tau_i^{(l)}, \tau_i^{(l+1)}\bm),~ l=0,1,\ldots,L_i-1}$, and $\textstyle{\bm X_i(t)=\bm x_i^{(L_i)}}$ for $\textstyle{t\geq\tau_i^{(L_i)}}$. 
%Readers may refer to literature in joint modeling of longitudinal and survival data \citep{ibrahim2010basic,wu2012analysis} for more complex formulations of $\bm X_i(t)$.
We use WDR to predict survival probabilities by the (time-varying) covariates, but not vice versa; how the time-varying covariates are influenced by the survival status or disease progression is out of our scope.

We model subjects surviving competing events $j=1,\ldots,J$ with time-varying covariates in the framework of WDR by assuming piecewise parametric hazard functions \citep{sparling2006parametric}. Specifically, for subject $i$, we assume a  $\textstyle{\mbox{Weibull}(a, \sum\nolimits_{j,k}\lambda_{ijk}^{(l)})}$ hazard in the $l$th interval $\bm[\tau_i^{(l)}, \tau_i^{(l+1)}\bm)$ for $l=0,1,\ldots,L_i-1$ and $\bm[\tau_i^{(l)}, \infty\bm)$ for $l=L_i$,  
  namely $\textstyle{h^{(l)}(t) =
  %h_{\tau^{(l)}}(t\given a, \sum_{j,k}\lambda_{jk}^{(l)}) =
  a\sum_{j,k}\lambda_{ijk}^{(l)}t^{a-1} }$ where $\textstyle{\lambda_{ijk}^{(l)}\sim \mbox{Gamma}(r_{jk}, \exp({\bm x_i^{(l)\T} \betav_{jk}}) )}$. Consequently, the survival function of subject $i$ at time $t$, $t>\tau_i^{(L_i)}$, after the last covariate measurement is 
\begin{align}
\textstyle{
S(t\given \{\tau_i^{(l)}, \lambda_{ijk}^{(l)}\}_{l=0}^{L_i})=S_{\tau_i^{(L_i)}}(t\given a, \sum\nolimits_{j,k}\lambda_{ijk}^{(L_i)})\prod\nolimits_{l=0}^{L_i-1} S_{\tau_i^{(l)}}(\tau_i^{(l+1)}\given a, \sum\nolimits_{j,k}\lambda_{ijk}^{(l)})
} \label{eq:tv_survf}   
\end{align}
and the density function is 
\begin{align}
\textstyle{
f(t\given \{\tau_i^{(l)}, \lambda_{ijk}^{(l)}\}_{l=0}^{L_i})= f_{\tau_i^{(L_i)}}(t\given a, \sum\nolimits_{j,k}\lambda_{ijk}^{(L_i)})\prod\nolimits_{l=0}^{L_i-1} S_{\tau_i^{(l)}}(\tau_i^{(l+1)}\given a, \sum\nolimits_{j,k}\lambda_{ijk}^{(l)})
}\label{eq:tv_pdf}  
\end{align}
where $\textstyle{S_{\tau_i^{(l)}}}$ and $\textstyle{f_{\tau_i^{(l)}}}$ are the survival and density functions, respectively, of a left-truncated Weibull distribution, as shown in Definition~\ref{def:tweibull} and Corollary~\ref{cor:tweibull}.

Equations~\eqref{eq:tv_survf} and~\eqref{eq:tv_pdf} solve the problem of time-varying covariates by dealing with left truncation and right censoring. By Equation \eqref{eq:tv_pdf}, the likelihood of subject $i$ having an event at time $t$ with covariate values $\xv_i^{(0)},\ldots,\xv_i^{(L_i)}$ measured at $\tau_i^{(0)},\ldots,\tau_i^{(L_i)}$ is equivalent to the joint likelihood of $L_i+1$ pseudo subjects: One pseudo subject has the left truncation time $\tau_i^{(L_i)}$ and the event time $t$ with fixed covariates $\xv_i^{(L_i)}$ and the other $L_i$ have the left truncation time $\tau_i^{(l)}$ and the right censoring time $\tau_i^{(l+1)}$ with fixed covariates $\xv_i^{(l)}$ for $l=0,\ldots,L_i-1$. Analogously, Equation~\eqref{eq:tv_survf} implies that a right censoring of the subject with time-varying covariates is equivalent to $L_i+1$ pseudo subjects with a left truncation and a right censoring.
The derivation of \eqref{eq:tv_survf} and~\eqref{eq:tv_pdf} is deferred to Appendix~\ref{app:proof}. Hereby, WDR handles time-varying covariates by equivalently modeling left truncation and right censoring of pseudo subjects with time-invariant covariates. Since left truncation is intrinsically accommodated by WDR, we focus on censoring and Bayesian inference in the presence of time-invariant covariates in the remainder of this section.

\subsection{Censoring, Missing Outcomes,  and WDR Classification}\label{sec:lh}
%It is rarely the case that both $y_i$ and $t_i$ are observed for each subject $i$, and one often needs to deal with censoring and missing data. 
We denote  by $\Psi$ a subject-specific censoring condition on a left-truncated event time $t_i\sim\mbox{Weibull}_{\tau_i}(a,\lambda)$. Specifically, $\Psi$ can be $\bm (T_{r.c.},\infty\bm )$ indicating a right censoring at  $T_{r.c.}>\tau_i$, $\bm (\tau_i, T_{l.c.}\bm )$ a left censoring at  $T_{l.c.}>\tau_i$, or $\bm (T_{1}, T_{2}\bm )$ an interval censoring with $T_2>T_1>\tau_i$. %The censoring condition $\Psi$ can differ among subjects. 
The density of $t_i$ with the constraint $t_i\in\Psi$ is $\textstyle{f_{\tau_i,\Psi}(t\given a, \lambda) = f_{\tau_i}(t\given a, \lambda)/\int_{\Psi}f_{\tau_i}(s\given a, \lambda)ds}$.

If $y_i$ or $t_i$ is missing or there exists censoring, we have two situations where auxiliary variables can be introduced to achieve a factorized likelihood of the Weibull racing model: (a) If we only observe $y_i$ (or $t_i$) without censoring, then we can draw $t_i$ (or $y_i$) %\sim\mbox{Exp}\left(\sum_{j=1}^S \lambda_{j}\right)$ 
 by~\eqref{eq:Cat} % shown in \eqref{eq:Cat}) 
as an auxiliary variable, leading to the factorized likelihood of $t_i$ and $y_i$ as in~\eqref{eq:joint_Weibull}. (b) If we do not observe $t_i$ but know $t_i\in\Psi$ with probability $\mbox{Pr}(t_i\in\Psi\given a, \{\lambda_{ij}\}_{j})=\int_{\Psi}f_{\tau_i}(s\given a, \sum_j\lambda_{ij}) ds$, then we draw $t_i$ by the density $f_{\tau_i,\Psi}(t\given a, \sum_j\lambda_{ij})$, resulting in the likelihood
\begin{align*}\vspace{-1mm}
\textstyle{
p(t_i, t_i\in \Psi\given a, \sum\nolimits_j \lambda_{ij},\tau_i)=f_{\tau_i,\Psi} (t_i\given a, \sum\nolimits_j\lambda_{ij} )\mbox{Pr} (t_i\in\Psi\given a, \sum\nolimits_j\lambda_{ij} )=f_{\tau_i}(t\given a, \sum_j\lambda_{ij}).
}
\end{align*}
%= a\left(\sum\nolimits_j\lambda_j\right)t^{a-1} e^{-(t^a-\tau^a)\sum\nolimits_j\lambda_j}.}
With $y_i$ that can be drawn by \eqref{eq:Cat} if missing, the likelihood $\textstyle{p(y_i,t_i,t_i\in\Psi\given a, \{\lambda_{ij}\}_{j},\tau_i)}$ becomes the same as the right hand side of \eqref{eq:joint_Weibull}. 
Therefore, under different censoring conditions or with missing event times or types, sampling  $t_i$ and/or $y_i$ gives Weibull racing the same factorized likelihood as in \eqref{eq:joint_Weibull}. 

\looseness -1 If all the event times are unobserved, that is, if $y_i$ is the only dependent variable, WDR survival analysis will be reduced to a classification model. Specifically, with latent times $\textstyle{t_{ijk}}$'s and $\textstyle{t_{ij}=\min_k t_{ijk}}$, the category $\textstyle{y_i=\mathop{\mathrm{argmin}}_{j} t_{ij}}$. This provides an alternative view of Weibull (delegate) racing from the perspective of discrete choice models \citep{hanemann1984discrete,greene2003econometric,train2009discrete,zhang2017permuted}. 
Specifically, the observed event type (or category) $y_i$ is equal to the one whose latent arrival time is earlier than all the others. Distinct from ordinary discrete choice models where $y_i$ corresponds to the category that brings the maximum latent utility and the utility values are not identifiable or of interest, in Weibull (delegate) racing, the event type $y_i$ is determined to minimize the waiting time for the first arrival, and the minimum waiting time $t_i$ can be either observed and studied in the survival model, or missing but imputed as an auxiliary variable in both the survival and classification models. This finding unifies Bayesian inference of the WDR classification and survival analysis with missing event times. See Appendix~\ref{app:discrete_choice} for more details on WDR classification.

\subsection{Hierarchical Model of WDR and MCMC}\label{sec:mcmc}
For the implementation of WDR, we follow \citet{zhou2015negative} to truncate the number of atoms of the gamma processes at $K$, which is a sufficiently large integer, by choosing a finite and discrete base measure $\textstyle{ G_{0j}=\sum_{k=1}^K \gamma_{0j}\delta_{\betav_{jk}}/K}$, $j=1,\ldots,J$. In this way, we allow up to $K$ latent sub-events within each competing event $j$.
To obtain a factorized likelihood of WDR, we first follow Section~\ref{sec:lh} to sample $y_i$ and $t_i$ if missing or censored. 
Next, WDR requires another auxiliary variable $\textstyle{ \kappa_{iy_i}\sim\mbox{Categorical}(\lambda_{iy_i1}/\sum_{k=1}^K \lambda_{iy_ik},\ldots,{\lambda_{iy_iK}}/{\sum_{k=1}^K \lambda_{iy_ik}})
}$, which is the label of the winning sub-event within competing event $y_i$ in the first phase of the race. Consequently,  the factorized likelihood of subject $i$ becomes
\begin{align*}
\textstyle{ p(t_i,y_i,\kappa_{iy_i}\given a, \{\lambda_{ijk}\}_{jk},\tau_i)%&=p(t_i,y_i\given a, \{\lambda_{ijk}\}_{j,k},\tau_i)p(\kappa_{iy_i}\given y_i, \{\lambda_{ijk}\}_{j,k})
%\\
=\lambda_{iy_i\kappa_{iy_i}} a t_i^{a-1} \exp({-(t_i^a-\tau_i^a) \sum\nolimits_{j,k}\lambda_{ijk}}).
}
\end{align*}
%Given subject $i$ with time-invariant covariates $\xv_i$, left truncation time $\tau$, event time $t_i$, and event type $y_i$ under risks of competing events $j=1,\ldots,J$, the hierarchical model of WDR is written as
We write the hierarchical model of WDR as
\begin{align}
&t_i= t_{iy_i},~y_i=\mathop{\mathrm{argmin}}\nolimits_{j\in\{1,\ldots,J\}} t_{ij},~ 
t_{ij}= t_{ij\kappa_{ij}},~\kappa_{ij}= \mathop{\mathrm{argmin}}\nolimits_{k\in\{1,\ldots,K\}} t_{ijk},
 \nonumber\\ %~k=0,1,\cdots,K, \nonumber\\
&t_{ijk} \sim \mbox{Weibull}_\tau(a, \lambda_{ijk}),~\lambda_{ijk} \sim \mbox{Gamma}(r_{jk}, \exp({\xv_{i}^\T\betav_{jk}})), %~k=1,\cdots, K,
\nonumber\\
&\betav_{jk}\sim \prod\nolimits_{v=1}^{V} \mbox{N}(0,\alpha_{vjk}^{-1}), ~
\alpha_{vjk}\sim \mbox{Gamma}(a_0,1/b_0),
 ~r_{jk}\sim \mbox{Gamma}(\gamma_{0j}/K, 1/c_{0j})\notag,
\end{align}
where  $i=1,\cdots, n$, $j=1,\cdots,J$, and $k=1,\cdots, K$. We choose non-informative hyperpriors $\gamma_{0j}\sim \mbox{Gamma}(d_0,1/e_0)$ and $c_{0j}\sim \mbox{Gamma}(d_1,1/e_1)$, with % and let 
$d_0=e_0=d_1=e_1=0.01$.
%to reduce the impact of the hyperpriors. 
%The Gamma prior on the precision parameter $\alpha$ for each element of $\betav$ is imposed to penalize large absolute values of $\betav$ to avoid overfit \citep{tipping2001sparse}. 

With the factorized likelihood and additional auxiliary variables from the P\'olya gamma \citep{polson2013bayesian} and Chinese restaurant table \citep{zhou2015negative} distributions, our MCMC algorithm updates all the parameters by Gibbs sampling except the Weibull shape parameter $a$. A possible solution is to update $a$ by Metropolis-Hastings, but the proposal distribution has to be tuned. Considering the fact that the full conditional distribution of $a$ is unimodal when left truncation times are 0 (see Step 4 of the MCMC algorithm in Appendix~\ref{app:mcmc}), we alternatively use slice sampling \citep{damlen1999gibbs,neal2003slice} that is more efficient and less sensitive to tuning parameters. To remove unnecessary modeling capacity, the gamma processes regulate the weights of the sub-events by pushing some of $r_{jk}$'s towards zero. In addition, we propose a scheme (Step 9 of the MCMC algorithm) based on latent count allocations to actively prune negligible latent sub-events during MCMC iterations and thus accelerate the convergence of the algorithm.  
Appendix~\ref{app:mcmc} shows the complete MCMC algorithm for WDR survival analysis including censoring and missing outcome imputation.

WDR assumes that competing events are conditionally independent given covariates. To relax this assumption, one can incorporate random effects in WDR, such as a random intercept concatenated to $\betav$ for each subject. In this way, flexible dependence among competing events and/or subjects is allowed, even if the covariates are given. Notably, incorporating random effects in WDR does not undermine the Gaussian conjugacy of $\betav_{jk}$'s using our data augmentation scheme and MCMC if Gaussian random effects are used.
Moreover, the Gaussian conjugacy admits easy and various regularizations on $\betav_{jk}$'s by imposing a prior distribution. We use  Gaussian-inverse-gamma priors, but other priors, such as the Laplace prior \citep{park2008bayesian} and the horseshoe prior \citep{carvalho2010horseshoe,johndrow2020scalable} also apply. We defer mixed-effects WDR and other priors on $\betav_{jk}$'s to future work.

\section{Synthetic Data Analysis and Model Comparison}\label{sec:experiments}

We validate WDR survival analysis on synthetic data by showing its parsimonious nonlinearity and comparing its prediction accuracy with benchmark models. In Section~\ref{sec:benchmark}, we introduce the synthetic data and the quantification of prediction accuracy. In Section~\ref{sec:synthetic}, we illustrate the estimation of nonlinear covariate effects by WDR. In Section~\ref{sec:synthetic_tv}, we introduce the benchmark models and compare the models on the data with competing events, left truncation, and time-varying covariates. Technical details and supplementary results are deferred to Appendix~\ref{app:additional_result}.
We show that WDR is an attractive approach for its interpretability, versatility, and prediction accuracy. 
%we compare the models on synthetic data with constant covariates and no left truncation and illustrate the parsimonious nonlinearity of WDR

% 
\subsection{Data and Model Evaluation}\label{sec:benchmark}

\begin{table}[t]
\centering
\renewcommand{\arraystretch}{1} % Default value: 1
\makebox[\linewidth]{
\resizebox{\linewidth}{!}{
\begin{tabular}{lll}%\vspace{-6.5mm}
\toprule 
{Data 1} & {Data 2} & {Data 3}\\
\midrule
%$\xv_i\in \mathbb{R}^{10},~\xv_i\sim \mbox{N}(\bm 0, \mathrm{\bf I})$  & $\xv_i\in \mathbb{R}^{10},~\xv_i\sim \mbox{N}(\bm 0, \mathrm{\bf I})$ & $\xv_i\in \mathbb{R}^{10},~\xv_i\sim \mbox{U}(-\bm{1}, \bm 1)$  \\
$t_{i1}\sim\mbox{Weibull}(0.8, \exp(\xv_i^\T\bv_1))$ & $t_{i1}\sim\mbox{Weibull}(3,|\sinh(\xv_i^\T\bv_1)|)$  & $t_{i1} \sim \mbox{Weibull}(1, \exp({ (\xv_i^\T\bv_1)^2}))$ \\
$t_{i2}\sim\mbox{Weibull}(0.8, \exp({\xv_i^\T\bv_2}))$ & $t_{i2}\sim\mbox{Weibull}(3, \cosh(\xv_i^\T\bv_2))$  & $t_{i2} \sim \mbox{Weibull}(1, \exp({ (\xv_i^\T\bv_2)^2}))$  \\
 $t_i=\min(t_{i1},t_{i2},T_{r.c.}=2)$  &$t_i=\min(t_{i1},t_{i2},T_{r.c.}=1.2)$ & $t_i=\min(t_{i1},t_{i2},T_{r.c.}=1.2)$\\
 Left truncation time: $0.05$ & Left truncation time: $0.3$ & Left truncation time: $0.05$ \\
Covariates update time: $0.15, 0.25$ &Covariates update time: $0.5, 0.7$ & Covariates update time: $0.1, 0.15$\\
Evaluation time: $.1,.3,.5,.7,.9$ &Evaluation time: $.4,.55,.7,.85,1$ &Evaluation time: $0.1,0.3,0.5,0.7,0.9$ \\
\midrule
{Data 4} &{Data 5} & {Data 6}\\
\midrule
%$\xv_i\in \mathbb{R}^{10},~\xv_i\sim \mbox{U}(-\bm{1}, \bm 1)$  & $\xv_i\in \mathbb{R}^{10},~\xv_i\sim \mbox{N}(\bm 0, \mathrm{\bf I})$ & $\xv_i\in \mathbb{R}^{10},~\xv_i\sim \mbox{N}(\bm 0, \mathrm{\bf I})$  \\
$t_{i1}\sim\mbox{logNormal}(\mu_{i1}, 0.2^2)$ & $t_{i1}\sim\mbox{logNormal}(\mu_{i1}, 0.2^2)$ & $t_{i1}\sim\mbox{logNormal}(\mu_{i1}, 0.05^2)$ \\
$\textstyle{\mu_{i1}=\frac{\xv_i^\T\bv_3\exp(\xv_i^\T\bv_1)- \xv_i^\T\bv_4\exp(\xv_i^\T\bv_2)}{\exp(\xv_i^\T\bv_1)+\exp(\xv_i^\T\bv_2)}}$  & $\mbox{logit}(\mu_{i1}) = \bv_1^\T \xv_i\xv_i^\T\bv_3 - \bv_2^\T \xv_i\xv_i^\T\bv_4 $ & $\mbox{logit}(\mu_{i1})=\cosh(\xv_i^\T \bv_1) - \cosh(\xv_i^\T \bv_2)$ \\
$t_{i2}\sim\mbox{logNormal}(\mu_{i2}, 0.2^2)$ &$t_{i2}\sim\mbox{logNormal}(\mu_{i2}, 0.2^2)$  & $t_{i2}\sim\mbox{logNormal}(\mu_{i2}, 0.05^2)$\\
$\textstyle{\mu_{i2}=\frac{\xv_i^\T\bv_4\exp(\xv_i^\T\bv_2)- \xv_i^\T\bv_3\exp(\xv_i^\T\bv_1)}{\exp(\xv_i^\T\bv_1)+\exp(\xv_i^\T\bv_2)}}$ & $\mbox{logit}(\mu_{i2}) = \bv_2^\T \xv_i\xv_i^\T\bv_4 - \bv_1^\T \xv_i\xv_i^\T\bv_3$ & $\mbox{logit}(\mu_{i2})=\cosh(\xv_i^\T \bv_2) - \cosh(\xv_i^\T \bv_1)$\\
$t_i=\min(t_{i1},t_{i2},T_{r.c.}=1)$  &$t_i=\min(t_{i1},t_{i2},T_{r.c.}=1.4)$ & $t_i=\min(t_{i1},t_{i2},T_{r.c.}=1.6)$\\
Left truncation time: $0.2$ & Left truncation time: $0.8$ & Left truncation time: $0.8$ \\ 
Covariates update time: $0.3, 0.4$ &Covariates update time: $0.9, 1$ & Covariates update time: $0.9, 1$\\
Evaluation time: $.3,.45,.6,.75,.9$ &Evaluation time: $0.9,1,1.1,1.2,1.3$ &Evaluation time: $1,1.1,1.2,1.3,1.4$ \\
 \bottomrule
\end{tabular}%\vspace{-0.5mm}
%\end{wraptable} 
}}
\caption{Data synthesis.}
\label{tab:syntheticnew}\vspace{-3.5mm}
\end{table}

We simulate six data sets with $J=2$ competing events for each. The data generating process is provided in Table \ref{tab:syntheticnew}, where each subject $i$ has covariates $\xv_i$ from a Gaussian or uniform distribution. 
%We denote $t_{ij}$ following a Weibull distribution as the latent time to event $j$ of subject $i$ and $t_i$ as the observed survival time.
%We let $t_{ij}$ following a Weibull distribution denote the latent time to event $j$ of subject $i$ and $t_i$ is the observed survival time. 
Times to competing events follow Weibull distributions in  data~1 to~3  and log-normal distributions in data~4 to~6. For data~1, the covariates have linear effects on survival times.
For data~2 to~6, we use nonlinear functions of $\xv_i$ as the distribution parameters. Specifically, we use the hyperbolic sine and cosine functions in data 2 and 6, quadratic functions in data 3 and 5, and weighted linear functions with the weights being covariate dependent in data 4.
Subject $i$ is right censored at time $T_{r.c.}$ if the latent event times $t_{i1}$ and $t_{i2}$ are greater than $T_{r.c.}$. 
%The right censoring time $T_{r.c.}$ is equal to $1.6$, $1.3$, and $0.6$ for data 1, 2, and 3, respectively.
The observed event type $y_i=\mathop{\mathrm{argmin}}_j t_{ij}$ if $t_i < T_{r.c.}$, and $y_i=0$ is used to denote right censoring if $t_i=T_{r.c.}$. 
In Section~\ref{sec:synthetic}, we simulate time-invariant covariates for each subject and let the left truncation equal to $0$. 
In  Section~\ref{sec:synthetic_tv}, we consider left truncation and time-varying covariates, and the covariate update times and left truncation times are given in Table~\ref{tab:syntheticnew}. The distribution of $\xv_i$, the values of $\bv$'s, and the simulation of time-varying covariates are deferred to Tables~\ref{tab:synthetic}  and~\ref{tab:xbet} and Algorithm~\ref{alg:tvsimulation}, respectively, in Appendix~\ref{sec:tvsimulation}. 

\looseness -1 We use the Brier score \citep{gerds2008performance,steyerberg2010assessing} to quantify the prediction accuracy. 
%Specifically, the cumulative incidence function of subject $i$ for event $j$ at time $t$ is defined as     $\mbox{CIF}_j(i, t) =\mbox{Pr}(t_i\leq t, y_i=j )$  \citep{fine1999proportional,kalbfleisch2011statistical,crowder2001classical}.
Specifically, with subjects $i=1,\ldots,n$, the Brier score (BS) for event $j$ at time $t$ is 
$
\mbox{BS}_j(t)=\frac{1}{n}\sum\nolimits_{i=1}^{n}\left[\bm 1(t_i\leq t, y_i=j)-\mbox{Pr}(t_i\leq t, y_i=j)\right]^2.
$
The Brier score represents the mean squared distances between the observed survival status and the estimated cumulative incidence function (CIF) that is equal to $\mbox{Pr}(t_i\leq t, y_i=j)$. Brier scores are between $0$ and $1$, and a smaller value indicates a more accurate prediction. 
In Section~\ref{sec:synthetic_tv}, we evaluate Brier scores at five time points whose range covers roughly the middle 80\% of the survival times in each data. The five evaluation time points are given in Table~\ref{tab:syntheticnew}. In all the experiments,
we set $K=10$ in WDR to allow up to 10 latent sub-events for each competing event, run 20,000 MCMC iterations, and collect the last 2,000 for posterior estimations.    %We derive the CIF for WDR in Appendix~\ref{app:mcmc}.

\subsection{
Parsimonious Nonlinearity of WDR
%Synthetic data analysis with constant covariates and no left truncation
}\label{sec:synthetic}

\begin{figure}[!t]\vspace{0mm}
 \centering
\begin{subfigure}[t]{0.32\textwidth}
 \centering
\includegraphics[width=1\linewidth]{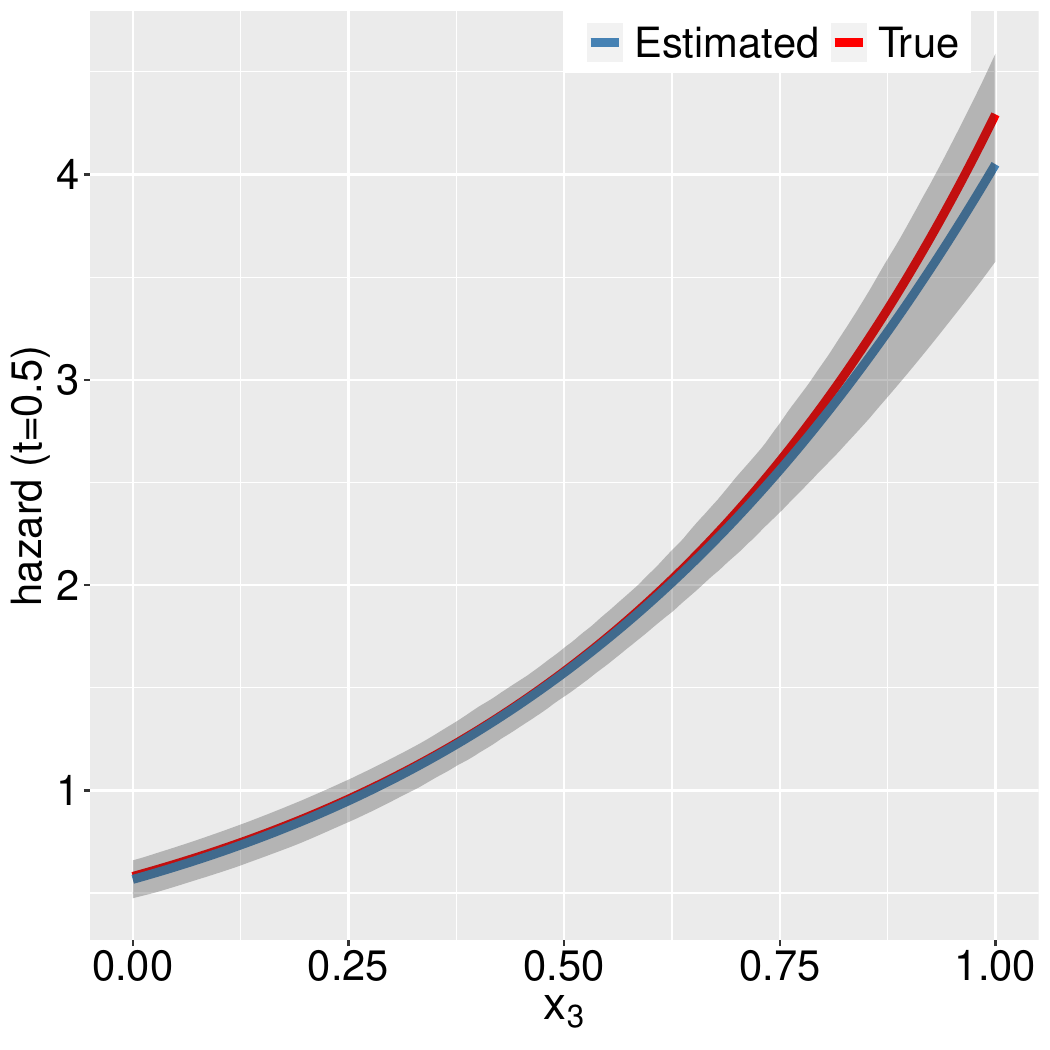}\vspace{-2.5mm}%\label{linear_r}
 \caption{Sythetic data 1.}\vspace{-1mm}
 \end{subfigure}\hfil%
\begin{subfigure}[t]{0.32\textwidth}
 \centering
\includegraphics[width=1\linewidth]{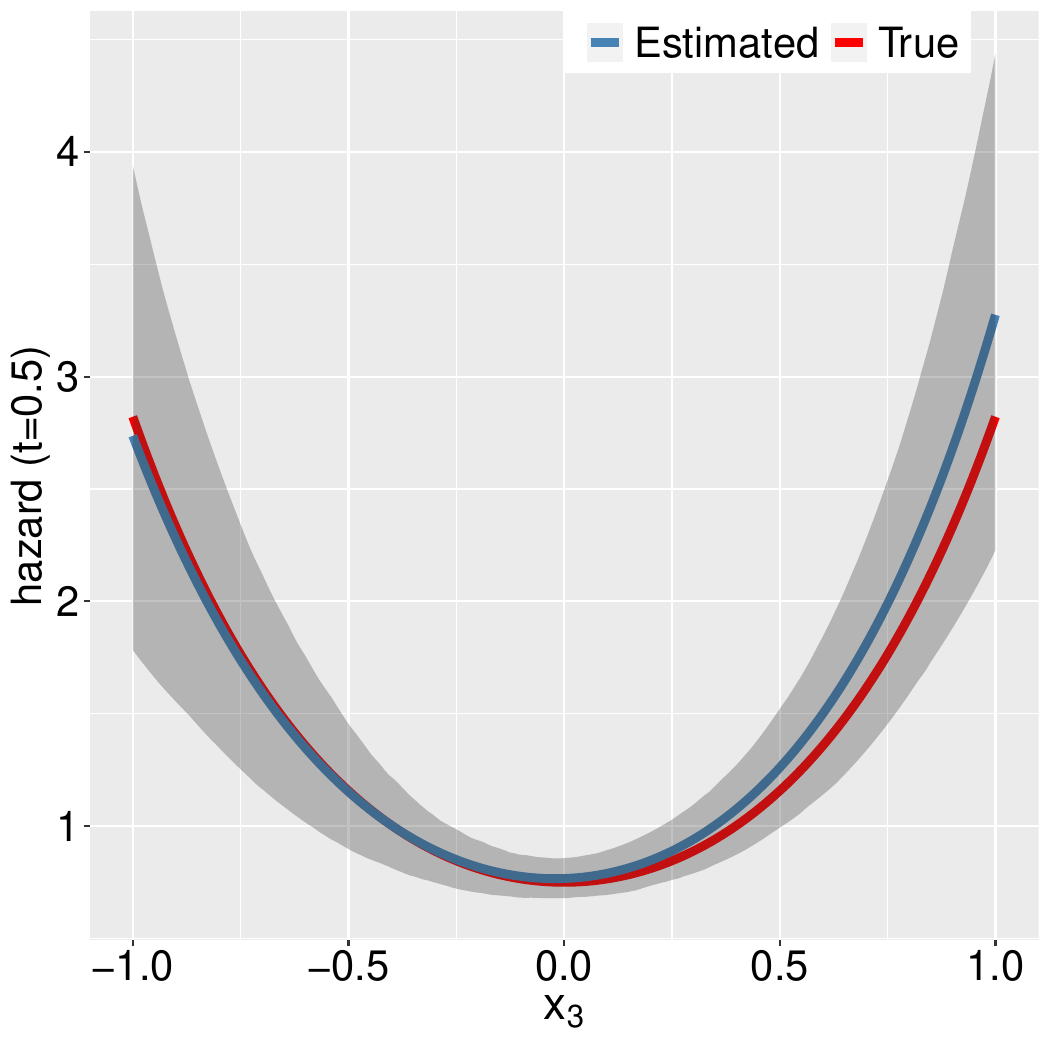}\vspace{-2.5mm}%\label{linear_r}
 \caption{Sythetic data 2.}\vspace{-1mm}
 \end{subfigure}\hfil%
 \begin{subfigure}[t]{0.32\textwidth}
 \centering
\includegraphics[width=1\linewidth]{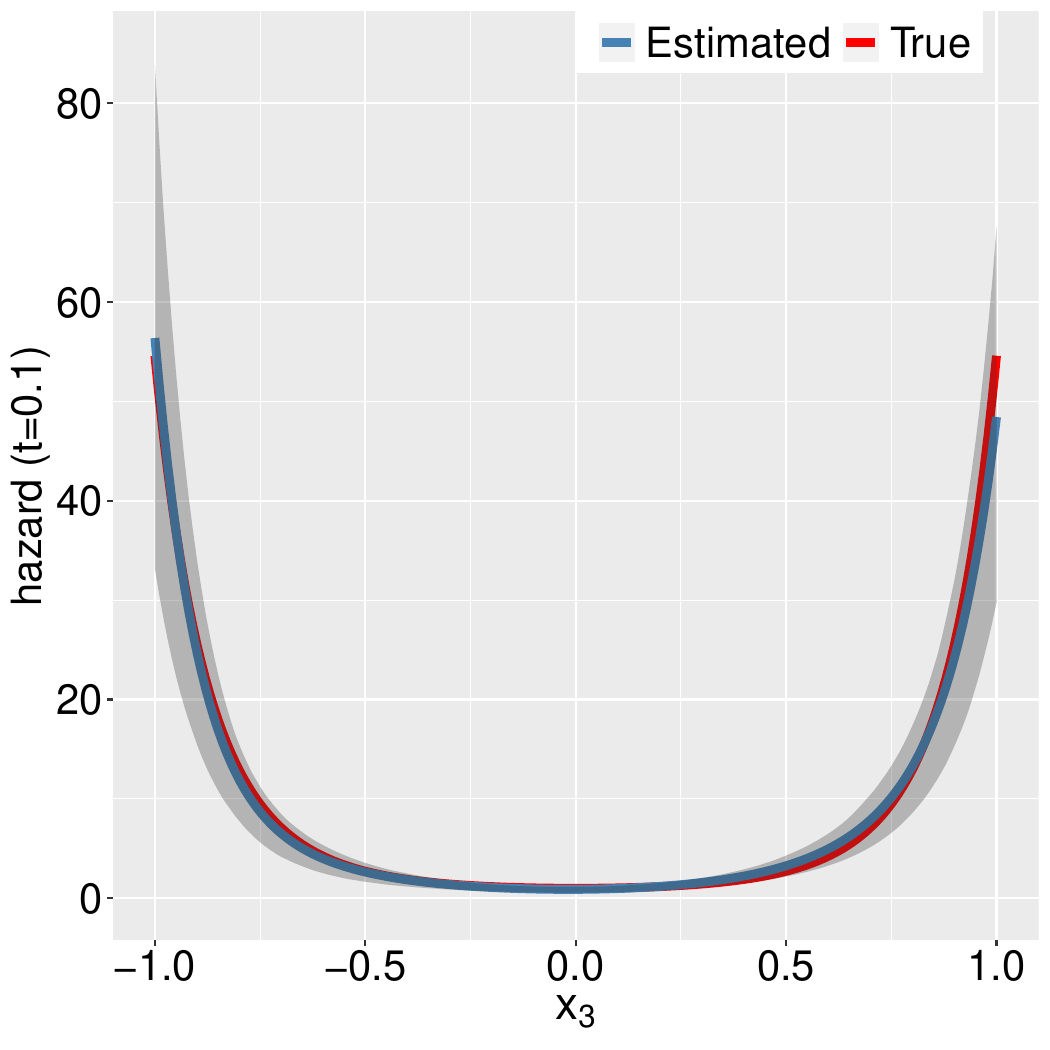}\vspace{-2.5mm}%\label{linear_r}
 \caption{Sythetic data 3.}\vspace{-1mm}
 \end{subfigure}\hfil%
\\
 \begin{subfigure}[t]{0.32\textwidth}
 \centering
\includegraphics[width=1\linewidth]{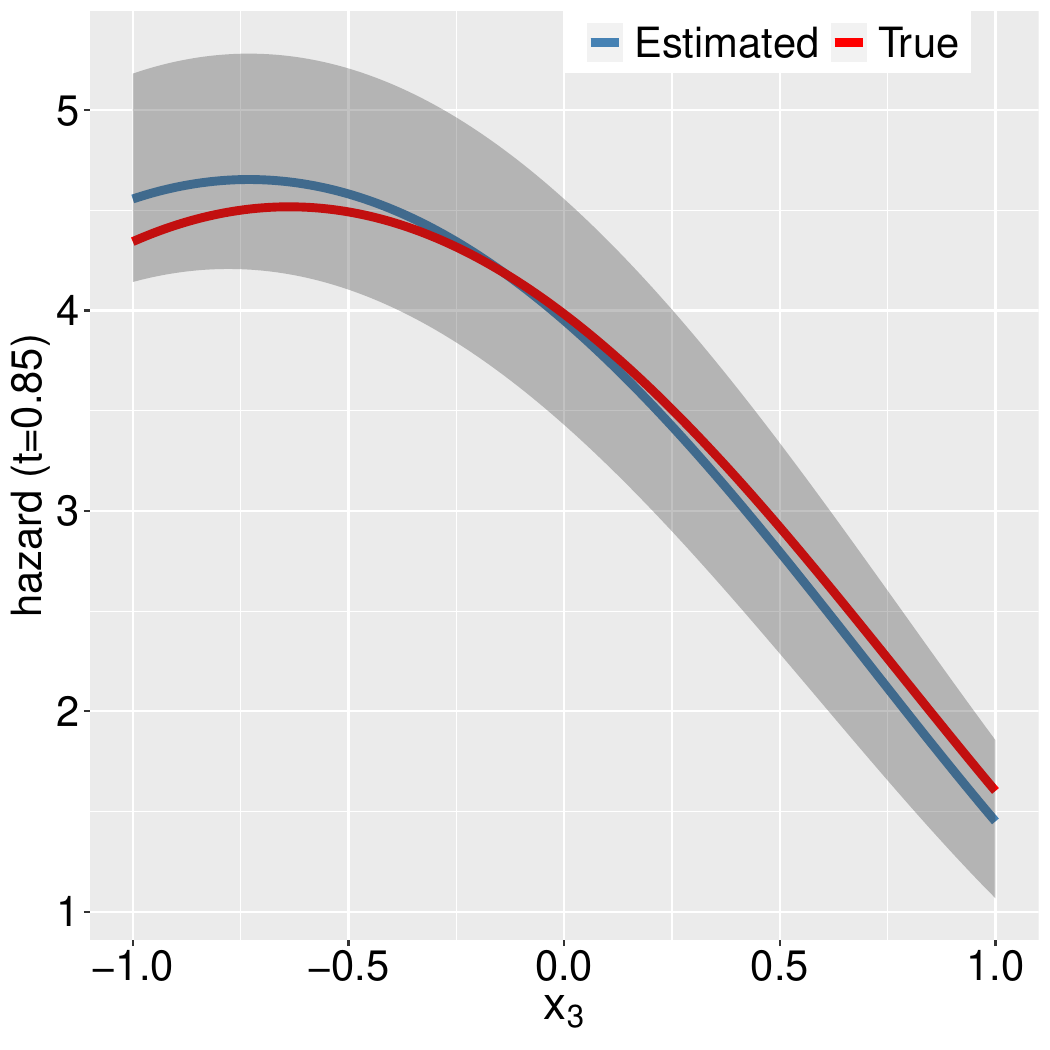}\vspace{-2.5mm}%\label{linear_r}
 \caption{Sythetic data 4.}\vspace{-1mm}
 \end{subfigure}\hfil%
 \begin{subfigure}[t]{0.32\textwidth}
 \centering
\includegraphics[width=1\linewidth]{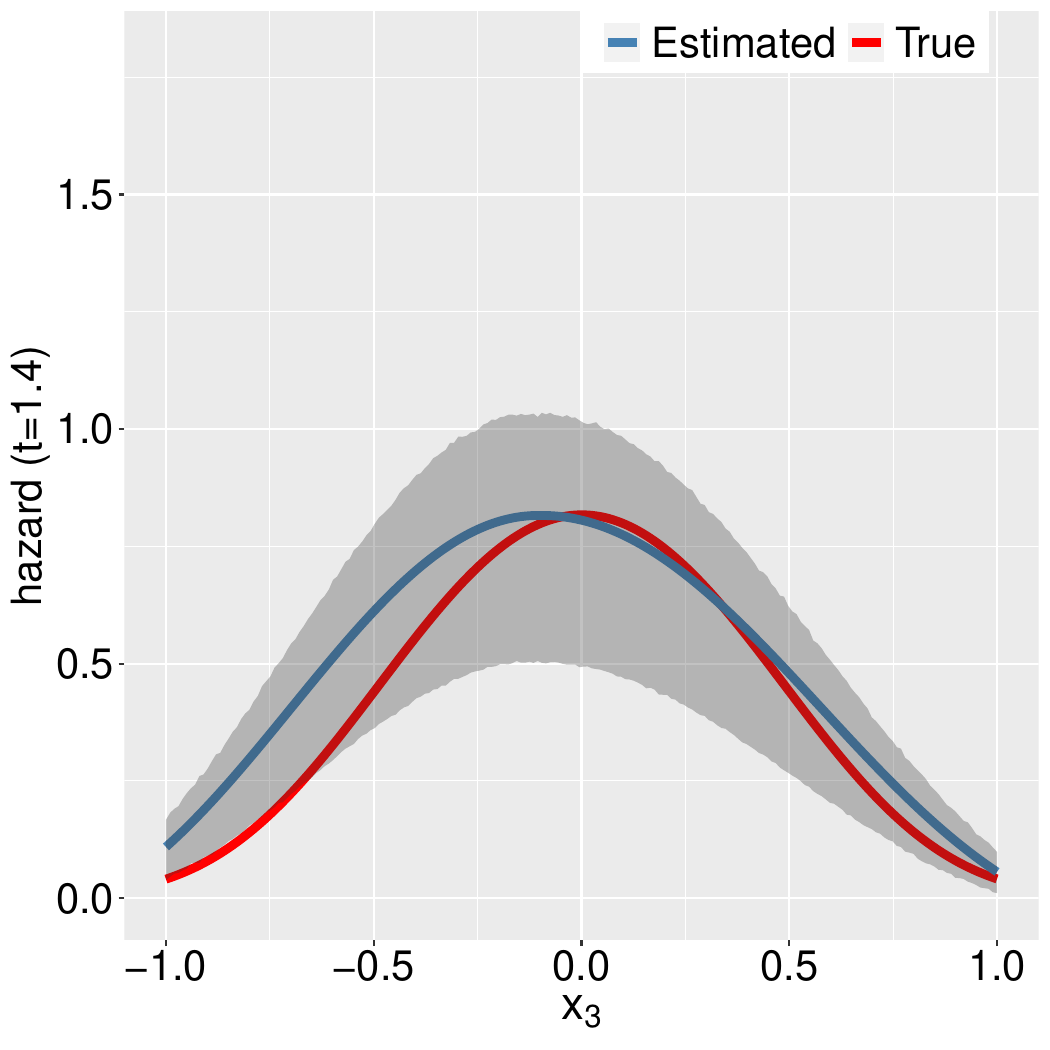}\vspace{-2.5mm}%\label{linear_r}
 \caption{Sythetic data 5.}\vspace{-1mm}
 \end{subfigure}\hfil%
 \begin{subfigure}[t]{0.32\textwidth}
 \centering
\includegraphics[width=1\linewidth]{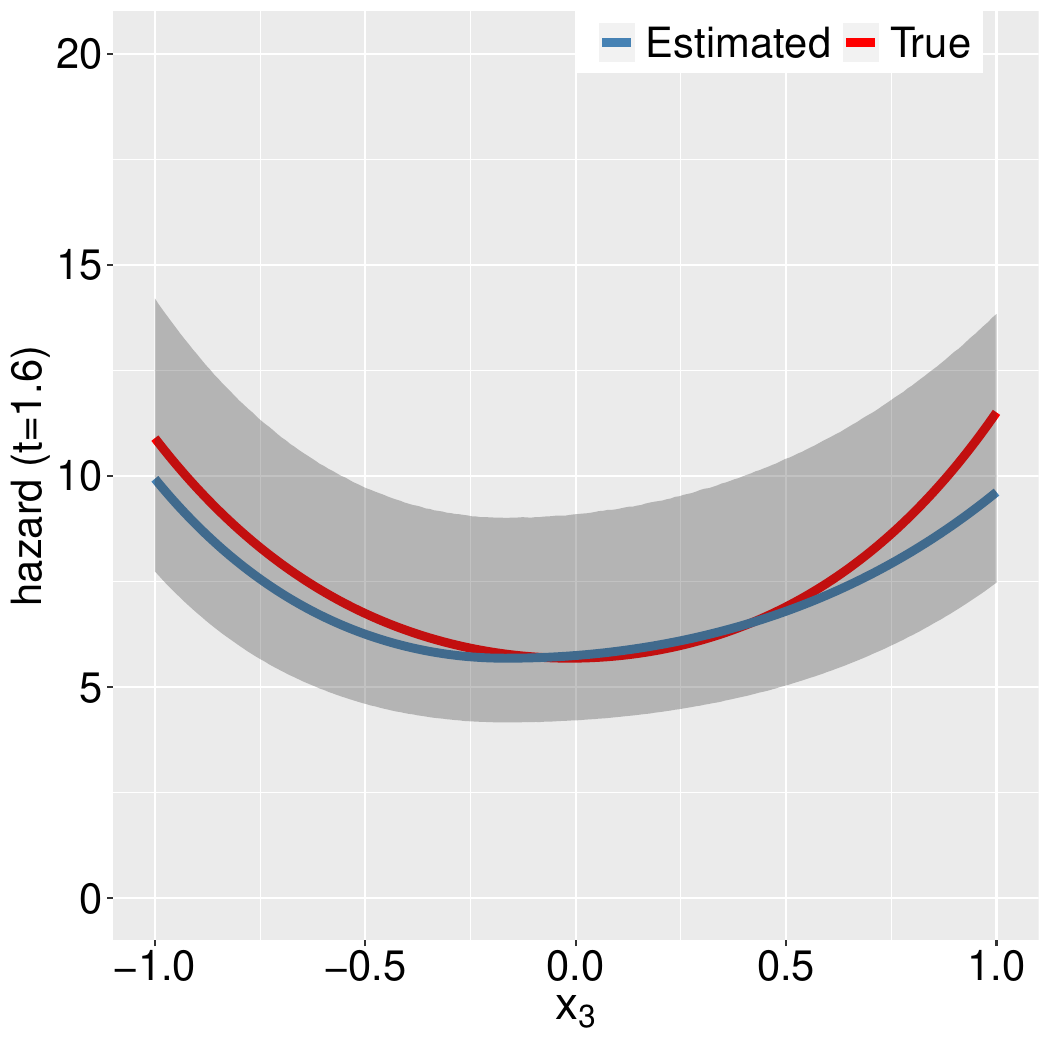}\vspace{-2.5mm}%\label{linear_r}
 \caption{Sythetic data 6.}\vspace{-1mm}
 \end{subfigure}\hfil%
 \vspace{-2mm}
\caption{WDR estimations (with 95\% credible intervals) of hazards of event $2$ against $x_3$.}\label{fig:hazardx} 
\end{figure}

We first illustrate the performance of WDR in estimating nonlinear covariate effects. Specifically, we simulate time-invariant covariates $\xv_i=(x_{i1},x_{i1},x_{i3})^\T$ from a uniform distribution (see Table~\ref{tab:synthetic} in the Appendix) and follow the data generating process in Table~\ref{tab:syntheticnew} to synthesize the six data sets without left truncation. 
We predict the hazards of event 2 for subjects with $x_1=0$, $x_2=0.5$ and $x_3\in\bm[0,1\bm]$ in data 1 at $t=0.5$, with $x_1=0$, $x_2=0$ and $x_3\in\bm[-1,1\bm]$ in data 2, 4, 5, and 6 at $t=0.5$, $0.85$, $1.4$, and $1.6$, respectively, and with $x_1=0.5$, $x_2=0.5$ and $x_3\in\bm[-1,1\bm]$ in data 3 at $t=0.1$.
The true and estimated hazards along with the $95\%$ credible intervals are shown in Figure~\ref{fig:hazardx}. We see WDR successfully recovers the log-linear (data 1) and non-monotonic (data 2 to 6) covariate effects on the hazards.

\begin{figure}[!t]\vspace{0mm}
 \centering
 \begin{subfigure}[t]{0.32\textwidth}
 \centering
\includegraphics[width=1\linewidth]{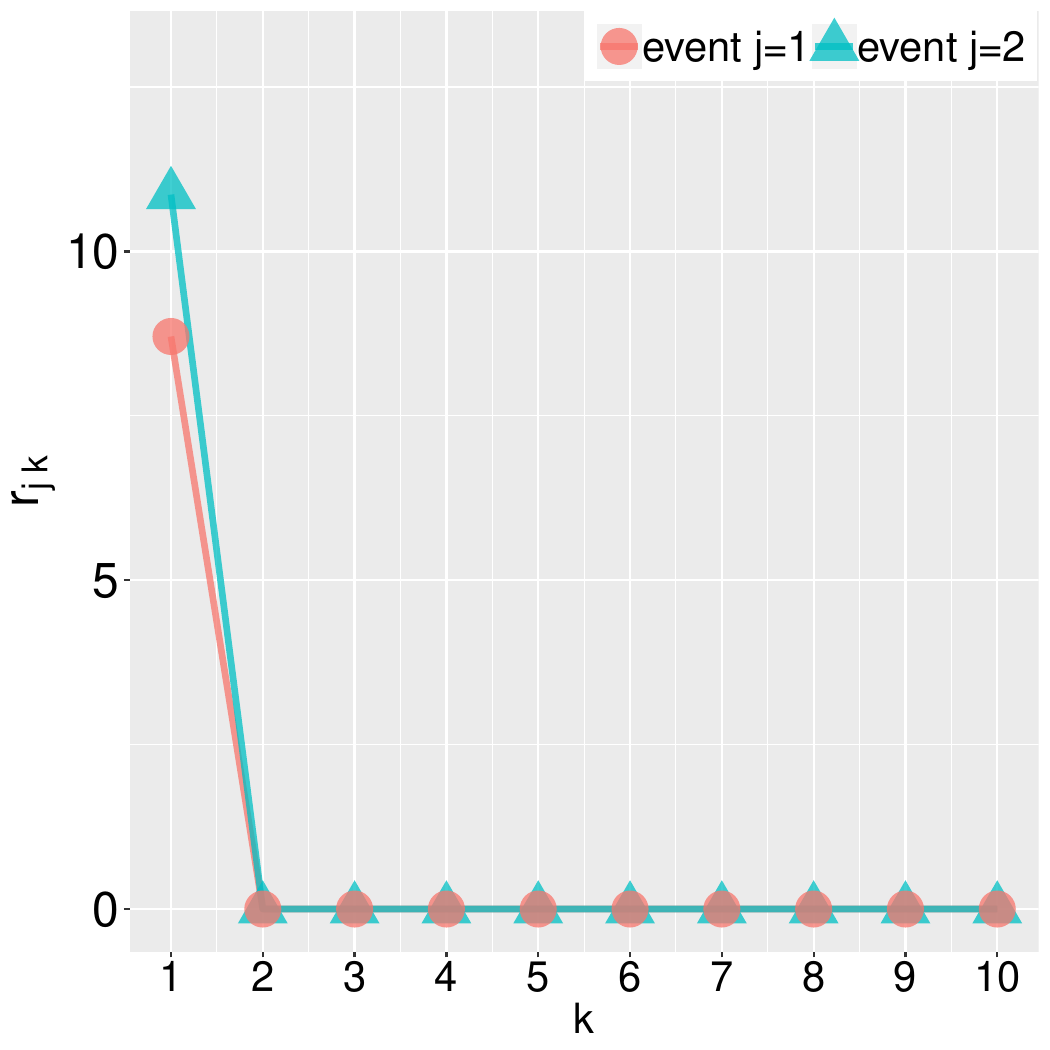}\vspace{-2.5mm}%\label{cosh_risk1}
\caption{Synthetic data 1.}\vspace{-1mm}
 \end{subfigure}\hfil%
\begin{subfigure}[t]{0.32\textwidth}
 \centering
\includegraphics[width=1\linewidth]{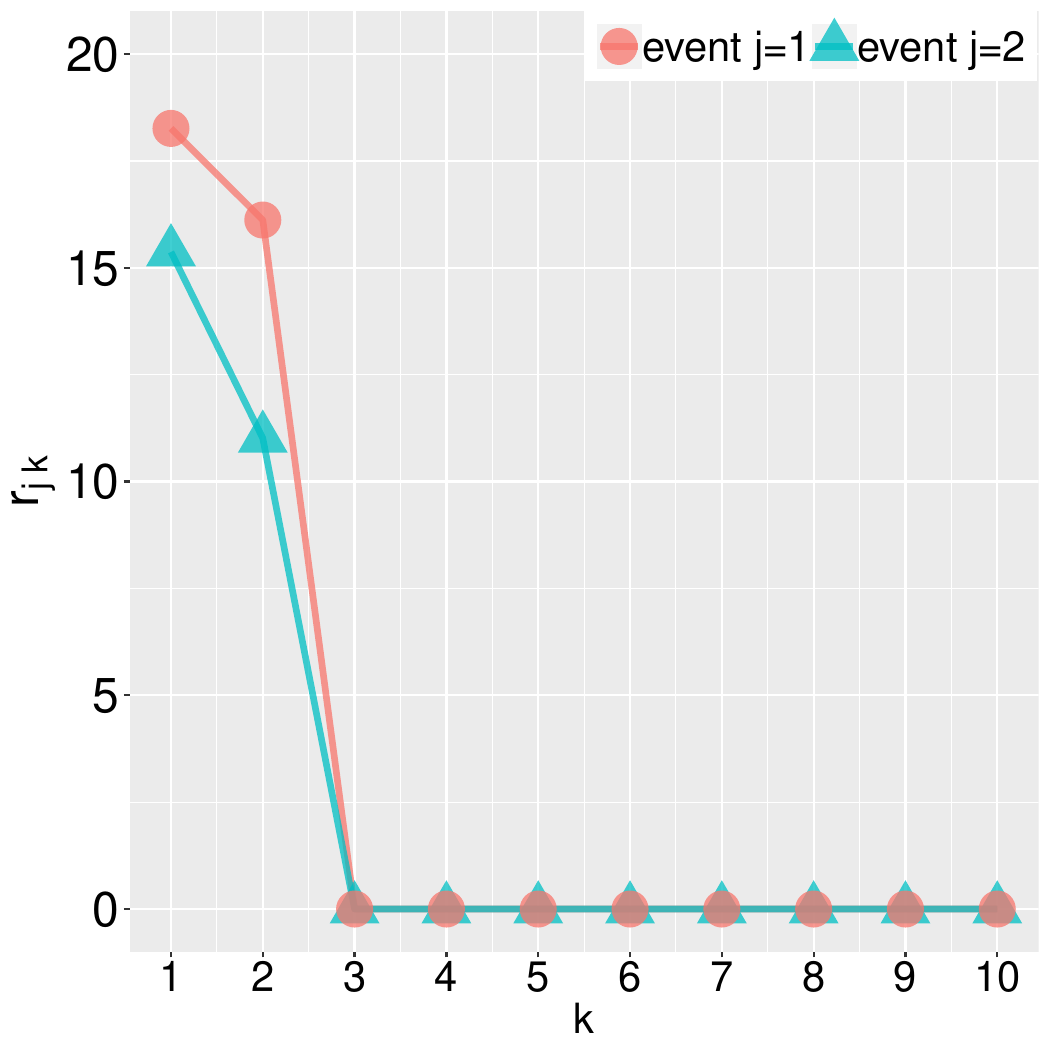}\vspace{-2.5mm}%\label{cosh_risk1}
\caption{Synthetic data 2.}\vspace{-1mm}
 \end{subfigure}\hfil%
\begin{subfigure}[t]{0.32\textwidth}
 \centering
\includegraphics[width=1\linewidth]{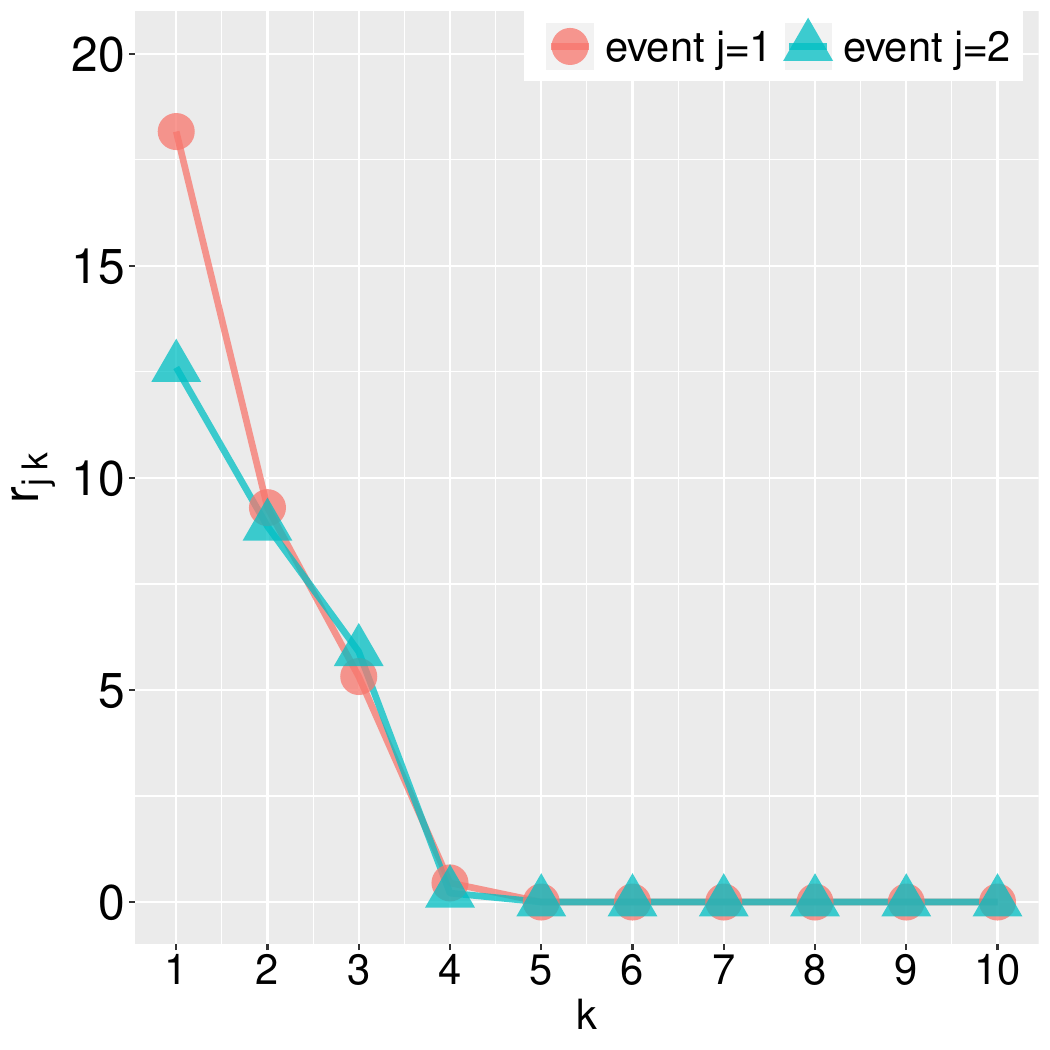}\vspace{-2.5mm}%\label{cosh_risk1}
\caption{Synthetic data 3.}\vspace{-1mm}
 \end{subfigure}\hfil%
\\
 \begin{subfigure}[t]{0.32\textwidth}
 \centering
\includegraphics[width=1\linewidth]{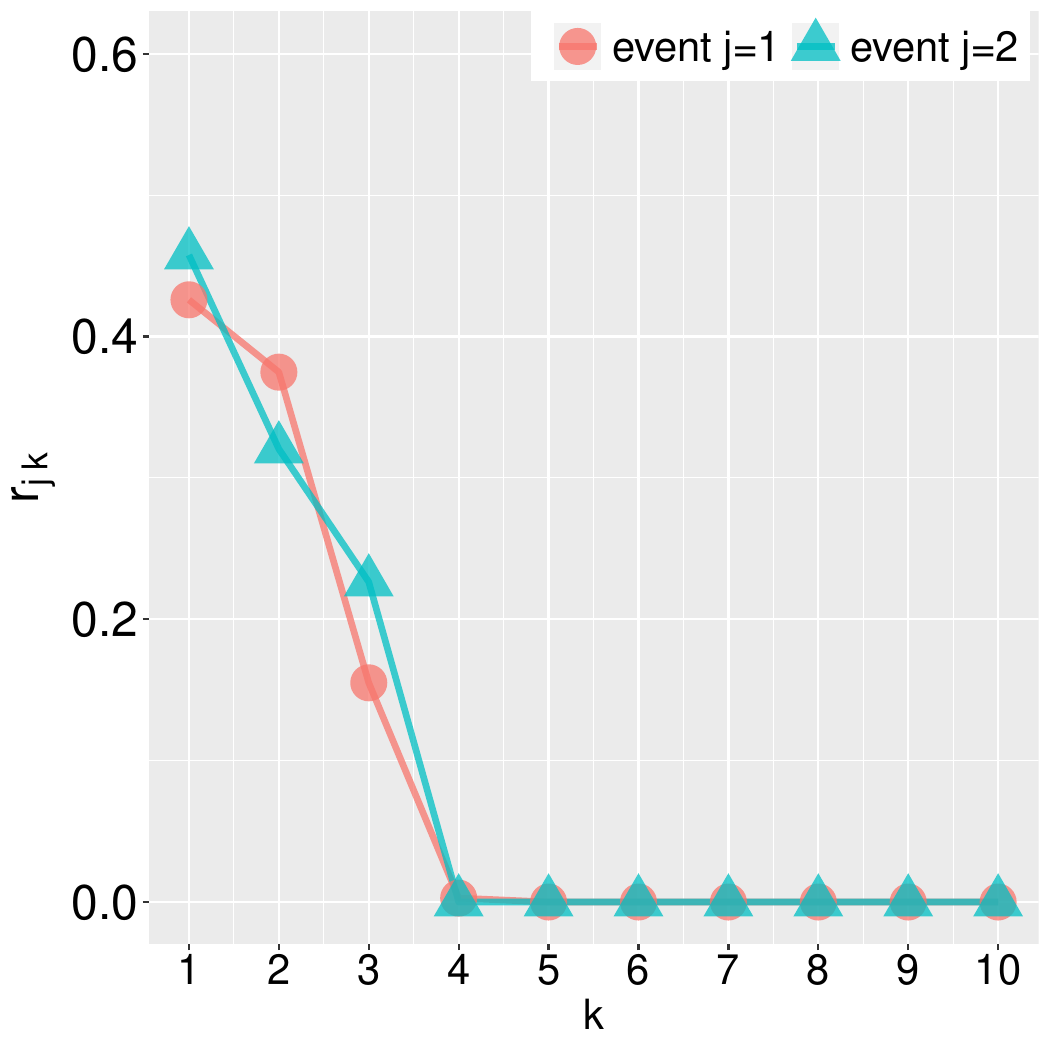}\vspace{-2.5mm}%\label{cosh_risk1}
\caption{Synthetic data 4.}\vspace{-1mm}
 \end{subfigure}\hfil%
\begin{subfigure}[t]{0.32\textwidth}
 \centering
\includegraphics[width=1\linewidth]{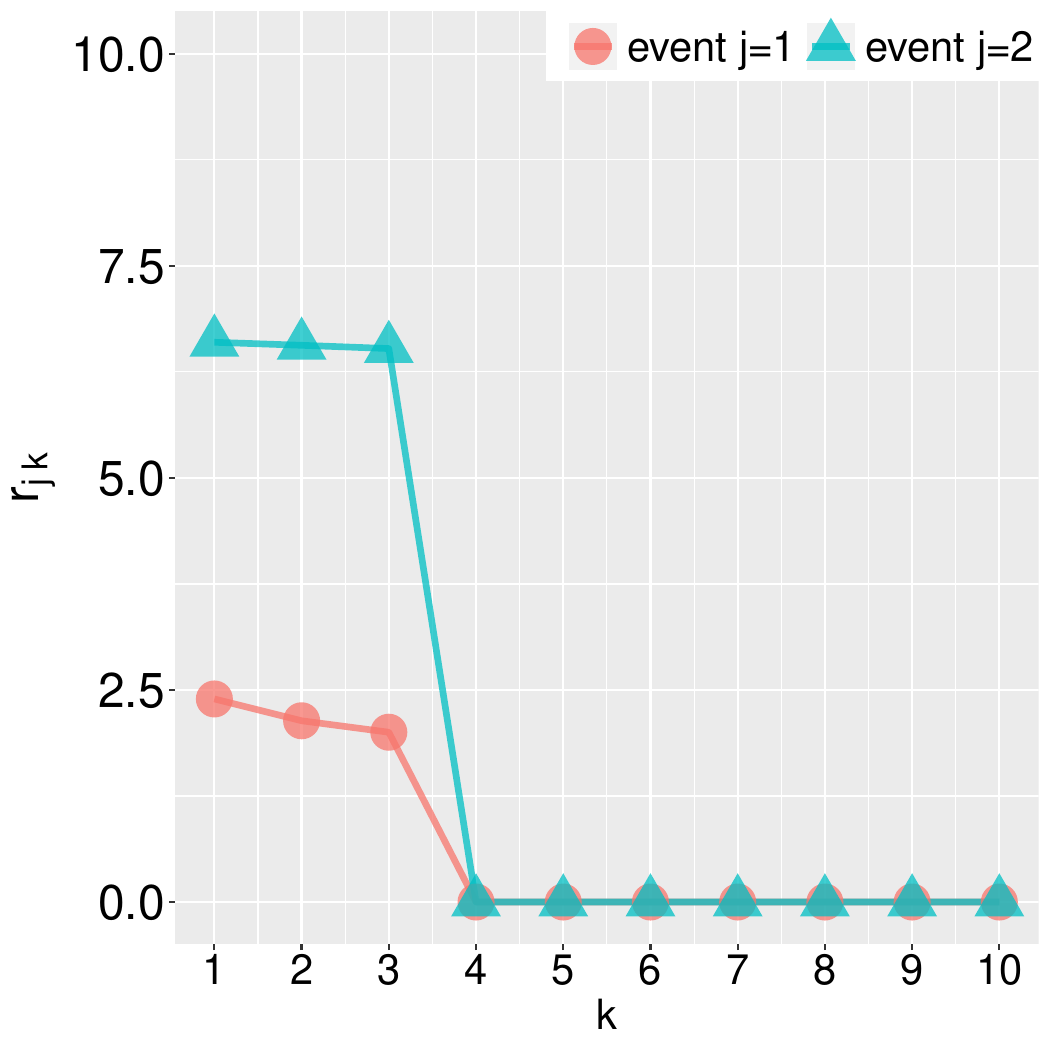}\vspace{-2.5mm}%\label{cosh_risk1}
\caption{Synthetic data 5.}\vspace{-1mm}
 \end{subfigure}\hfil%
\begin{subfigure}[t]{0.32\textwidth}
 \centering
\includegraphics[width=1\linewidth]{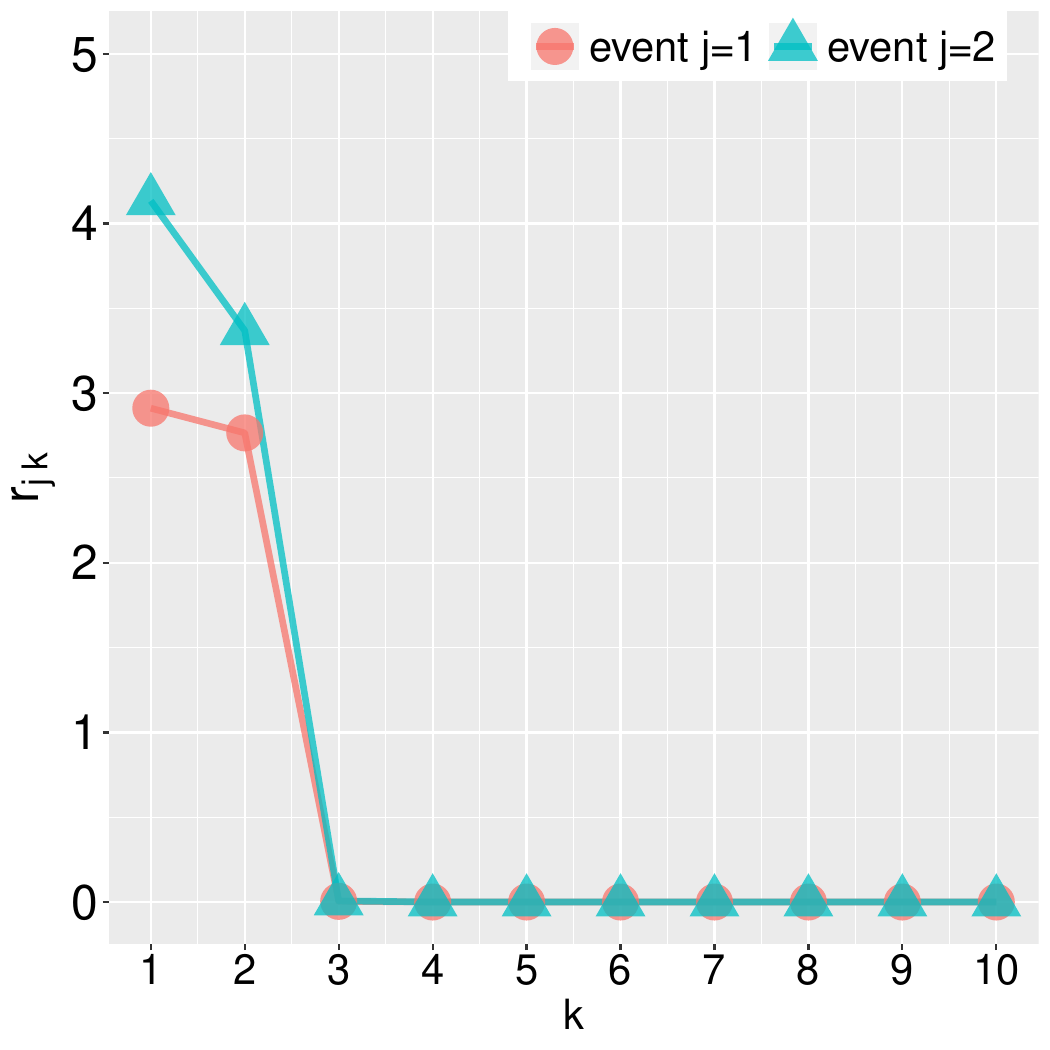}\vspace{-2.5mm}%\label{cosh_risk1}
\caption{Synthetic data 6.}\vspace{-1mm}
 \end{subfigure}\hfil% 
 \vspace{-2mm}
 \caption{Sub-event weights $\{r_{jk}\given k=1,\ldots,10\}$ in descending order for $j=1$ and $2$.}\label{fig:r} 
\end{figure}

Plotted in Figure \ref{fig:r} are $r_{jk}$'s that reflect the data-adaptive, parsimonious nonlinearity of WDR. For data 1, only one sub-event is discovered for competing events 1 and 2, respectively, suggesting the linear covariate effects in data 1. 
Racing between two sub-events within each competing event can approximate the generating process of data 2 and~6. Furthermore, WDR uses racing among three latent sub-events to model the quadratic and interacting covariate effects in data 3, 4, and 5. The other prespecified sub-events are redundant as their weights $r_{jk}$'s are very close to zero.

\subsection{Model Comparison}\label{sec:synthetic_tv}

Considering the lack of nonlinear benchmarks accommodating competing events, left truncation, and time-varying covariates, we develop a kernel-based Fine-Gray (KFG) model that is similar to the kernel Cox models for the analysis of a single event \citep{li2002kernel,evers2008sparse}. Specifically, KFG replaces the covariate matrix of the Fine-Gray model with a radial basis function kernel matrix and uses the gradient boosting method of \citet{binder2009boosting} for sparsity. Note that KFG is a combination of existing approaches and is used for the purpose of benchmarking; we do not claim a contribution to this model. We also compare with four other models: the Fine-Gray model (FG) \citep{fine1999proportional} that is a linear model, random survival forests (RF) \citep{ishwaran2014random}, DeepHit \citep{lee2018deephit}, and the piecewise constant hazards (PCH) method \citep{kvamme2021continuous}.

We compare the model capability in Table~\ref{tab:models}. WDR, FG, and KFG are able to deal with competing events, left truncation, and time-varying covariates. RF and DeepHit accommodate competing events but are incapable of left truncation or time-varying covariates. DeepHit transforms survival analysis into a classification problem by time discretization and models the class probabilities by a neural network. PCH is proposed for single-event survival analysis without left truncation or time-varying covariates, assumes piecewise constant hazards that are modeled by a neural network, and uses  interpolation for continuous-time prediction. We use the censoring trick to adapt PCH to competing-event analysis. Concretely, we analyze one competing event at a time and treat subjects having other events as right censored. To adapt RF, DeepHit, and PCH to the scenario with time-varying covariates, we use the left-truncated and right-censored pseudo subjects as described in Section~\ref{sec:timevarying}. Accommodating 
%or developing heuristics for
left truncation using RF, DeepHit, or PCH has not been investigated and is out of the scope of this paper. So for these three models, we pretend any left truncation is at time $0$ and will show that overlooking left truncation results in poor predictions. 
For DeepHit and PCH, we discretize the continuous survival time into 20 intervals of an equal length,  in each of which the survival or hazard function is constant and modeled by a neural network. Detailed experiment settings are provided in Appendix~\ref{sec:tvsimulation}.
Overall, WDR is versatile compared to RF, DeepHit, and PCH, and its nonlinearity and interpretability can be more appealing than the FG or KFG model. 

\begin{table}[t]
\centering
\renewcommand{\arraystretch}{1} % Default value: 1
% \makebox[\linewidth]{
% \resizebox{\linewidth}{!}{
\begin{tabular}{lcccccc}%\vspace{-6.5mm}
\toprule 
& WDR & FG & KFG & RF & DeepHit & PCH\\
\midrule
Nonlinear   &  \cmark  &  \xmark  &  \cmark  & \cmark   & \cmark   & \cmark\\
Easy to interpret  & \cmark   &    \cmark   &  \xmark   &  \xmark   &   \xmark  &  \xmark \\
Continuous time    &  \cmark    &  \cmark    &   \cmark   &   \cmark   & \xmark   &  \cmark  \\
Competing events    &  \cmark  & \cmark   &  \cmark  & \cmark   &  \cmark  & \xmark \\
Left truncation    &  \cmark  &  \cmark  & \cmark   &  \xmark   &   \xmark  & \xmark \\
Time-varying covariates    &  \cmark  &  \cmark  &  \cmark  & \xmark    &  \xmark   & \xmark \\
\midrule
\multirow{ 2}{*}{Heuristics}  & \multirow{ 2}{*}{-} & \multirow{ 2}{*}{-} & \multirow{ 2}{*}{-} & \multirow{ 2}{*}{\shortstack{Pseudo\\subjects}} & \multirow{ 2}{*}{\shortstack{Pseudo\\subjects}} & \multirow{ 2}{*}{\shortstack{Censoring and\\pseudo subjects}} \\
 & & & & & & \\
\bottomrule
\end{tabular}%\vspace{-0.5mm}
%\end{wraptable} 
% }}
\caption{Model capability.}
\label{tab:models}\vspace{-3.5mm}
\end{table}

\subsubsection{Time-Invariant Covariates and No Left Truncation}\label{sec:synthetic:constant}
We first provide the model comparison on the synthetic data without left truncation or time-varying covariates. 
Specifically, we simulate covariates $\xv_i\in\mathbb{R}^{10}$ from a Gaussian or uniform distribution (see Table~\ref{tab:xbet} in the Appendix) and follow  Table~\ref{tab:syntheticnew} to generate the six data sets but without left truncation or covariate updates. 
In this scenario, all the models compared (the censoring heuristics is used for PCH) are able to handle the survival analysis. For each data set, we simulate 2000 subjects and take 20 random partitions into a training set of 1800 and a testing set of 200. We evaluate the Brier scores at the five time points as in Table~\ref{tab:syntheticnew} and report in Table~\ref{tab:comparison:constant} the average score over the partitions and the time evaluated. The Brier scores at each specific time are deferred to  Tables~\ref{tab:event1:constant} and ~\ref{tab:event2:constant} in the Appendix. 
 
\begin{table}[!t]
\centering
\begin{tabular}{llcccccc}
  \toprule
& & Data 1 & Data 2 & Data 3 & Data 4 & Data 5 & Data 6 \\ 
  \midrule
\multirow{ 6}{*}{Event 1} & WDR & \textbf{0.190} & 0.127 & 0.170 & 0.073 & \textbf{0.132} & 0.093 \\ 
& FG & 0.192 & 0.150 & 0.209 & 0.117 & 0.202 & 0.193 \\ 
& KFG & 0.192 & 0.124 & \textbf{0.166} & \textbf{0.066} & 0.160 & 0.130 \\ 
&  RF & 0.200 & \textbf{0.122} & 0.168 & 0.086 & 0.147 & 0.130 \\ 
&  DeepHit & 0.227 & 0.143 & 0.247 & 0.089 & 0.154 & \textbf{0.081} \\ 
&  PCH & 0.205 & 0.148 & 0.192 & 0.076 & 0.147 & 0.085 \\ 
\midrule
\multirow{ 6}{*}{Event 2} &  WDR & \textbf{0.184} & 0.181 & \textbf{0.168} & {0.069} & \textbf{0.131} & 0.101 \\ 
&  FG  & 0.185 & 0.193 & 0.213 & 0.124 & 0.202 & 0.208 \\ 
&  KFG  & 0.186 & 0.179 & \textbf{0.168} & 0.070 & 0.162 & 0.136 \\ 
&  RF  & 0.193 & \textbf{0.176} & 0.173 & 0.088 & 0.152 & 0.140 \\ 
&  DeepHit  & 0.218 & 0.195 & 0.253 & 0.082 & 0.151 & 0.088 \\ 
&  PCH & 0.200 & 0.193 & 0.180 & \textbf{0.066} & 0.140 & \textbf{0.072} \\ 
   \bottomrule
\end{tabular}
\caption{Brior scores for synthetic data with \textit{constant} covariates and \textit{no} left truncation.} 
\label{tab:comparison:constant}\vspace{-3.5mm}
\end{table}

On data 1 where the covariates have linear effects, WDR, FG, and KFG have similar performances in survival prediction and are slightly better than RF, DeepHit, and PCH. On data~2 to~6 with nonlinear covariate effects, FG does not work well, and WDR is among the best-performing models. Note that DeepHit has a larger variation in prediction accuracy across data or time points than the other models, likely because it models survival probabilities in discrete time and the performance is sensitive to time discretization \citep{kvamme2021continuous}. This implies the importance of continuous-time modeling in survival analysis.

\subsubsection{Time-Varying Covariates and Left Truncation}\label{sec:synthetic:TV}
We compare the model performance on synthetic data with left truncation and time-varying covariates as described in Table~\ref{tab:syntheticnew}. 
In each data set, we simulate covariates $\xv_i\in\mathbb{R}^{10}$ from a Gaussian or uniform distribution (see Table~\ref{tab:xbet} in the Appendix) for each subject and allow up to two covariate updates after the first measurement. In this case, WDR, FG, and KFG can handle the survival analysis with competing events, left truncation, and time-varying covariates. We use PCH with the censoring heuristics for competing events and RF, DeepHit, and PCH with the pseudo subjects heuristics for time-varying covariates. 
For each data set, we simulate 1000 subjects and take 20 random partitions into a training set of 900 and a testing set of 100.  We report in Tables~\ref{tab:event1:TV}  and~\ref{tab:event2:TV} the Brier scores (mean$\,\pm\,$standard error) for events~1 and~2, respectively, by the six models. 

\begin{table}[!t]
\centering
\begin{tabular}{lccccc}
  \hline
Data 1 & $t=0.1$ & $t=0.3$ & $t=0.5$ & $t=0.7$ & $t=0.9$ \\ 
  \hline
WDR & \textbf{0.069}$\pm$0.006 & \textbf{0.176}$\pm$0.005 & \textbf{0.202}$\pm$0.004 & 0.217$\pm$0.003 & 0.223$\pm$0.003 \\ 
  FG & 0.071$\pm$0.006 & 0.181$\pm$0.006 & 0.203$\pm$0.005 & \textbf{0.215}$\pm$0.004 & \textbf{0.219}$\pm$0.003 \\ 
  KFG & 0.071$\pm$0.006 & 0.182$\pm$0.006 & 0.204$\pm$0.005 & 0.216$\pm$0.004 & 0.221$\pm$0.003 \\ 
  RF & 0.071$\pm$0.006 & 0.188$\pm$0.006 & 0.216$\pm$0.006 & 0.231$\pm$0.004 & 0.237$\pm$0.003 \\ 
  DeepHit & 0.075$\pm$0.007 & 0.184$\pm$0.006 & 0.221$\pm$0.007 & 0.231$\pm$0.004 & 0.236$\pm$0.003 \\ 
  PCH & 0.072$\pm$0.006 & 0.204$\pm$0.008 & 0.226$\pm$0.008 & 0.237$\pm$0.005 & 0.250$\pm$0.006 \\ 
   \hline
Data 2 & $t=0.4$ & $t=0.55$ & $t=0.7$ & $t=0.85$ & $t=1$ \\ 
  \hline
WDR & \textbf{0.030}$\pm$0.004 & \textbf{0.086}$\pm$0.004 & \textbf{0.131}$\pm$0.005 & \textbf{0.166}$\pm$0.005 & \textbf{0.182}$\pm$0.004 \\ 
  FG & 0.032$\pm$0.004 & 0.090$\pm$0.005 & 0.153$\pm$0.006 & 0.189$\pm$0.005 & 0.210$\pm$0.004 \\ 
  KFG & 0.032$\pm$0.004 & 0.091$\pm$0.005 & 0.143$\pm$0.006 & 0.177$\pm$0.005 & 0.197$\pm$0.004 \\ 
  RF & 0.032$\pm$0.004 & 0.089$\pm$0.005 & 0.149$\pm$0.006 & 0.184$\pm$0.005 & 0.203$\pm$0.004 \\ 
  DeepHit & 0.032$\pm$0.004 & 0.089$\pm$0.005 & 0.154$\pm$0.007 & 0.186$\pm$0.005 & 0.205$\pm$0.004 \\ 
  PCH & 0.032$\pm$0.004 & 0.091$\pm$0.005 & 0.157$\pm$0.006 & 0.199$\pm$0.007 & 0.232$\pm$0.008 \\ 
   \hline
Data 3 & $t=0.1$ & $t=0.3$ & $t=0.5$ & $t=0.7$ & $t=0.9$ \\ 
  \hline
WDR & \textbf{0.079}$\pm$0.004 & \textbf{0.190}$\pm$0.003 & \textbf{0.202}$\pm$0.003 & \textbf{0.207}$\pm$0.003 & \textbf{0.210}$\pm$0.003 \\ 
  FG & 0.091$\pm$0.005 & 0.219$\pm$0.005 & 0.236$\pm$0.004 & 0.242$\pm$0.003 & 0.245$\pm$0.002 \\ 
  KFG & 0.090$\pm$0.005 & 0.193$\pm$0.004 & 0.207$\pm$0.003 & 0.213$\pm$0.003 & 0.216$\pm$0.002 \\ 
  RF & 0.089$\pm$0.005 & 0.21$\pm$0.005 & 0.227$\pm$0.004 & 0.233$\pm$0.003 & 0.236$\pm$0.003 \\ 
  DeepHit & 0.096$\pm$0.005 & 0.312$\pm$0.01 & 0.373$\pm$0.012 & 0.398$\pm$0.012 & 0.416$\pm$0.013 \\ 
  PCH & 0.096$\pm$0.005 & 0.286$\pm$0.015 & 0.333$\pm$0.02 & 0.351$\pm$0.023 & 0.21$\pm$0.005 \\ 
   \hline
Data 4 & $t=0.3$ & $t=0.45$ & $t=0.6$ & $t=0.75$ & $t=0.9$ \\ 
  \hline
WDR & \textbf{0.028}$\pm$0.003 & \textbf{0.076}$\pm$0.003 & \textbf{0.091}$\pm$0.003 & \textbf{0.100}$\pm$0.003 & \textbf{0.094}$\pm$0.003 \\ 
  FG & 0.039$\pm$0.005 & 0.119$\pm$0.005 & 0.153$\pm$0.004 & 0.166$\pm$0.003 & 0.171$\pm$0.002 \\ 
  KFG & 0.039$\pm$0.006 & 0.082$\pm$0.006 & 0.098$\pm$0.004 & 0.101$\pm$0.003 & 0.105$\pm$0.003 \\ 
  RF & 0.039$\pm$0.006 & 0.114$\pm$0.006 & 0.145$\pm$0.005 & 0.160$\pm$0.004 & 0.163$\pm$0.002 \\ 
  DeepHit & 0.036$\pm$0.005 & 0.129$\pm$0.007 & 0.118$\pm$0.005 & 0.128$\pm$0.005 & 0.126$\pm$0.004 \\ 
  PCH & 0.032$\pm$0.005 & 0.086$\pm$0.003 & 0.091$\pm$0.004 & 0.108$\pm$0.006 & 0.138$\pm$0.009 \\ 
   \hline
Data 5 & $t=0.9$ & $t=1$ & $t=1.1$ & $t=1.2$ & $t=1.3$ \\ 
  \hline
WDR & \textbf{0.087}$\pm$0.004 & \textbf{0.167}$\pm$0.005 & 0.177$\pm$0.004 & \textbf{0.161}$\pm$0.003 & \textbf{0.153}$\pm$0.003 \\ 
  FG &  0.090$\pm$0.005 & 0.200$\pm$0.005 & 0.240$\pm$0.004 & 0.247$\pm$0.003 & 0.250$\pm$0.002 \\ 
  KFG & \textbf{0.087}$\pm$0.006 & 0.168$\pm$0.006 & 0.187$\pm$0.003 & 0.188$\pm$0.002 & 0.188$\pm$0.001 \\ 
  RF & \textbf{0.087}$\pm$0.006 & 0.176$\pm$0.006 & 0.200$\pm$0.004 & 0.199$\pm$0.003 & 0.199$\pm$0.002 \\ 
  DeepHit & 0.093$\pm$0.007 & 0.218$\pm$0.009 & 0.210$\pm$0.009 & 0.180$\pm$0.013 & 0.171$\pm$0.014 \\ 
  PCH & \textbf{0.087}$\pm$0.006 & 0.185$\pm$0.007 & \textbf{0.170}$\pm$0.005 & 0.175$\pm$0.008 & 0.198$\pm$0.01 \\ 
   \hline
Data 6 & $t=1$ & $t=1.1$ & $t=1.2$ & $t=1.3$ & $t=1.4$ \\ 
  \hline
WDR & \textbf{0.101}$\pm$0.005 & \textbf{0.150}$\pm$0.005 & \textbf{0.130}$\pm$0.004 & \textbf{0.118}$\pm$0.003 & \textbf{0.120}$\pm$0.002 \\ 
  FG & 0.117$\pm$0.005 & 0.227$\pm$0.005 & 0.241$\pm$0.004 & 0.244$\pm$0.003 & 0.246$\pm$0.002 \\ 
  KFG & 0.102$\pm$0.006 & 0.174$\pm$0.003 & 0.177$\pm$0.002 & 0.177$\pm$0.002 & 0.178$\pm$0.002 \\ 
  RF & 0.109$\pm$0.006 & 0.182$\pm$0.004 & 0.187$\pm$0.003 & 0.191$\pm$0.003 & 0.195$\pm$0.002 \\ 
  DeepHit & 0.116$\pm$0.007 & 0.215$\pm$0.005 & 0.140$\pm$0.003 & 0.138$\pm$0.003 & 0.139$\pm$0.003 \\ 
  PCH & 0.108$\pm$0.004 & 0.194$\pm$0.005 & 0.186$\pm$0.005 & 0.132$\pm$0.008 & 0.185$\pm$0.011 \\ 
   \hline
\end{tabular}
\caption{Brier scores for event 1 of the synthetic data (mean$\,\pm\,$stander error).} 
\label{tab:event1:TV}\vspace{-3.5mm}
\end{table}

\begin{table}[!t]
\centering
\begin{tabular}{lccccc}
  \hline
Data 1 & $t=0.1$ & $t=0.3$ & $t=0.5$ & $t=0.7$ & $t=0.9$ \\ 
  \hline
WDR & \textbf{0.060}$\pm$0.004 & \textbf{0.172}$\pm$0.005 & 0.201$\pm$0.004 & 0.214$\pm$0.004 & 0.222$\pm$0.004 \\ 
  FG & 0.062$\pm$0.004 & 0.175$\pm$0.006 & \textbf{0.198}$\pm$0.004 & \textbf{0.210}$\pm$0.004 & \textbf{0.218}$\pm$0.004 \\ 
  KFG & 0.062$\pm$0.004 & 0.178$\pm$0.007 & 0.201$\pm$0.005 & 0.211$\pm$0.004 & 0.219$\pm$0.004 \\ 
  RF & 0.062$\pm$0.004 & 0.179$\pm$0.006 & 0.209$\pm$0.004 & 0.221$\pm$0.003 & 0.230$\pm$0.003 \\ 
  DeepHit & 0.066$\pm$0.005 & 0.180$\pm$0.006 & 0.218$\pm$0.005 & 0.224$\pm$0.003 & 0.232$\pm$0.002 \\ 
  PCH & 0.063$\pm$0.004 & 0.201$\pm$0.008 & 0.236$\pm$0.007 & 0.256$\pm$0.007 & 0.266$\pm$0.008 \\ 
   \hline
Data 2 & $t=0.4$ & $t=0.55$ & $t=0.7$ & $t=0.85$ & $t=1$ \\ 
  \hline
WDR & \textbf{0.042}$\pm$0.004 & \textbf{0.137}$\pm$0.005 & \textbf{0.213}$\pm$0.005 & \textbf{0.238}$\pm$0.002 & \textbf{0.227}$\pm$0.002 \\ 
  FG & 0.043$\pm$0.004 & 0.144$\pm$0.007 & 0.241$\pm$0.007 & 0.256$\pm$0.002 & 0.247$\pm$0.001 \\ 
  KFG & 0.043$\pm$0.004 & 0.144$\pm$0.007 & 0.230$\pm$0.007 & 0.243$\pm$0.002 & 0.234$\pm$0.001 \\ 
  RF & \textbf{0.042}$\pm$0.004 & 0.143$\pm$0.007 & 0.238$\pm$0.007 & 0.255$\pm$0.003 & 0.248$\pm$0.001 \\ 
  DeepHit & 0.043$\pm$0.004 & 0.144$\pm$0.007 & 0.251$\pm$0.007 & 0.265$\pm$0.003 & 0.253$\pm$0.002 \\ 
  PCH & 0.043$\pm$0.004 & 0.147$\pm$0.007 & 0.256$\pm$0.008 & 0.276$\pm$0.006 & 0.255$\pm$0.002 \\ 
   \hline
Data 3 & $t=0.1$ & $t=0.3$ & $t=0.5$ & $t=0.7$ & $t=0.9$ \\ 
  \hline
WDR & \textbf{0.074}$\pm$0.004 & \textbf{0.207}$\pm$0.002 & \textbf{0.218}$\pm$0.002 & \textbf{0.218}$\pm$0.002 & \textbf{0.219}$\pm$0.003 \\ 
  FG & 0.076$\pm$0.005 & 0.237$\pm$0.004 & 0.250$\pm$0.002 & 0.252$\pm$0.002 & 0.253$\pm$0.001 \\ 
  KFG & 0.076$\pm$0.005 & 0.209$\pm$0.004 & 0.220$\pm$0.002 & 0.222$\pm$0.002 & 0.223$\pm$0.001 \\ 
  RF & 0.076$\pm$0.005 & 0.230$\pm$0.004 & 0.243$\pm$0.002 & 0.245$\pm$0.002 & 0.246$\pm$0.001 \\ 
  DeepHit & 0.080$\pm$0.005 & 0.376$\pm$0.011 & 0.445$\pm$0.011 & 0.471$\pm$0.011 & 0.484$\pm$0.012 \\ 
  PCH & 0.080$\pm$0.005 & 0.340$\pm$0.016 & 0.391$\pm$0.021 & 0.405$\pm$0.023 & 0.228$\pm$0.007 \\ 
   \hline
Data 4 & $t=0.3$ & $t=0.45$ & $t=0.6$ & $t=0.75$ & $t=0.9$ \\ 
  \hline
WDR & \textbf{0.020}$\pm$0.002 & \textbf{0.061}$\pm$0.004 & \textbf{0.091}$\pm$0.003 & \textbf{0.088}$\pm$0.003 & \textbf{0.085}$\pm$0.003 \\ 
  FG &   {0.031}$\pm$0.005 & 0.114$\pm$0.004 & 0.154$\pm$0.002 & 0.167$\pm$0.002 & 0.166$\pm$0.001 \\ 
  KFG & {0.029}$\pm$0.003 & 0.072$\pm$0.005 & 0.096$\pm$0.003 & 0.099$\pm$0.003 & 0.102$\pm$0.003 \\ 
  RF & {0.030}$\pm$0.004 & 0.105$\pm$0.006 & 0.144$\pm$0.004 & 0.160$\pm$0.003 & 0.161$\pm$0.003 \\ 
  DeepHit & {0.025}$\pm$0.003 & 0.113$\pm$0.007 & 0.124$\pm$0.005 & 0.118$\pm$0.005 & 0.118$\pm$0.006 \\ 
  PCH & {0.022}$\pm$0.002 & 0.079$\pm$0.005 & 0.102$\pm$0.003 & 0.098$\pm$0.004 & 0.121$\pm$0.007 \\ 
   \hline
Data 5 & $t=0.9$ & $t=1$ & $t=1.1$ & $t=1.2$ & $t=1.3$ \\ 
  \hline
WDR & \textbf{0.091}$\pm$0.005 & {0.156}$\pm$0.005 & \textbf{0.172}$\pm$0.005 & \textbf{0.157}$\pm$0.003 & \textbf{0.151}$\pm$0.003 \\ 
  FG & {0.098}$\pm$0.005 & 0.191$\pm$0.005 & 0.232$\pm$0.002 & 0.245$\pm$0.002 & 0.250$\pm$0.001 \\ 
  KFG & 0.096$\pm$0.006 & 0.163$\pm$0.006 & 0.184$\pm$0.004 & 0.188$\pm$0.002 & 0.190$\pm$0.001 \\ 
  RF & 0.096$\pm$0.006 & 0.174$\pm$0.006 & 0.195$\pm$0.004 & 0.198$\pm$0.002 & 0.200$\pm$0.002 \\ 
  DeepHit & 0.102$\pm$0.007 & 0.203$\pm$0.008 & 0.217$\pm$0.007 & 0.161$\pm$0.013 & 0.163$\pm$0.013 \\ 
  PCH & 0.094$\pm$0.007 & \textbf{0.153}$\pm$0.006 & 0.177$\pm$0.006 & 0.158$\pm$0.004 & 0.163$\pm$0.008 \\ 
   \hline
Data 6 & $t=1$ & $t=1.1$ & $t=1.2$ & $t=1.3$ & $t=1.4$ \\ 
  \hline
WDR & 0.109$\pm$0.005 & \textbf{0.160}$\pm$0.004 & \textbf{0.131}$\pm$0.003 & \textbf{0.125}$\pm$0.003 & \textbf{0.125}$\pm$0.003 \\ 
  FG & 0.120$\pm$0.005 & 0.225$\pm$0.004 & 0.234$\pm$0.002 & 0.239$\pm$0.002 & 0.245$\pm$0.001 \\ 
  KFG & \textbf{0.104}$\pm$0.007 & 0.172$\pm$0.004 & 0.173$\pm$0.003 & 0.175$\pm$0.002 & 0.177$\pm$0.002 \\ 
  RF & 0.112$\pm$0.006 & 0.183$\pm$0.004 & 0.184$\pm$0.003 & 0.189$\pm$0.003 & 0.195$\pm$0.002 \\ 
  DeepHit & 0.122$\pm$0.008 & 0.213$\pm$0.005 & 0.136$\pm$0.002 & 0.139$\pm$0.003 & 0.142$\pm$0.003 \\ 
  PCH & 0.111$\pm$0.005 & 0.182$\pm$0.003 & 0.170$\pm$0.004 & 0.147$\pm$0.004 & 0.169$\pm$0.006 \\ 
   \hline
\end{tabular}
\caption{Brier scores for event 2 of the synthetic data (mean$\,\pm\,$stander error).} 
\label{tab:event2:TV}\vspace{-3.5mm}
\end{table}

\looseness-1 On data 1 where the covariate effects are linear, WDR, FG, and KFG have comparable performance. But on all the other data, WDR consistently outperforms the other models, suggesting its excellent prediction accuracy. 
Notably, RF, DeepHit, and PCH deliver comparable performance to WDR if the covariates are constant and there are no left truncations (see Table~\ref{tab:comparison:constant}), but significantly underperform WDR in the presence of left truncation and time-varying covariates. This is likely because of the bias from overlooking left truncation. In addition,  using the pseudo-subjects approach to handle time-varying covariates is underpinned by  Equations~\eqref{eq:tv_survf} and~\eqref{eq:tv_pdf} for WDR, but it serves as a heuristic method for RF, DeepHit, and PCH, and to our knowledge, no theoretical guarantees have been provided. This may be another reason for their underperformance. Overall, the experiments and the model comparison justify the advantage of WDR and its adaptability to various applications.

\section{Biomedical Data Analysis}
\label{sec:real}
We apply WDR to biomedical data analysis and demonstrate its clinical utility. First, we use WDR on a diffuse large B-cell lymphoma data set where the covariates are time-invariant gene expression and there is no left truncation. In this example, we illustrate the non-monotonic change of hazards with covariates and provide insights into potential new disease subtypes. Second, we study cognitive decline among normal individuals using a data set with left truncation, right censoring, and time-varying covariates. We find that three biomarkers are significantly associated with the progression of Alzheimer’s disease.

\subsection{Surviving Diffuse Large B-Cell Lymphoma}
\label{sec:DLBCL}
\looseness-1 We apply WDR to the diffuse large B-cell lymphoma (DLBCL) data\footnote{The data is publicly accessible at \url{https://llmpp.nih.gov/DLBCL/}. Last access in July 2022.} \citep{rosenwald2002use} where the covariates are time-invariant.  Multiple unsuccessful treatments to increase the survival rate suggest that there may exist several subtypes of DLBCL that differ in responsiveness to chemotherapy. \cite{rosenwald2002use} identify three subgroups of gene expression in the DLBCL data, activated B-cell-like (ABC), germinal-center B-cell-like (GCB), and type-3 (T3) DLBCL, which may be related to three different diseases as a result of different mechanisms of malignant transformation. They also conjecture that T3 might be associated with more than one such mechanism. In our analysis, we treat the death of ABC, GCB, or T3 as three competing events and study the time to death due to a specific subtype. Right censoring applies to those who were alive at the end of the study. The data set contains a microarray gene expression profile of 7399 genes from 240 patients, and only 434 genes have no missing values in all the patients. Since missing covariate imputation is beyond our scope and the number of patients is small, we use
seven of the 434 genes as covariates, as they have been reported to be related to clinical phenotypes \citep{li2005boosting}. We focus on the WDR estimation of covariate effects and defer the model comparison to Table~\ref{tab:dlbcl:BS} in the Appendix, where WDR consistently delivers accurate predictions.  

\begin{figure}[!t]\vspace{0mm}
 \centering
 \begin{subfigure}[t]{0.25\textwidth}
 \centering
\includegraphics[width=1\linewidth]{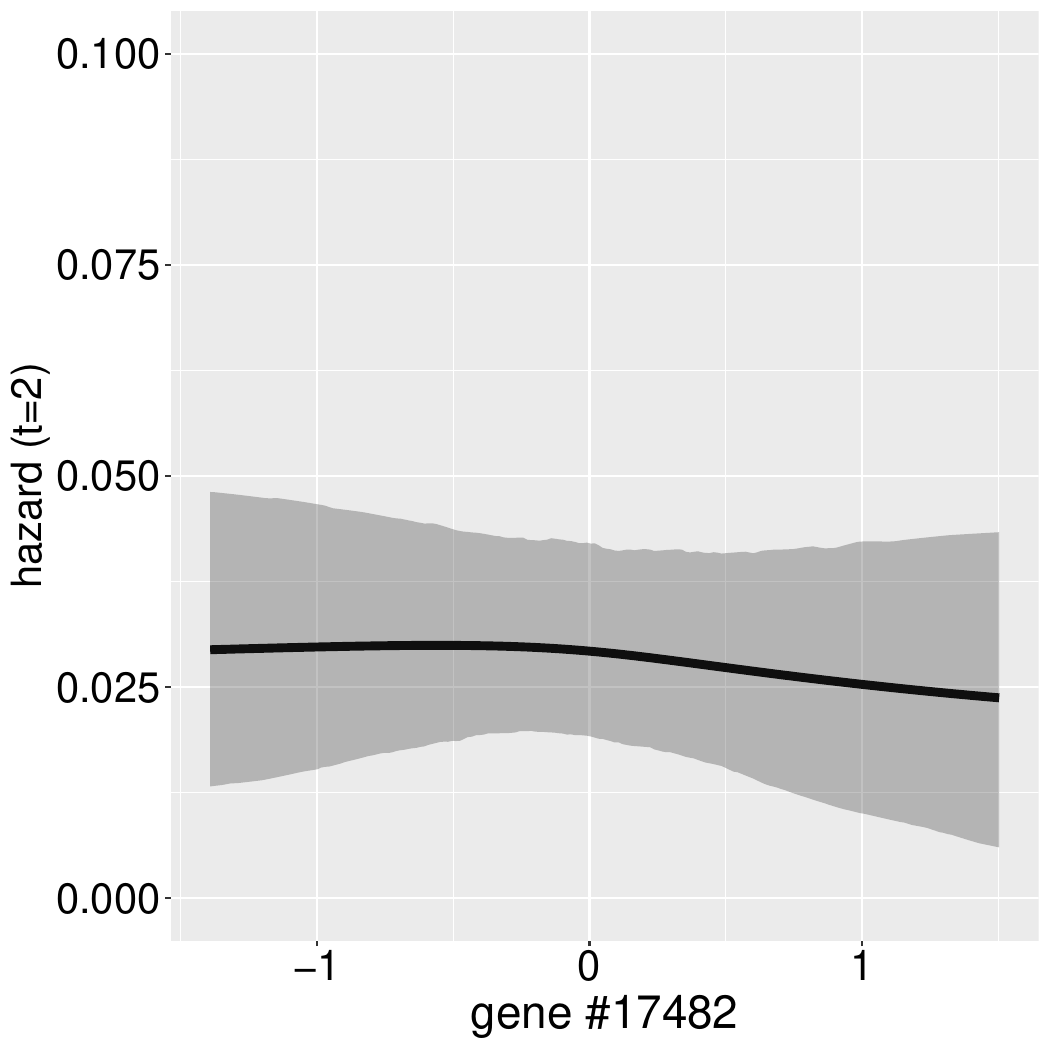}\vspace{-2.5mm}%\label{cosh_risk1}
\caption{Gene 17482.}\vspace{-1mm}
 \end{subfigure}\hfil%
\begin{subfigure}[t]{0.25\textwidth}
 \centering
\includegraphics[width=1\linewidth]{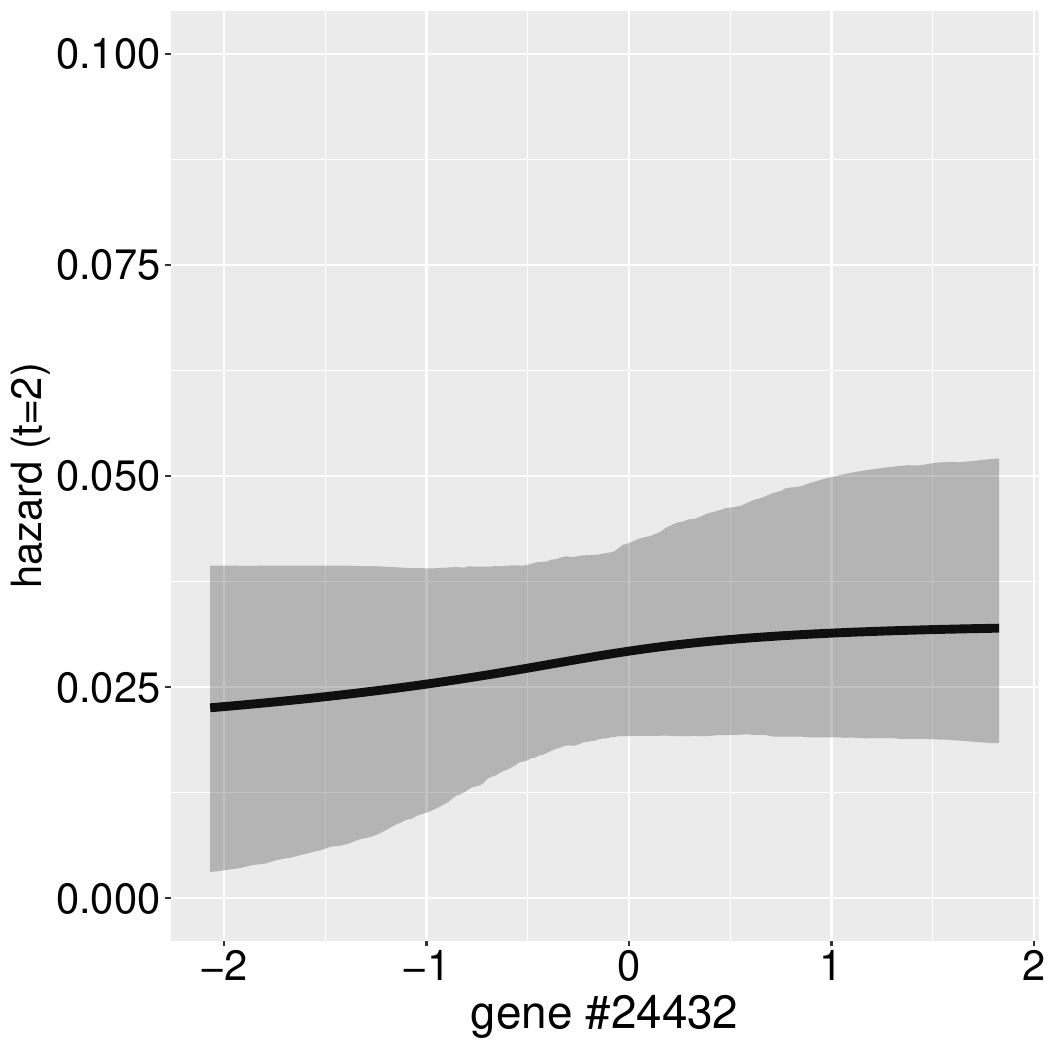}\vspace{-2.5mm}%\label{cosh_risk1}
\caption{Gene 24432.}\vspace{-1mm}
 \end{subfigure}\hfil%
\begin{subfigure}[t]{0.25\textwidth}
 \centering
\includegraphics[width=1\linewidth]{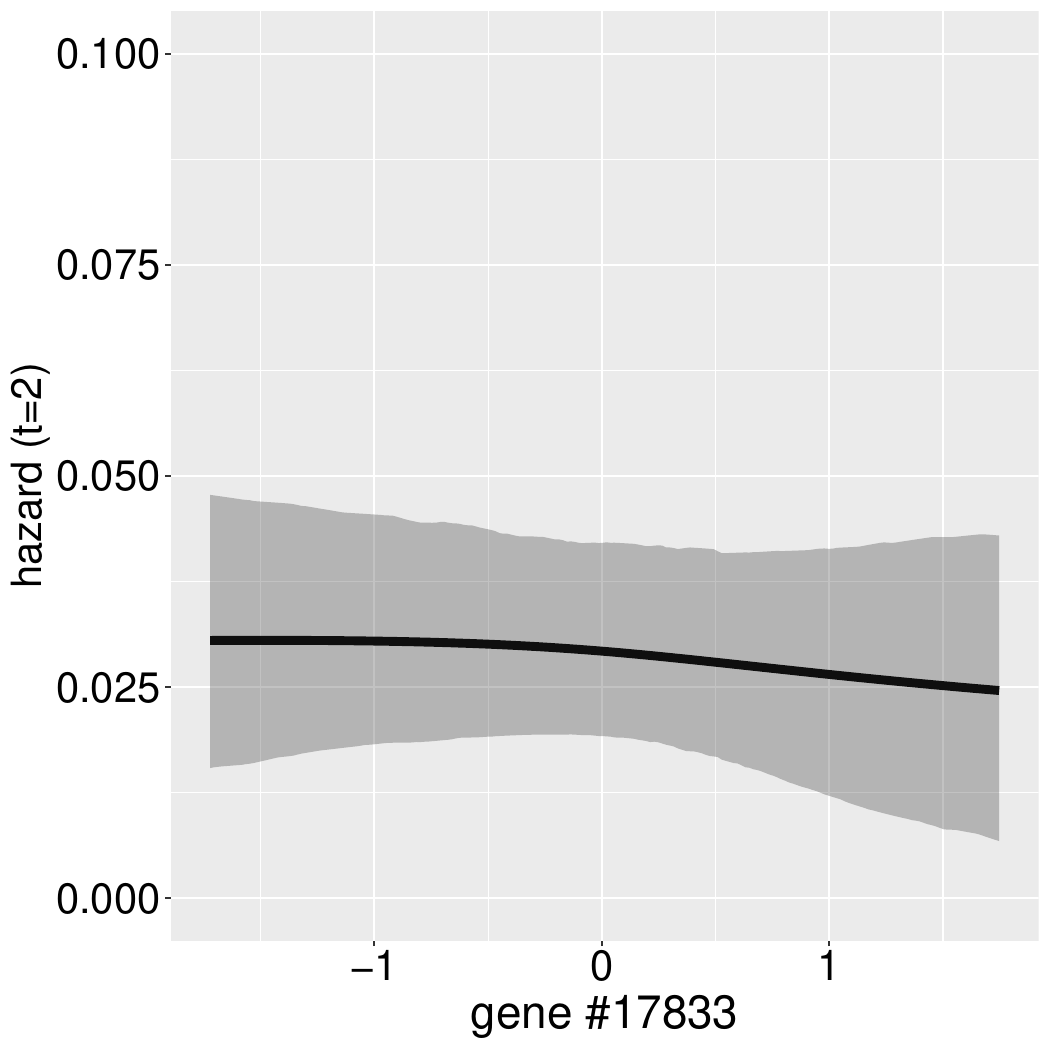}\vspace{-2.5mm}%\label{cosh_risk1}
\caption{Gene 17833.}\vspace{-1mm}
 \end{subfigure}\hfil%
\begin{subfigure}[t]{0.25\textwidth}
 \centering
\includegraphics[width=1\linewidth]{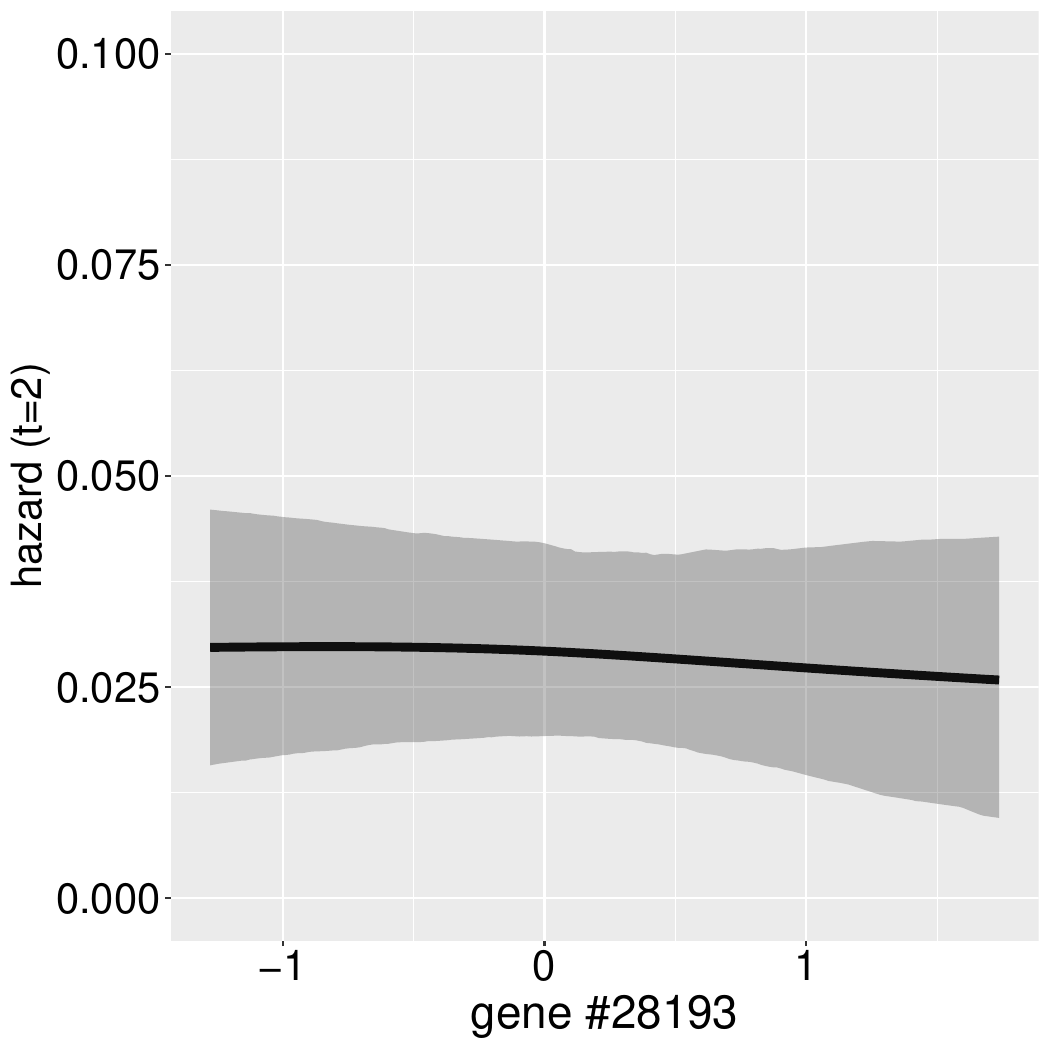}\vspace{-2.5mm}%\label{cosh_risk1}
\caption{Gene 28193.}\vspace{-1mm}
 \end{subfigure}\hfil%
\medskip
 
\begin{subfigure}[t]{0.25\textwidth}
 \centering
\includegraphics[width=1\linewidth]{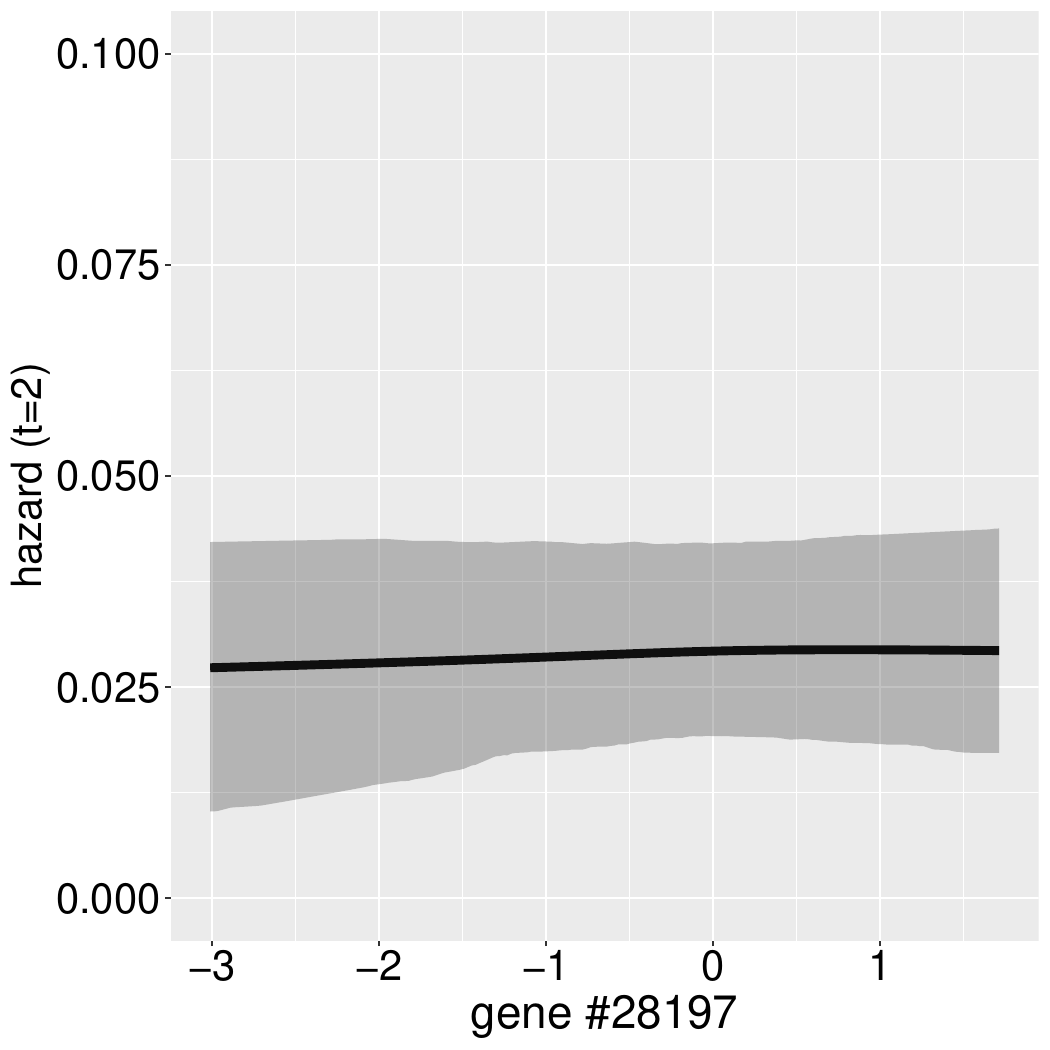}\vspace{-2.5mm}%\label{cosh_risk1}
\caption{Gene 28197.}\vspace{-1mm}
 \end{subfigure}\hfil%
\begin{subfigure}[t]{0.25\textwidth}
 \centering
\includegraphics[width=1\linewidth]{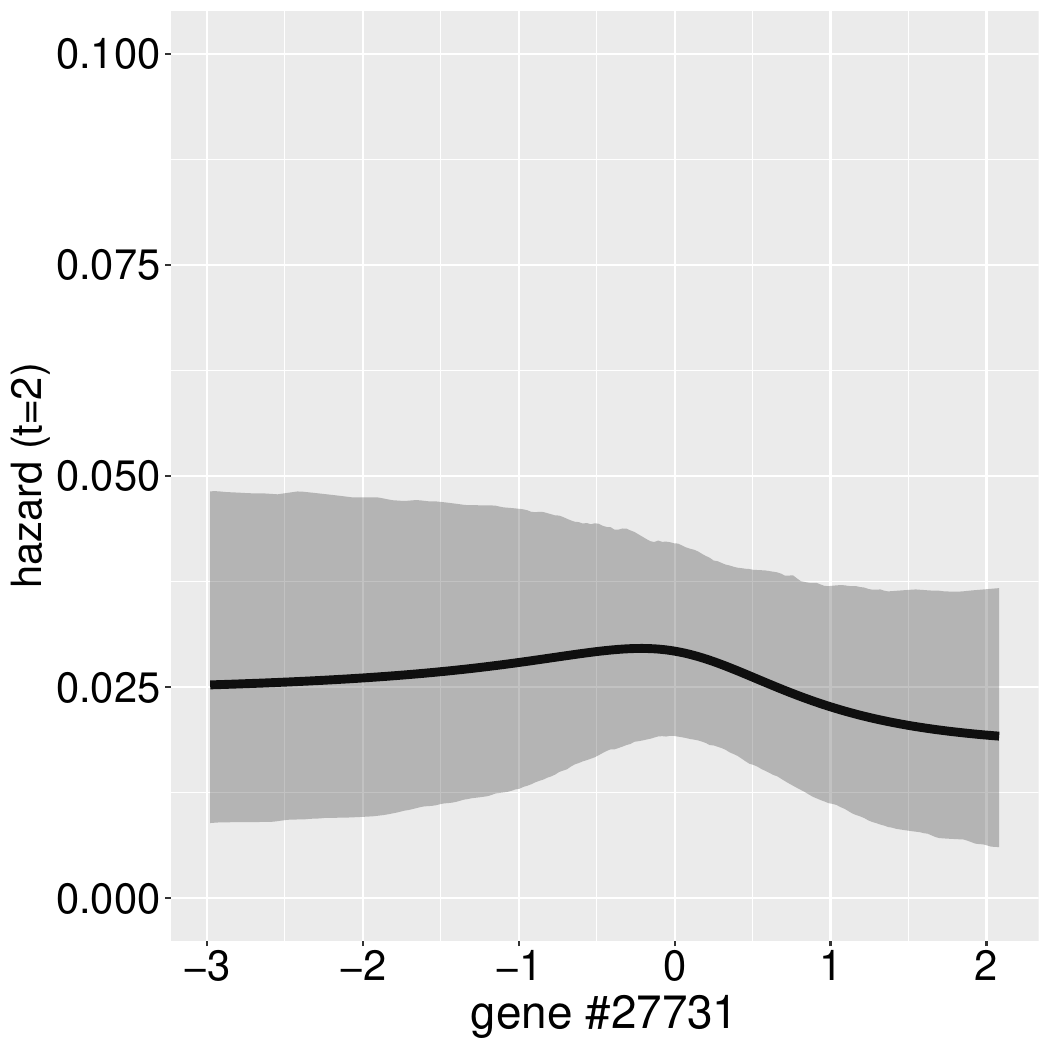}\vspace{-2.5mm}%\label{cosh_risk1}
\caption{Gene 27731.}\vspace{-1mm}
 \end{subfigure}\hfil%
\begin{subfigure}[t]{0.25\textwidth}
 \centering
\includegraphics[width=1\linewidth]{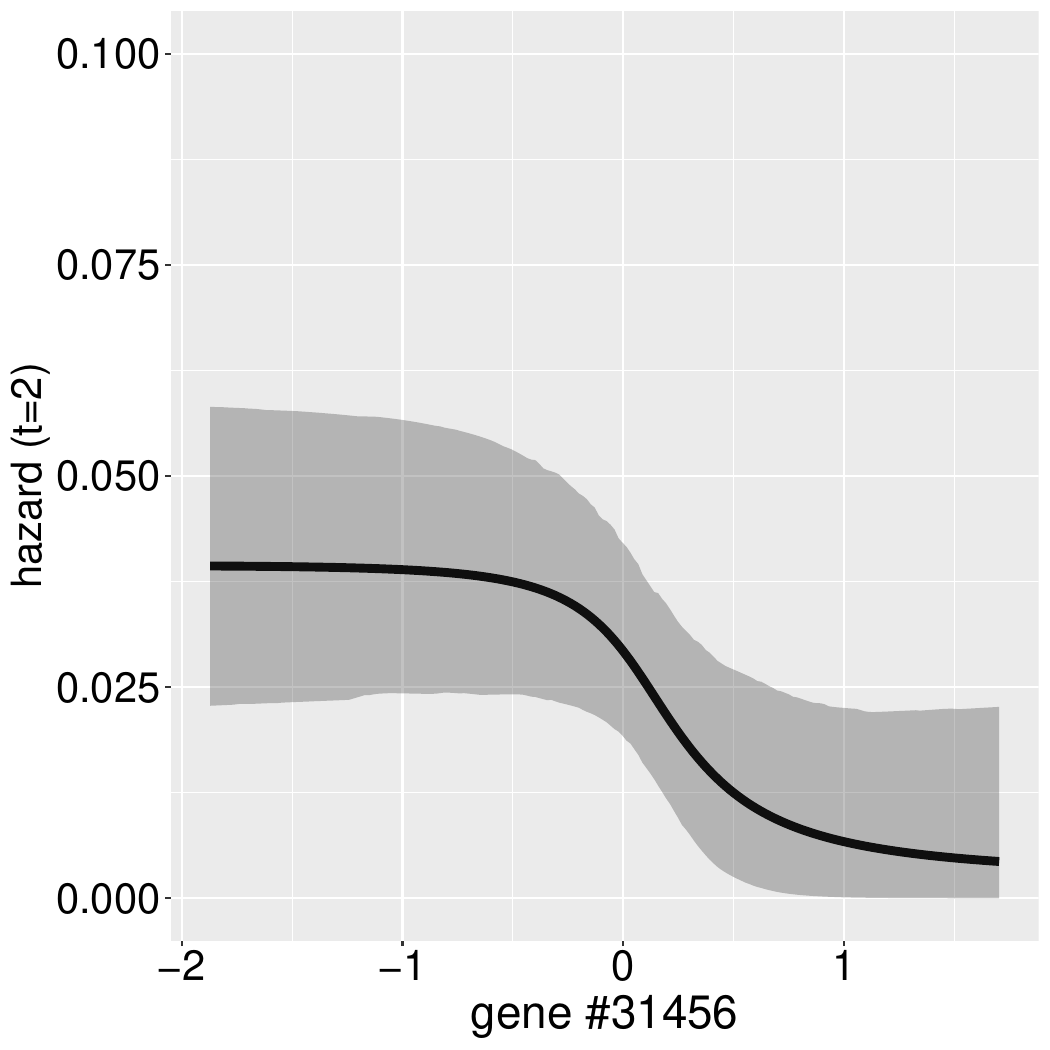}\vspace{-2.5mm}%\label{cosh_risk1}
\caption{Gene 31456.}\vspace{-1mm}
 \end{subfigure}\hfil% 
 \vspace{-2mm}
\caption{Hazards of T3 at time $t=2$ against gene expressions.}\label{fig:dlbcl} \end{figure}

%The estimated Weibull shape parameter $\hat a=1.38$ with the 95\% credible interval $(1.12, 1.71)$, suggesting that the death hazard first increase then decrease over time. Moreover, 
We find by WDR one sub-event for ABC and GCB, respectively, implying that the gene expression is linearly associated with the time to death of ABC or GCB. Meanwhile, two sub-events are found under T3, implying nonlinear effects of the genes. Concretely, we show in Figure~\ref{fig:dlbcl} how the hazard function of T3 changes with the gene expression. In each panel, we plot the hazard function at $t=2$ against one gene that varies between its minimum and maximum values (the data has been centered) as in the data with the expression of other genes being zero. The figures only show the relative change of hazards whose absolute values are arbitrary. The WDR parameter estimations are provided in Table~\ref{tab:dlbcl} in Appendix~\ref{app:additional_result}.

The decreasing hazard functions in panels~(a), (c), (d), and~(g) imply that high expression of genes 17482, 17833, 28193, and 31456 is associated with late death of T3 (at $t=2$), whereas high  expression of genes 24432 and 28197 (panels (b) and (e)) is associated with early death. Interestingly noted that panel (g) suggests an inverse S-shaped effect of gene 31456 on the hazard of T3, which is apparently not log-linear (as shown in Figure~\ref{fig:hazardx}~(a)). Specifically, a plateau is observed between gene expression values~$-2$  and~$-0.5$.  Then the hazard sharply goes down until the expression of~$0.5$ and mildly decreases afterward. Moreover, gene 24432 seems to deliver an S-shaped effect. 
In contrast to either increasing or decreasing effects, panel (f) reveals a non-monotonic effect of gene 27731 such that the hazard is first increasing and then decreasing as the gene expression grows.  
The two sub-events found potentially imply two different mechanisms of malignant transformation of type-3 DLBCL whose progression is linearly associated with the genes. One can use WDR to find a patient's sub-event type of T3 ($\kappa_{iy_i}$ for $y_i=3$), study the T3 DLBCL progression by $\betav_{3k}$'s, and investigate individualized treatment. 

\subsection{Alzheimer's Disease Data Analysis}
\label{sec:AD}
To demonstrate the applicability of WDR to left-truncated and right-censored data with intermittently updated covariates, we apply WDR to a data set from the Biomarkers of Cognitive Decline Among Normal Individuals (BIOCARD) cohort study. It was administrated by the National Institute of Health from 1995 to 2005 and re-established at Johns Hopkins School of Medicine after being stopped for four years. The BIOCARD study recruited participants who were %cognitively  normal   
MCI-free at baseline and collected their cognitive performance testing scores along with other biomarkers that are potentially related to Alzheimer’s disease (AD) annually or semiannually during the study, aiming to identify biomarkers associated with the development of AD progression. 

In the BIOCARD study, we study the failure time defined as the age at the onset of symptoms of mild cognitive impairment (MCI). The data are left-truncated and right-censored, where the truncation time is an individual’s age at the time when she or he entered the study. Death is a competing event since subjects can die due to cancer or other diseases without symptoms of MCI. 
%before the onset of clinical symptoms of AD and the death will terminate the observation of the onset of MCI. 
We consider three time-invariant biomarkers: sex, education, and the $\epsilon$4 allele of the apolipoprotein E (ApoE-4) gene, and 10 time-varying continuous biomarkers: 
%education,\red{(note: education is a time-invariant covariate in this study )}
5 cerebrospinal fluid (CSF) variables and 5 cognitive measures. These biomarkers are selected since they are potentially important measures related to the time to onset of AD clinical symptoms based on the findings from previous literature \citep{albert2014cognitive}. The CSF biomarkers include Abeta 42, Abeta 40, total tau (t-tau), phosphorylated tau 181 (p-tau181), and ptau\_amyloid. Cognitive biomarkers were measured by the annual, comprehensive neuropsychological battery test for participants in the BIOCARD to examine whether they are significantly associated with the time to onset of clinical symptoms, which were a harbinger of a diagnosis of MCI. We include logmem (Logical Memory IIA - Delayed), paired (Paired Associates I - New Learning), DSST (WAIS-R Digit Symbol), CVLTTOTL (Number correct Trials 1-5), and MMscore (Total Mini-Mental State Examination score) in the analyses. Of the 291 subjects included in our analyses, 35 subjects were observed death during the study, 82 subjects were diagnosed with MCI or dementia due to AD, and 209 subjects remained cognitively normal at their last visits. Table~\ref{tab:AD_data}  gives a brief summary of the participants in our analysis.

\begin{table}[!t] 
\centering
\begin{tabular}{p{10cm}r}
\hline
\multicolumn{1}{l}{Age, mean years (SD)} &  56.9 (9.3) \\
\multicolumn{1}{l}{Sex, \% females} &  58.8\%\\
\multicolumn{1}{l}{Education, mean years (SD)} & 17.1 (2.4) \\
\multicolumn{1}{l}{\% ApoE-4 carriers} &  33.7\%\\
\hline
\end{tabular}
\caption{Baseline characteristics of the participants in the analysis of the BIOCARD study.}\label{tab:AD_data}\vspace{-3.5mm}
\end{table}

\begin{table}[!t]
\centering
\renewcommand{\arraystretch}{1} % Default value: 1
  \begin{tabular}{lll}
    \hline
     &   ~MCI ($j=1$) &  ~Death ($j=2$)\\ \hline
    sex & -0.010 (-0.225, 0.225) & -0.110 (-0.755, 0.463) \\ 
    education & \textbf{-0.102}  (-0.124, -0.057) & \textbf{-0.124}  (-0.203, -0.064) \\  
    ApoE-4 & ~0.041 (-0.148, 0.319) & ~0.265 (-0.523, 0.896)  \\ 
    p-tau181 & ~0.114  (-0.256, 0.562) & -0.009  (-0.451, 0.471) \\ 
    t-tau & ~0.080 (-0.090, 0.287) & -0.011 (-0.391, 0.220) \\ 
    Abeta 42 & -0.097 (-0.447, 0.188) & ~0.037 (-0.494, 0.598) \\  
    Abeta 40 & ~0.007 (-0.222, 0.236) & ~0.098 (-0.377, 0.622) \\ 
    ptau\_amyloid & ~0.154 (-0.237, 0.574) & ~0.029 (-0.470, 0.431) \\ 
    paired & \textbf{-0.402} (-0.646, -0.228) & -0.057 (-0.373, 0.151)\\  
    logmem & -0.023 (-0.227, 0.248) & -0.012 (-0.287, 0.227) \\ 
    DSST & \textbf{-0.454} (-0.674, -0.207) & -0.139 (-0.466, 0.185)\\  
    CVLTTOTL & -0.142 (-0.417, 0.024) & -0.211 (-0.811, 0.047) \\  
   MMSCORE& -0.083 (-0.278, 0.05) & -0.105 (-0.363, 0.089) \\ \hline
  \end{tabular}
  \caption{Posterior means and 95\% credible intervals of the coefficients $\betav_{jk}$, $k=1$.}\label{tab:AD_beta}\vspace{-3.5mm}
\end{table}

We find one sub-event under MCI and death of other causes, respectively, by WDR, implying that the covariates are linearly related to the time to either competing event. Specifically, we run 10,000 MCMC iterations with a
burn-in of 8,000 iterations, collect 2,000 post-burn-in MCMC samples, and calculate the posterior means and the 95\% credible intervals of the coefficients $\betav_{jk}$ for $j=1,2$ and $k=1$ as reported in Table~\ref{tab:AD_beta}. We find a decelerated progression of MCI significantly associated with higher values of \textit{education}, \textit{paired}, and \textit{DSST}. These results are corroborated by existing literature. A high level of education has been reported to reduce the risk of MCI and AD \citep{sattler2012cognitive}. It has also been reported that the increased risk of progressing from normal cognition to the onset of clinical symptoms is associated with lower scores of paired and DSST \citep{albert2014cognitive}. On the other hand, a higher education level is associated with lower death hazards of other causes, while the other ten AD-related biomarkers do not significantly contribute to explaining the time to death. In addition, we randomly split the subjects into 80\% of training data and 20\% of testing data to assess the performance of WDR and defer the model comparison to  Table~\ref{tab:AD:BS} in the 
Appendix.

 \section{Conclusion}\label{sec_conclusion}  
Assuming a two-phase racing among latent sub-events, the proposed Weibull delegate racing (WDR) survival model not only accommodates non-monotonic covariate effects but also preserves interpretability. We use a gamma process to support a potentially infinite number of sub-events, rely on its inherent shrinkage property to remove unneeded model capacity, and thus enable WDR to explore mechanisms of surviving competing events. Moreover, WDR can handle left truncation, time-varying covariates, missing event times or types, and different censoring. To the best of our knowledge, WDR is the first nonlinear survival model for competing events, left truncation, and time-varying covariates without using covariate transformations. Simulation studies have shown excellent performance and favorable properties of WDR compared to the Fine-Gray models, the random survival forests, the DeepHit, and the piecewise constant hazards model. 
The analysis of the lymphoma and Alzheimer's disease data shows intriguing findings that help researchers discover new disease types or interpret covariate effects related to disease progression. 
%The interpretation of WDR as an accelerated failure time model allows us to both qualitatively and quantitatively summarize the feature effects. 
Overall, WDR is an attractive alternative to existing models for various applications that require interpretable nonlinearity.

%\section*{Acknowledgments}
%Yanxun Xu is supported in part by NSF 1918854.  Mei-Cheng Wang is supported in part by NIH Grant U19 AG033655. Mingyuan Zhou is supported in part by NSF 1812699 and NSF 2212418.

\acks{We thank the editor and the reviewers for their dedicated time and constructive comments. Yanxun Xu is supported in part by NSF 1918854.  Mei-Cheng Wang is supported in part by NIH Grant U19 AG033655. Mingyuan Zhou is supported in part by NSF 1812699 and NSF 2212418.}

%\newpage
\appendix
\section{Theorem and Proof}\label{app:proof}
\subsection{Proof of the Weibull Racing Property}
We prove Property~\ref{prop:truncated_weibull_race} as follows.
\begin{proof}%[Proof of Property 1]%\ref{prop:weibull_race}.] 
~~Since $\mbox{Pr}(t>t_0)=\prod_j \mbox{Pr}(t_j>t_0)=\prod_j S_\tau(t_0\given a, \lambda_j) = \frac{\exp({-t_0^a\sum_j \lambda_j})}{\exp({-\tau^a\sum_j \lambda_j})}$ for $t_0\geq \tau$, then $t=\min_j t_j \sim \mbox{Weibull}\tau(a, \sum_j \lambda_j)$. Assuming $t_h=\min_j t_j$ and for arbitrary $t_0\geq \tau$, we have
\begin{align*}
\mbox{Pr}(\min_j t_j>t_0,~t_h=\min_j t_j)&=\prod_{j\neq h} \mbox{Pr}(t_0<t_h<t_j)\\
&=\int_{t_0}^\infty f_\tau(t_h \given a, \lambda_h)\prod_{j\neq h} \mbox{Pr}(t_j>t_h)dt_h\\
&=\int_{t_0}^\infty  \frac{f(t_h\given a,\lambda_h)}{S(\tau\given a, \lambda_h)}\prod_{j\neq h}\frac{S(t_h\given a,\lambda_j)}{S(\tau\given a, \lambda_j)}
dt_h\\
&=\frac{\lambda_h}{\sum_j \lambda_j} \frac{\exp({-t_0^a \sum_j \lambda_j})}{\exp({-\tau^a  \sum_j \lambda_j})}.
\end{align*}
Let $t_0=\tau$. We have $\mbox{Pr}(t_h=\min_j t_j)=\frac{\lambda_h}{\sum_j \lambda_j}$. This proves the categorical distribution of $y=\mathop{\mathrm{argmin}}\limits_{j} t_j$. Consequently, 
\begin{equation*}
\mbox{Pr}(\min t_j>t_0,~t_h=\min_j t_j)=\mbox{Pr}(t>t_0,~y=h)=\mbox{Pr}(t>t_0)\mbox{Pr}(y=h).
\end{equation*}
This proves the independence of $t$ and $y$.
\end{proof}

\subsection{WDR Survival Function with Time-Varying Covariates}
We derive equations~\eqref{eq:tv_survf} and~\eqref{eq:tv_pdf}. Omitting the subject index $i$ for brevity and given the piecewise Weibull hazards $$h^{(l)}(t) = h_{\tau^{(l)}}(t\given a, \sum_{j,k}\lambda_{jk}^{(l)}) = a\sum_{j,k}\lambda_{jk}^{(l)}t^{a-1}$$ for $t\in \bm[\tau^{(l)}, \tau^{(l+1)}\bm)$, $l=0,1,\ldots,L-1$ and $\bm[\tau^{(l)},\infty\bm)$ for $l=L$, 
the cumulative hazard function for $t>\tau^{(L)}$ is 
\begin{align*}
   H(t\given  \{\tau^{(l)}\}_{l=0}^L)  &= \int_0^{t} \Big[\sum_{l=0}^{L-1} h^{(l)}(s) \bm 1(s\in (\tau^{(l)}, \tau^{(l+1)}]) +  h^{(L)}(s)\bm 1(s\in (\tau^{(L)}, \infty))  \Big]ds \\
    &= \sum_{l=0}^{L-1} \int_{\tau^{(l)}}^{\tau^{(l+1)}}  a\sum_{j,k}\lambda_{jk}^{(l)}s^{a-1} ds+\int_{\tau^{(L)}}^t  a\sum_{j,k}\lambda_{jk}^{(L)}s^{a-1} ds\\
    &=\sum_{l=0}^{L-1}\Big[\sum_{j,k}\lambda_{jk}^{(l)} (\tau^{(l+1)})^a-\sum_{j,k}\lambda_{jk}^{(l)}(\tau^{(l)})^a\Big] +\sum_{j,k}\lambda_{jk}^{(L)} t^a-\sum_{j,k}\lambda_{jk}^{(L)}(\tau^{(L)})^a.
\end{align*}
Consequently, the survival function at time $t>\tau^{(L)}$ is
\begin{align*}
S(t\given \{\tau^{(l)}, \lambda_{jk}^{(l)}\}_{l=0}^{L}) &= \exp({ H(t\given  \{\tau^{(l)}\}_{l=0}^L)} )
%&= e^{-\sum_{l=0}^{L-1}[\lambda(x_l)t_{l+1}^a-\lambda(x_l)t_{l}^a] - \lambda(x_L)t^a+\lambda(x_L)t_L^a} \\
%&=\frac{S(t\given a, \lambda(x_L))}{S(t_L\given a, \lambda(x_L))}\prod_{l=0}^{L-1}\frac{S(t_{l+1}\given a, \lambda(x_l))}{S(t_{l}\given a, \lambda(x_l))}\\
=S_{\tau^{(L)}}(t\given a, \sum_{j,k}\lambda_{jk}^{(L)})\prod_{l=0}^{L-1} S_{\tau^{(l)}}(\tau^{(l+1)}\given a, \sum_{j,k}\lambda_{j,k}^{(l)}). 
\end{align*}  
The density function is obtained by taking the derivative of $1-S(t\given \{\tau^{(l)}, \lambda_{jk}^{(l)}\}_{l=0}^{L})$.

\subsection{Marginal Distribution of the WDR Event Time}\label{sec:marginal}
\begin{theorem}\label{thm:t_marginal}
If $t_i\sim \mbox{Weibull}_\tau(a, \lambda_{i\bullet\bullet})$ with $\lambda_{i\bullet\bullet}=\sum_{j,k}\lambda_{ijk}$ and $\lambda_{ijk}\sim \mbox{Gamma}(r_{jk}, 1/b_{ijk})$, the density function of $t_i$ given $a$, $\{r_{jk}\}$, and $\{b_{ijk}\}$ and $t_i>\tau$ is 
\begin{align*}
f(t_i\given \{r_{jk}\}_{j,k},\{b_{ijk}\}_{j,k})=a t_i^{a-1} c_i\sum_{m=0}^\infty\frac{(\rho_i+m) \delta_{im}b_{i(1)}^{\rho_i+m}}{(t_i^a-\tau^a+b_{i(1)})^{1+\rho_i+m}},
\end{align*}
and the cumulative density function is
\begin{align}
\mbox{Pr}(t_i<q\given  \{r_{jk}\}_{j,k},\{b_{ijk}\}_{j,k}, t_i>\tau)=1-c_i\sum_{m=0}^\infty\frac{\delta_{im}b_{i(1)}^{\rho_i+m}}{(q^a-\tau^a+b_{i(1)})^{\rho_i+m}},\label{cdf_convolution}
\end{align}
where $c_{i}=\prod_{j,k}\left(\frac{b_{ijk}}{b_{i(1)}}\right)^{r_{jk}}$, $b_{i(1)}=\max_{j,k} b_{ijk}$, $\rho_i=\sum_{j,k}r_{jk}$, $\delta_{i0}=1$,\\ $\delta_{im+1}=\frac{1}{m+1}\sum_{h=1}^{m+1}h\gamma_{ih}\delta_{im+1-h}$ for $m\geq 0$, and $\gamma_{ih}=\sum_{j,k}\frac{r_{jk}}{h}\left(1-\frac{b_{ijk}}{b_{i(1)}}\right)^h$.
\end{theorem}
It is difficult to use the density or cumulative distribution functions of $t_i$ in the form of series, but we can use a finite truncation as an approximate. Concretely, as $\mbox{Pr}(t_i<\infty\given a, \{r_{jk}\}_{j,k},\{b_{ijk}\}_{j,k})=c_i\sum_{m=0}^\infty\delta_{im}=1$, we find an $M$ so large that $c_i\sum_{m=0}^M\delta_{im}$ close to $1$ (say no less than $0.9999$), and use $1-c_i\sum_{m=0}^M\frac{\delta_{im}b_{i(1)}^{\rho_i+m}}{(q^a-\tau^a+b_{i(1)})^{\rho_i+m}}$ as an approximation. Consequently, sampling $t_i$ is feasible by inverting the approximated cumulative distribution function for general cases. We have tried prediction by finite truncation on some synthetic data where $\tau=0$ and $a=1$ and found that $M$ is mostly between 10 and 30, which is computationally feasible. 
\begin{proof}%[Proof of Theorem \ref{noninteger-shape}.]
~~We first study the distribution of gamma convolution. Specifically, if $\lambda_s\overset{ind}{\sim} \mbox{Gamma}(r_s, 1/b_s)$ with $r_s,b_s\in \mathbb R_+$, then the density function of $\lambda=\sum_{s=1}^S\lambda_s$ can be written in a form of series \citep{moschopoulos1985distribution} as 
\begin{align*}
f(\lambda\given r_1, b_1,\cdots,r_S, b_S)=\begin{cases}
c\sum_{m=0}^\infty\frac{\delta_{m} \lambda^{\rho+m-1}\exp({-\lambda b_{(1)}})}{\Gamma(\rho+m)/b_{(1)}^{\rho+m}} &\mbox{ if } \lambda>0,\\
0 &\mbox{ otherwise}, 
\end{cases}
\end{align*} 
where $c=\prod_{s=1}^S\left(\frac{b_s}{b_{(1)}}\right)^{r_s}$, $b_{(1)}=\max_s b_s$, $\rho=\sum_{s=1}^S r_s$, $\delta_{0}=1$, $\delta_{m+1}=\frac{1}{m+1}\sum_{h=1}^{m+1} h \gamma_{h} \delta_{m+1-h}$ and $\gamma_{h}=\sum_{t=1}^T r_{t}\left(1-\frac{b_{t}}{b_{(1)}}\right)^{h}/h$. 
\citet{moschopoulos1985distribution} has proved that $0<\gamma_{ih}\leq \rho_i b_{i0}^h/h$ and $0<\delta_{im}\leq \frac{\Gamma(\rho_i+m)b_{i0}^m}{\Gamma(\rho_i)m!}$ where $b_{i0}=max_{j,k}(1-\frac{b_{ijk}}{b_{i(1)}})$. We want to show the PDF of $t_i$,
\begin{align}
&f(t_i\given \{r_{jk}\}_{j,k},\{b_{ijk}\}_{j,k})\nonumber\\
=&\int_0^{\infty} f(t_i\given \lambda_{i\bullet\bullet})f(\lambda_{i\bullet\bullet}\given \{r_{jk}\}_{j,k},\{b_{ijk}\}_{j,k}) d\lambda_{i\bullet\bullet}\nonumber\\
=&\int_0^{\infty}\sum_{m=0}^\infty \frac{c_i\delta_{im}at_i^{a-1} \lambda_{i\bullet\bullet}^{\rho_i+m}\exp (-(t_i^a-\tau^a)\lambda_{i\bullet\bullet}-b_{i(1)}\lambda_{i\bullet\bullet})}{\Gamma(\rho_i+m)/b_{i(1)}^{\rho_i+m}} d\lambda_{i\bullet\bullet} \nonumber\\
=&\sum_{m=0}^\infty \int_0^{\infty} \frac{c_i\delta_{im}at_i^{a-1} \lambda_{i\bullet\bullet}^{\rho_i+m}\exp (-(t_i^a-\tau^a)\lambda_{i\bullet\bullet}-b_{i(1)}\lambda_{i\bullet\bullet})}{\Gamma(\rho_i+m)/b_{i(1)}^{\rho_i+m}} d\lambda_{i\bullet\bullet}  \label{pdf_interchange_new}\\
=&a t_i^{a-1} c_i\sum_{m=0}^\infty\frac{(\rho_i+m) \delta_{im}b_{i(1)}^{\rho_i+m}}{(t_i^a-\tau^a+b_{i(1)})^{1+\rho_i+m}},\nonumber
\end{align}
which suffices to prove the equality in \eqref{pdf_interchange_new}. Specifically,
\begin{align*}
&f(t_i\given n_i,\lambda_{i\bullet\bullet})f(\lambda_{i\bullet\bullet}\given \{r_{jk}\}_{j,k},\{b_{ijk}\}_{j,k})\\
=&a t_i^{a-1} c_i\lambda_{i\bullet\bullet}^{\rho_i} b_{i(1)}^{\rho_i}\exp (-(t_i^a-\tau^a)\lambda_{i\bullet\bullet}-b_{i(1)}\lambda_{i\bullet\bullet})\sum_{m=0}^\infty \frac{\delta_{im}b_{i(1)}^m\lambda_{i\bullet\bullet}^m}{\Gamma(\rho_i+m)}\\
\leq &a t_i^{a-1}c_i\lambda_{i\bullet\bullet}^{\rho_i} b_{i(1)}^{\rho_i}\exp (-(t_i^a-\tau^a)\lambda_{i\bullet\bullet}-b_{i(1)}\lambda_{i\bullet\bullet}) \sum_{m=0}^\infty \frac{(b_{i0}b_{i(1)}\lambda_{i\bullet\bullet})^m}{\Gamma(\rho_i)m!}\\
=& a t_i^{a-1}c_i\lambda_{i\bullet\bullet}^{\rho_i} b_{i(1)}^{\rho_i} \exp (-(t_i^a-\tau^a)\lambda_{i\bullet\bullet}-b_{i(1)}\lambda_{i\bullet\bullet} +b_{i0}b_{i(1)}\lambda_{i\bullet\bullet} ),
\end{align*}
which shows the uniform convergence of $f(t_i\given n_i,\lambda_{i\bullet\bullet})f(\lambda_{i\bullet\bullet}\given \{r_{jk}\}_{j,k},\{b_{ijk}\}_{j,k})$. So, the integration and countable summation are interchangeable, and consequently \eqref{pdf_interchange_new} holds. Next, we want to calculate the CDF of $t_i$,
\begin{align}
\mbox{Pr}(t_i<q\given n_i, \{r_{jk}\}_{j,k},\{b_{ijk}\}_{j,k})=&\int_0^q a t_i^{a-1} c_i\sum_{m=0}^\infty\frac{(\rho_i+m) \delta_{im}b_{i(1)}^{\rho_i+m}}{(t_i^a-\tau^a+b_{i(1)})^{1+\rho_i+m}} dt_i \nonumber\\
=& \sum_{m=0}^\infty \int_0^q a t_i^{a-1} c_i \frac{(\rho_i+m) \delta_{im}b_{i(1)}^{\rho_i+m}}{(t_i^a-\tau^a+b_{i(1)})^{1+\rho_i+m}}  dt_i. \label{cdf_interchange_new}
\end{align}
It suffices to show \eqref{cdf_interchange_new}. Specifically,
\begin{align*}
&\sum_{m=0}^\infty\frac{(\rho_i+m) \delta_{im}b_{i(1)}^{\rho_i+m}}{(t_i^a-\tau^a+b_{i(1)})^{1+\rho_i+m}} \\
=&\sum_{m=0}^\infty\frac{\Gamma(1+\rho_i+m) \delta_{im}b_{i(1)}^{\rho_i+m}}{\Gamma(\rho_i+m)(t_i^a-\tau^a+b_{i(1)})^{n_i+\rho_i+m}}\\
\leq & \sum_{m=0}^\infty \frac{\Gamma(1+\rho_i+m)b_{i(1)}^{\rho_i+m}\Gamma(1+\rho_i)}{\Gamma(\rho_i+m)(t_i^a-\tau^a+b_{i(1)})^{1+\rho_i+m}\Gamma(\rho_i)m!}\\
=&\frac{\Gamma(\rho_i+1)b_{i(1)}^{\rho_i} }{\Gamma(\rho_i)(t_i^a-\tau^a+b_{i(1)})^{1+\rho_i} } \sum_{m=0}^\infty\left[\frac{\Gamma(1+\rho_i+m)}{\Gamma(1+\rho_i)m!}\left(\frac{b_{i(1)}}{t_i^a-\tau^a+b_{i(1)}}\right)^m \right]\\
=&\frac{ \Gamma(\rho_i+1)b_{i(1)}^{\rho_i}(t_i^a-\tau^a)^{1+\rho_i} }{\Gamma(\rho_i)(t_i^a-\tau^a+b_{i(1)})^{2(1+\rho_i)}}.
\end{align*}
The last equation holds because the summation of a negative binomial probability mass function is 1. Therefore, $f(t_i\given a, \{r_{jk}\}_{j,k},\{b_{ijk}\}_{j,k})$ is uniformly convergent and \eqref{cdf_interchange_new} holds. Calculating the integration, we obtain the CDF of $t_i$.
\end{proof}

\section{WDR Inference}\label{app:mcmc}
\subsection{MCMC}
Let $\tau_i$, $T_i$ and $T_{ic}$ denote the left truncation time, the observed failure time, and the right censoring time, respectively, for subject $i=1,\ldots,n$, with $\tau_i$ less than $T_i$ or $T_{ic}$. Since left censoring is uncommon and not shown in our real data, we only consider right censoring in our inference and leave to readers other types of censoring which can be analogously done.
The inference by MCMC accommodating missing event time or missing event types proceeds by iterating the following steps.
\begin{enumerate} [Step 1:]
\item If $y_i$ is observed, we first sample $\kappa_{iy_i}$ by
%\label{infer_lc_k*}\\
$$\mbox{Pr}(\kappa_{iy_i}=k\given y_i,\cdots)=\frac{\lambda_{iy_ik}}{\sum_{k'=1}^K \lambda_{iy_ik'}}.$$
If $y_i$ is unobserved which means a missing event type, we sample $(y_i,\kappa_{iy_i})$ by 
$$\mbox{Pr}(y_i=j, \kappa_{iy_i}=k\given \cdots)=\frac{\lambda_{ijk}}{\sum_{j'=1}^S \sum_{k'=1}^K \lambda_{ij'k'}}.$$
We then denote $m_{jk}=\sum_{i:y_i=j} \bm 1 (\kappa_{iy_i}=k)$. Define $n_{ijk}=1$ if $y_i=j$ and $\kappa_{iy_i}=k$, and otherwise $n_{ijk}=0$.
The above sampling procedure means that given the event type $y_i$, we sample the index of the sub-event that has the minimum event time.
\item Update $t_{i}$ for $i=1,\cdots,n$, $j=1,\cdots, J$ and $k=1,\cdots,K$.\label{sample_t}
\begin{enumerate}
\item If the event time $T_i$ is observed, we set $t_i=T_i$. 
\item Otherwise, we sample $t_i \sim \mbox{Weibull}_{(T_{ic},\infty)}(a, \sum_{j=1}^S\sum_{k=1}^K \lambda_{ijk})$ where \\ $\mbox{Weibull}_{(T_{ic},\infty)}(\cdot,\cdot)$ is a truncated Weibull distribution so that $t_i\in (T_{ic},\infty)$. Note $T_{ic}=0$ if both event time and censoring time are missing for observation~$i$. 
\end{enumerate}

\item Sample ($\lambda_{ijk}\given-)\sim \mbox{Gamma}\left(r_{jk}+n_{ijk},\frac{\exp({\xv_i^\T \betav_{jk}})}{1+(t_i^a-\tau_i^a) \exp({\xv_i^\T \betav_{jk}})} \right)$, for $i=1,\cdots,n$, $j=1,\cdots,J$, and $k=1,\cdots,K$.

\item Sample $a$ by slice sampling. $a$ determines how the hazard varies over time, and we assume an improper prior of $a\in \mathbb{R}_+$, $p(a)\propto \bm{1}(a>0)$ to reduce the impact of the prior on the posterior. The full conditional distribution of $a$ is 
\begin{align*}
p(a\given \ldots)\propto 
%p(a) p({t_i,y_i}_i\given a,\ldots)=
a^n \prod_i \left[t_i^{a-1} \prod_{j,k}\left(1+(t_i^a-\tau_i^a) \exp({\xv_i^\T \betav_{jk}})\right)^{-n_{ijk}-r_{jk}}\right].
\end{align*}
If $\tau_i=0$, the unimodality of $p(a\given \ldots)$ can be shown so that slice sampling can be implemented without tuning parameters. Concretely, when $\tau_i=0$,
\begin{align*}
\frac{d \log p(a|\ldots)}{da}= \frac{n}{a} + \sum_i \log t_i-\sum_{i,j,k}(n_{ijk}+r_{jk})\frac{\exp({a\log t_i + \xv_i^\T \betav_{jk}})\log t_i }{1+\exp({a\log t_i + \xv_i^\T \betav_{jk}})}.
\end{align*}
Since $\frac{n}{a}$ is decreasing in $a$ while $\sum_{i,j,k}(n_{ijk}+r_{jk})\frac{\exp({a\log t_i + \xv_i^\T \betav_{jk}})\log t_i }{1+\exp({a\log t_i + \xv_i^\T \betav_{jk}})}$ is increasing in $a$, there must be at most one $a\in \mathbb{R}_+$ satisfying $\frac{d \log p(a|\ldots)}{da}=0$. So, $p(a|\ldots)$ is unimodal. We use the \texttt{mcmc} function in the \texttt{R} package \texttt{diversitree} \citep{diversitree}.

\item Sample $\betav_{jk}$, for $j=1,\cdots,J$ and $k=1,\cdots,K$, by P\'olya Gamma (PG) data augmentation.
First sample $(\omega_{ijk}\given-)\sim \mbox{PG}(r_{jk}+n_{ijk}, \xv_i^\T\betav_{jk}+\log( t_i^a-\tau_i^a) )$. Then sample $(\betav_{jk}\given-)\sim \mbox{MVN}(\bm \mu_{jk}, \bm\Sigma_{jk})$ where $\bm\Sigma_{jk}=\left(V_{jk}+\bm X^\T\Omega_{jk}\bm X\right)^{-1}$,\\ $\bm X=[\xv_1,\cdots,\xv_N]^\T$, $\Omega_{jk}=\mbox{diag}(\omega_{1jk},\cdots,\omega_{njk})$ and\\ $\bm\mu_{jk}=\bm\Sigma_{jk}\left[-\sum_{i=1}^{N}\left(\omega_{ijk}\log (t_i^a-\tau_i^a) +\frac{r_{jk}-n_{ijk}}{2}\right)\xv_i\right]$.
To sample from the P\'olya-Gamma distribution, we use a fast and accurate approximate sampler \citep{SoftplusReg_2016} that matches the first two moments of the original distribution; we set the truncation level of that sampler as five. 

\item Sample $(\alpha_{vjk}\given-)\sim \mbox{Gamma}\left(a_0+0.5,1/(b_0+0.5\beta_{vjk}^2)\right)$ for $v=0,\cdots,V$, $j=1,\cdots,J$ and $k=1,\cdots,K$.
\item Sample $r_{jk}$ and $\gamma_{0j}$, for $j=1,\cdots,J$ and $k=1,\cdots,K$, by Chinese restaurant table (CRT) data augmentation \citep{zhou2015negative}.

First sample $(n^{(2)}_{ijk}\given-)\sim \mbox{CRT}(n_{ijk},r_{jk})$, and $(l_{jk}\given-)\sim \mbox{CRT}(\sum_{i=1}^N n^{(2)}_{ijk}, \gamma_{0j}/K)$. Then sample $(r_{jk}\given-)\sim \mbox{Gamma}\left(\sum_{i=1}^N n^{(2)}_{ijk}+\gamma_{0j}/K, \frac{1}{c_{0j}+\sum_{i=1}^N \log(1+(t_i^a-\tau_i^a) \exp({\xv_i^\T \betav_{jk}}))} \right)$, and  
$(\gamma_{0j}\given-)\sim \mbox{Gamma}\left(d_0+\sum_{k=1}^K l_{jk}, \frac{1}{e_0-\frac{1}{K}\sum_{k=1}^K \log(1-p_{jk})}\right)$,\\
 where $p_{jk}=\frac{\sum_{i=1}^N \log(1+(t_i^a-\tau_i^a) \exp({\xv_i^\T \betav_{jk}}))}{c_{0j}+\sum_{i=1}^N \log(1+(t_i^a-\tau_i^a)\exp({\xv_i^\T \betav_{jk}}))}$.
\item Sample $(c_{0j}\given-)\sim \mbox{Gamma}\left(d_1+\gamma_{0j},\frac{1}{e_1+\sum_{k=1}^K r_{jk}}\right)$ for $j=1,\cdots,J$.
\item (Optional.) Prune unneeded sub-events. For $j=1,\cdots,J$ and $k=1,\cdots, K$, if $m_{jk}$ is small enough, say $m_{jk}=0$ or $m_{jk}\leq 1\%\times n$, prune sub-event $k$ of competing event $j$ by setting $\lambda_{ijk} = 0$ and $t_{ijk}= \infty$ for $i=1,\ldots,n$. 
\end{enumerate}
Steps 1 through 8 are MCMC updates of parameters. Although the item weights $\{r_{jk}\}$ in the gamma processes are almost surely positive, the latent count allocation of $n_{ijk}$ and $n^{(2)}_{ijk}$ in Steps 1 and 7 makes it possible to prune a sub-event that is not an important component in a phase-one race within a competing event. Step 9 is used to explicitly prune the unneeded nonlinear modeling capacity. Specifically, $m_{jk}$ counts the number of observations that have sub-event $k$ winning the race within competing event $j$. The sub-event $k$ is redundant if $m_{jk}$ is small compared to the total number of observations $n$. For small data sets, we prune it if $m_{jk}=0$. For larger data sets with $n\geq$10,000,  we suggest pruning it if $m_{jk}\leq (n\times 1\%)$.

Step~1 tries to find to which subevent the event of type $y_i$ belongs by sampling $\kappa_{iy_i}$ from a multinomial distribution. This latent counts allocation is similar to those in \citet{SoftplusReg_2016} and \citet{zhou2016augmentable}, where a deep gamma hierarchy is used. Using a similar deep hierarchical model does not remarkably increase the capacity of WDR, possibly because of the following reason.
In \citet{SoftplusReg_2016} and \citet{zhou2016augmentable}, a Bernoulli-Poisson link is used, and the latent counts can be greater than one and are allocated to different layers of the deep gamma hierarchy.  
But in our context, the latent count is always $1$ or $0$, indicating whether a subject suffers from a latent subevent $k$ under competing event $j$ or not. Effectively, a deep gamma hierarchy is analogous to WDR where we have one latent layer with a potentially infinite number of nodes (subevents). We refer the readers to the two aforementioned papers for details. In addition, another reason we do not consider such a deep hierarchy is that it may sabotage the interpretability of WDR.

\subsection{Maximum a Posteriori Estimation}\label{sec:map}
We propose a Maximum a posteriori (MAP) estimation of the WDR parameters for big data applications. Since the MAP estimations do not involve latent count allocations as in Step 1 of the MCMC algorithm, and thus the sub-events cannot be explicitly pruned, we can determine the number of latent sub-events $K$ by cross-validation and model selection rules, like AIC or BIC. We show the estimations for data with right-censored subjects. Other types of censoring can be done analogously. 

With the reparameterization that $\lambda_{ijk}=\tilde{\lambda}_{ijk}\exp({\bm x_i^\T\bm\beta_{jk}})$ where $\tilde{\lambda}_{ijk}\overset{iid}{\sim}\mbox{Gamma}(r_{jk},1)$ we first find $p_i$, the likelihood of observation $i$ having event type $y_i$ at event time $t_i$.
\begin{align*}
p_i=\E\left(P(t_i,y_i\given \bm \lambda_i)\right)= \int \left(p_{t_i}\times p_{y_i}\right) p(\tilde {\bm \lambda}_i\given \bm r) d \tilde {\bm \lambda}_i
\end{align*}
where $\tilde{\bm\lambda}_i=\{\tilde{\lambda}_{ijk}\}_{j,k}$,
$p(\tilde {\bm \lambda}_i\given \bm r)=\prod_{j,k}\mbox{Gamma}(\tilde\lambda_{ijk}\given r_{jk},1)$, $\bm r=\{r_{jk}\}_{j,k}$, $\mbox{Gamma}(.\given r_{jk},1)$ is the pdf of a gamma distribution with shape $r_{jk}$ and scale $1$, and  
\begin{align*}
p_{t_i}=& \begin{cases}at_i^{a-1} (\sum_{j,k}\tilde{\lambda}_{ijk}\exp({\bm x_i^\T\bm\beta_{jk}})) \exp\big(-(t_i^a-\tau_i^a)\sum_{jk} \tilde{\lambda}_{ijk}\exp({\bm x_i^\T\bm\beta_{jk}})\big) \mbox{ if } t_i \mbox{ is uncensored,}\\% and observed,}\\
\exp\big(-T_{ic}^a \sum_{jk} \tilde{\lambda}_{ijk}\exp({\bm x_i^\T\bm\beta_{jk}})\big) ~~~~ \mbox{ if } t_i \mbox{ is right-censored at } T_{ic}\\%,\mbox{ i.e., } t_i>T_{ic},\\
1~~~~ \mbox{ if } t_i \mbox{ is missing, but } y_i \mbox{ is not,}
\end{cases} \\
p_{y_i}=& \begin{cases} \frac{\sum_k \tilde{\lambda}_{iy_ik}\exp({\bm x_i^\T\bm\beta_{y_ik}})}{\sum_{j,k}\tilde{\lambda}_{ijk}\exp({\bm x_i^\T\bm\beta_{jk}})} & \mbox{ if } y_i \mbox{ is not missing,}\\
1 & \mbox{ if } y_i \mbox{ is missing, but } t_i \mbox{ is not.}
\end{cases}
\end{align*}
We do not define the likelihood of subject $i$ if both $t_i$ and $y_i$ are missing and remove such observations from the data. For brevity, we write $p_t(\tilde{\bm\lambda}_i\given \bm r)$ as $p_{t_i}$ and $p_y(\tilde{\bm\lambda}_i\given \bm r)$ as $p_{y_i}$.

\looseness -1 Imposing a prior $p(a)$ on $a$, $p(\bm\beta_{jk})$ on $\bm\beta_{jk}$ and $p(r_{jk})$ on $r_{jk}$, the log posterior is 
\begin{align}
\log P &= \sum_{i} \log p_{i} + \log p(a) + \sum_{j,k}\log p(\bm\beta_{jk}) +\sum_{j,k}\log p(r_{jk})+C \label{map_posterior}
\end{align}
where $C$ is a constant function of $a$, $\{\bm\beta_{jk}\}$ and $\{r_{jk}\}$. In practice we assume an improper prior $p(a)\propto 1$ on $a$, a Student's $t$ distribution with degrees of freedom $\alpha$ on each element of $\bm\beta_{jk}$ and a Gamma$(0.01/K, 1/0.01)$ prior on $r_{jk}$. We also find that a Gamma$(1/K, 1)$ prior on $r_{jk}$ or an $L_2$-regularizer, $0.001||\bm r||_2$, is more numerically stable. Subsequently, we have 
\begin{align*}
\log P = \sum_{i} \log p_i
+\sum_{v,j,k}-\frac{\alpha+1}{2}\log\left(1+\beta_{vjk}^2/\alpha\right) +\sum_{j,k}\left[(0.01/K-1)\log r_{jk}-0.01 r_{jk}\right]+c
\end{align*}
where $c$ is also a constant function of $a$, $\{\bm\beta_{jk}\}$ and $\{r_{jk}\}$. For simplicity, we define $\bm\beta=\{\bm\beta_{jk}\}_{j,k}$. We want to maximize $\log P$ with respect to $\bm\beta$ and $\bm r$. The difficulty lies in $p_i$ being the expectation of $p_{t_i}\times p_{y_i}$ over $\tilde{\bm \lambda}_i$ which is a random variable parameterized by $\bm r$. Now we show how to approximate the derivatives of $\log p_i$ by Monte Carlo simulation, score function gradients, and self-normalization. Specifically,
\begin{align}
\nabla_{a, \bm\beta}\log p_i=\frac{\int \left[\nabla_{a,\bm\beta} \left(p_{t_i}\times p_{y_i}\right) \right] p(\tilde {\bm \lambda}_i\given \bm r) d \tilde {\bm \lambda}_i}{\int \left(p_{t_i}\times p_{y_i}\right) p(\tilde {\bm \lambda}_i\given \bm r) d \tilde {\bm \lambda}_i}
\approx \frac{\frac{1}{M}\sum_{m=1}^M \nabla_{a, \bm\beta}\left[p_t(\tilde{\bm\lambda}_i^{(m)}\given \bm r)\times p_y(\tilde{\bm\lambda}_i^{(m)}\given \bm r) \right] }{\frac{1}{M}\sum_{m=1}^M \left[p_t(\tilde{\bm\lambda}_i^{(m)}\given \bm r)\times p_y(\tilde{\bm\lambda}_i^{(m)}\given \bm r) \right] }\label{eq:partial_beta}
\end{align}
where $M$ is a reasonably large number, say $10$, $\tilde{\bm \lambda}_i^{(m)}=\{\tilde{\lambda}_{ijk}^{(m)}\}_{jk}$ and $\tilde{\lambda}_{ijk}^{(m)}\overset{iid}{\sim}\mbox{Gamma}(r_{jk},1)$, $i=1,\cdots,n$ and $m=1,\cdots,M$. With the fact that $\nabla_{\bm r} p(\tilde {\bm \lambda}_i\given \bm r) =p(\tilde {\bm \lambda}_i\given \bm r) \nabla_{\bm r} \log p(\tilde {\bm \lambda}_i\given \bm r)$, 
\begin{align}
\nabla_{\bm r}\log p_i &=\frac{\int \nabla_{\bm r}\left[ \left(p_{t_i}\times p_{y_i}\right) p(\tilde {\bm \lambda}_i\given \bm r) \right] d \tilde {\bm \lambda}_i}{\int \left(p_{t_i}\times p_{y_i}\right) p(\tilde {\bm \lambda}_i\given \bm r) d \tilde {\bm \lambda}_i}\nonumber\\
&=\frac{\int \left(p_{t_i}\times p_{y_i}\right) \nabla_{\bm r}\log p(\tilde {\bm \lambda}_i\given \bm r) p(\tilde {\bm \lambda}_i\given \bm r) d \tilde {\bm \lambda}_i}{\int \left(p_{t_i}\times p_{y_i}\right) p(\tilde {\bm \lambda}_i\given \bm r) d \tilde {\bm \lambda}_i}\nonumber\\
&\approx \frac{\frac{1}{M}\sum_{m=1}^M p_t(\tilde{\bm\lambda}_i^{(m)}\given \bm r)\times p_y(\tilde{\bm\lambda}_i^{(m)}\given \bm r) \nabla_{\bm r} \log p(\tilde {\bm \lambda}_i^{(m)}\given \bm r) }{\frac{1}{M}\sum_{m=1}^M \left[p_t(\tilde{\bm\lambda}_i^{(m)}\given \bm r)\times p_y(\tilde{\bm\lambda}_i^{(m)}\given \bm r) \right] }\notag\\
&=\sum_{m=1}^M \frac{ p_t(\tilde{\bm\lambda}_i^{(m)}\given \bm r)\times p_y(\tilde{\bm\lambda}_i^{(m)}\given \bm r) } {{\sum_{m'=1}^M \left[p_t(\tilde{\bm\lambda}_i^{(m')}\given \bm r)\times p_y(\tilde{\bm\lambda}_i^{(m')}\given \bm r) \right] }}\nabla_{\bm r} \log p(\tilde {\bm \lambda}_i^{(m)}\given \bm r) .\label{eq:partial_r}
\end{align}
Therefore, we can approximate the derivatives of $\log P$ with respect to $a$, $\bm\beta$ and $\bm r$ by plugging in \eqref{eq:partial_beta} and \eqref{eq:partial_r}, respectively, and maximize $\log P$ by (stochastic) gradient ascent.
%\section{Data description}

\subsection{Cumulative Incidence Function for WDR }\label{app:quantification}
The cumulative incidence function (CIF) of subject $i$ for event $j$ at time $t$ is $\mbox{CIF}_j(i, t) =\mbox{Pr}(t_i\leq t, y_i=j )$   \citep{fine1999proportional,kalbfleisch2011statistical,crowder2001classical}, indicating the probability of event $j$ by time $t$. Given time-invariant $\xv_i$, a left truncation time $\tau$, $a$, $\{r_{jk}\}_{j,k}$ and $\{\betav_{jk}\}_{j,k}$, WDR has
\begin{align*}%\textstyle
\mbox{CIF}_j(i,t)=\mbox{Pr}(t_i\leq t, y_i=j)=E\left[ \frac{\sum_{k}\lambda_{ijk}}{\sum_{j',k}\lambda_{ij'k}}\Big(1-\exp\big({-(t^a-\tau^a) \sum_{j',k}\lambda_{ij'k} }\big) \Big)
\right],
\end{align*}
where the expectation is taken over  $\lambda_{ijk}\sim \mbox{Gamma}(r_{jk},\exp({\xv_i^\T\betav_{jk}}))$. The CIF for WDR can be evaluated using Monte Carlo estimation 
if we have point estimates or a collection of post-burn-in MCMC samples of $r_{jk}$ and $\betav_{jk}$. 

Given time-varying covariates $\bm X_i$ whose values are updated to $\xv_i^{(0)},\ldots,\xv_i^{(L)}$ at times $\tau^{(0)},\ldots,\tau^{(L)}$ before $t$, the cumulative incidence function of WDR is 
\begin{align*}
&\mbox{CIF}_j(i,t)\\
=&\mbox{Pr}(t_i\leq \tau^{(1)}, y_i=j)+\mbox{Pr}(\tau^{(1)}<t_i\leq \tau^{(2)}, y_i=j)+\cdots+\mbox{Pr}(\tau^{(L)}<t_i\leq t, y_i=j)\nonumber\\
=&E\left[ \frac{\sum_{k}\lambda_{ijk}^{(0)}}{\sum_{j',k}\lambda_{ij'k}^{(0)}}\left(1-\exp({-[(\tau^{(1)})^a-(\tau^{(0)})^a] \sum_{j',k}\lambda_{ij'k}^{(0)} }) \right)\right] + \nonumber\\
&E\left[ \frac{\sum_{k}\lambda_{ijk}^{(1)}}{\sum_{j',k}\lambda_{ij'k}^{(1)}}\left(1-\exp({-[(\tau^{(2)})^a-(\tau^{(1)})^a] \sum_{j',k}\lambda_{ij'k}^{(1)} }) \right)\exp({-[(\tau^{(1)})^a-(\tau^{(0)})^a] \sum_{j',k}\lambda_{ij'k}^{(0)} })\right]  \nonumber\\
&+\cdots+\nonumber\\
&E\left[ \frac{\sum_{k}\lambda_{ijk}^{(L)}}{\sum_{j',k}\lambda_{ij'k}^{(L)}}\left(1-\exp({-[t^a-(\tau^{(L)})^a] \sum_{j',k}\lambda_{ij'k}^{(L)} }) \right)\prod_{l=0}^{L-1}\exp({-[(\tau^{(l+1)})^a-(\tau^{(l)})^a] \sum_{j',k}\lambda_{ij'k}^{(l)} })\right]
\end{align*}
where the expectations are taken over $\lambda_{ijk}^{(l)}$'s with $\lambda_{ijk}^{(l)}\sim \mbox{Gamma}(r_{jk}, \exp({\bm x_i^{(l)\T} \betav_{jk}}))$.

\section{Data Synthesis and Experiment Settings}\label{sec:tvsimulation}
\begin{table}[!t]
\centering
\renewcommand{\arraystretch}{1} % Default value: 1
\makebox[\linewidth]{
\resizebox{\linewidth}{!}{
\begin{tabular}{lll}%\vspace{-6.5mm}
\toprule
Data 1 & Data 2 & Data 3\\
\midrule
$x_{i1}\sim \mbox{U}(-1,1)$   &$x_{i1}\sim \mbox{U}(-1,1)$, $i=1,\ldots, 2000$ & $x_{i1}\sim \mbox{U}(-1,1)$, $i=1,\ldots, 2000$ \\
$x_{i2}\sim \mbox{U}(0,1)$    &  $x_{i2}\sim \mbox{U}(0,1)$, $x_{i3}\sim \mbox{U}(-1,0)$, $i\leq 1000$ & $x_{i2}\sim \mbox{U}(0,1)$, $x_{i3}\sim \mbox{U}(-1,0)$, $i\leq 1000$\\
$x_{i3}\sim \mbox{U}(0,1)$    &  $x_{i2}\sim \mbox{U}(-1,0)$, $x_{i3}\sim \mbox{U}(0,1)$, $i> 1000$ & $x_{i2}\sim \mbox{U}(-1,0)$, $x_{i3}\sim \mbox{U}(0,1)$, $i> 1000$ \\
%$\xv_i=(x_{i1},x_{i2},x_{i3})^\T$ & $\xv_i=(x_{i1},x_{i2},x_{i3})^\T$ & $\xv_i=(x_{i1},x_{i2},x_{i3})^\T$ \\
$\bv_1 =(1,2,-1)^\T$, $\bv_2 =(1,-1,2)^\T$ & $\bv_1 =(1,2,4)^\T$, $\bv_2 =(1,4,2)^\T$ & $\bv_1 =(1,-2,1)^\T$, $\bv_2 =(1,-1,2)^\T$\\
%$t_{i1}\sim\mbox{Weibull}(0.8, \exp(\xv_i^\T\bv_1))$ & $t_{i1}\sim\mbox{Weibull}(3,|\sinh(\xv_i^\T\bv_1)|)$  & $t_{i1} \sim \mbox{Weibull}(1, \exp({ (\xv_i^\T\bv_1)^2}))$ \\
%$t_{i2}\sim\mbox{Weibull}(0.8, \exp({\xv_i^\T\bv_2}))$ & $t_{i2}\sim\mbox{Weibull}(3, \cosh(\xv_i^\T\bv_2))$  & $t_{i2} \sim \mbox{Weibull}(1, \exp({ (\xv_i^\T\bv_2)^2}))$  \\
% $t_i=\min(t_{i1},t_{i2}, T_{r.c.}=1.6)$  &$t_i=\min(t_{i1},t_{i2}, T_{r.c.}=1.3)$ & $t_i=\min(t_{i1},t_{i2}, T_{r.c.}=0.6)$\\
Right censoring: $T_{r.c.}=1.6$ &Right censoring: $T_{r.c.}=1.3$ &Right censoring: $T_{r.c.}=0.6$ \\
 \midrule
Data 4 & Data 5 & Data 6\\ 
 \midrule
$x_{i1}, x_{i2}, x_{i3}\sim \mbox{U}(-1,1)$ & $x_{i1}, x_{i2}, x_{i3}\sim \mbox{U}(-1,1)$ &$x_{i1}, x_{i2}, x_{i3}\sim \mbox{U}(-1,1)$ \\
$\bv_1 =(-0.6,0.2,-0.8)^\T$ & $\bv_1 =(-0.9,0.2,1.6)^\T$ & $\bv_1 =(-0.6,0.2,-0.8)^\T$\\
$\bv_2 =(-0.8,0.2,-0.6)^\T$ & $\bv_2 =(1.6,0.2,-0.9)^\T$ & $\bv_2 =(-0.8,0.2,-0.6)^\T$\\
$\bv_3 =(0.9,0.7,0.6)^\T$ &$\bv_3 =(-1.1,-0.1,0.1)^\T$ & \\
$\bv_4 =(0.6,0.7,0.9)^\T$ &$\bv_4 =(0.1,-0.1,-1.1)^\T$ & \\
%$t_{i1}\sim\mbox{logNormal}(\mu_{i1}, 0.2^2)$ & $t_{i1}\sim\mbox{logNormal}(\mu_{i1}, 0.2^2)$ & $t_{i1}\sim\mbox{logNormal}(\mu_{i1}, 0.05^2)$ \\
%$\textstyle{\mu_{i1}=\frac{\xv_i^\T\bv_3\exp(\xv_i^\T\bv_1)- \xv_i^\T\bv_4\exp(\xv_i^\T\bv_2)}{\exp(\xv_i^\T\bv_1)+\exp(\xv_i^\T\bv_2)}}$  & $\mbox{logit}(\mu_{i1}) = \bv_1^\T \xv_i\xv_i^\T\bv_3 - \bv_2^\T \xv_i\xv_i^\T\bv_4 $ & $\mbox{logit}(\mu_{i1})=\cosh(\xv_i^\T \bv_1) - \cosh(\xv_i^\T \bv_2)$ \\
%$t_{i2}\sim\mbox{logNormal}(\mu_{i2}, 0.2^2)$ &$t_{i2}\sim\mbox{logNormal}(\mu_{i2}, 0.2^2)$  & $t_{i2}\sim\mbox{logNormal}(\mu_{i2}, 0.05^2)$\\
%$\textstyle{\mu_{i2}=\frac{\xv_i^\T\bv_4\exp(\xv_i^\T\bv_2)- \xv_i^\T\bv_3\exp(\xv_i^\T\bv_1)}{\exp(\xv_i^\T\bv_1)+\exp(\xv_i^\T\bv_2)}}$ & $\mbox{logit}(\mu_{i2}) = \bv_2^\T \xv_i\xv_i^\T\bv_4 - \bv_1^\T \xv_i\xv_i^\T\bv_3$ & $\mbox{logit}(\mu_{i2})=\cosh(\xv_i^\T \bv_2) - \cosh(\xv_i^\T \bv_1)$\\
%$t_i=\min(t_{i1},t_{i2},T_{r.c.}=1.1)$  &$t_i=\min(t_{i1},t_{i2},T_{r.c.}=1.7)$ & $t_i=\min(t_{i1},t_{i2},T_{r.c.}=1.7)$\\
Right censoring: $T_{r.c.}=1.1$ & Right censoring: $T_{r.c.}=1.7$ & Right censoring: $T_{r.c.}=1.7$ \\
\bottomrule
\end{tabular}%\vspace{-0.5mm}
%\end{wraptable} 
}}
\caption{$\xv_i$ and $\bv$ for synthetic data with $\xv_i=(x_{i1}, x_{i2}, x_{i3})\in\mathbb{R}^3$.}
\label{tab:synthetic} 
\end{table}

The distribution of $\xv_i$ and the values of $\bv$'s for data synthesis are provided in Tables~\ref{tab:synthetic} and~\ref{tab:xbet}. We simulate synthetic data 1 to 6 with time-varying covariates using Algorithm~\ref{alg:tvsimulation}. Concretely, we allow up to $L$ updates of covariates and $L$ can differ among subjects. With updated covariates at time $\tau^{(l)}$,  we simulate the event time from a Weibull or log-normal distribution that is left truncated at $\tau^{(l)}$. If the event time is greater than $\tau^{(l+1)}$, we update the covariates at $\tau^{(l+1)}$ and repeat this procedure. Otherwise, we stop. Right censoring is also allowed.

\begin{table}[t]
\centering
\renewcommand{\arraystretch}{1} % Default value: 1
\makebox[\linewidth]{
\resizebox{\linewidth}{!}{
\begin{tabular}{lll}%\vspace{-6.5mm}
\toprule 
{Data 1} &  {Data 2} & {Data 3}\\
\midrule
$\xv_i\in \mathbb{R}^{10},~\xv_i\sim \mbox{N}(\bm 0, \mathrm{\bf I})$  & $\xv_i\in \mathbb{R}^{10},~\xv_i\sim \mbox{N}(\bm 0, \mathrm{\bf I})$ & $\xv_i\in \mathbb{R}^{10},~\xv_i\sim \mbox{U}(-\bm{1}, \bm 1)$  \\
$\bv_1=(-.1, -.1,  0,  .2, -.2,  .2,  .3,  .1,  .1, -.3)$ & $\bv_1=(-.2, -.1,  .1,  .4, -.3,  .4,  .4,  .2,  .1, -.4)$ & $\bv_1=(.5, .7, 1.1, 1.8, .4, 1.8, 1.9, 1.3, 1.3, .1)$\\
$\bv_2=(-.2, -.2,  .1, -.1,  .2,  0,  .1, .3, -.1,  .2)$ & $\bv_2=(-.3, -.3,  .2, -.1,  .3,  0,  .2,  .5, -.1,  .3)$ & $\bv_2=(.1, 1.3, 1.3, 1.9, 1.8, .4, 1.8, 1.1, .7, .5)$\\
\midrule
{Data 4} &{Data 5} & {Data 6}\\
\midrule
$\xv_i\in \mathbb{R}^{10},~\xv_i\sim \mbox{U}(-\bm{1}, \bm 1)$  & $\xv_i\in \mathbb{R}^{10},~\xv_i\sim \mbox{N}(\bm 0, \mathrm{\bf I})$ & $\xv_i\in \mathbb{R}^{10},~\xv_i\sim \mbox{N}(\bm 0, \mathrm{\bf I})$  \\
$\bv_1=( -.6,  .2, -.8,  1.6,  .3, -.8,  .5,  .7,  .6, -.3)$ & $\bv_1=(-.6,  .2, -.8,  1.6,  .3, -.8, .5,  .7,  .6, -.3)$ & $\bv_1=(-.6,  .2, -.8,  1.6,  .3, -.8, .5,  .7,  .6, -.3)$\\
$\bv_2=(-.3,  .6,  .7,  .5, -.8,  .3,  1.6, -.8,  .2, -.6)$   &$\bv_2=(-.3,  .6,  .7,  .5, -.8,  .3,  1.6, -.8,  .2, -.6)$ & $\bv_2=(-.3,  .6,  .7,  .5, -.8,  .3,  1.6, -.8,  .2, -.6)$ \\
 $\bv_3=(.9, .2, .7, .1, .3, .4, 0, .4, .9, .3)$   &$\bv_3=(1.5,  .4, -.6, -2.2,  1.1, 0,  0,  .9,  .8,  .6)$ & \\
 $\bv_4=( .3, .9, .4, 0, .4, .3, .1, .7, .2, .9)$      & $\bv_4=( .6,  .8,  .9, 0,  0,  1.1, -2.2, -.6, .4,  1.5)$  & \\
 \bottomrule
\end{tabular}%\vspace{-0.5mm}
%\end{wraptable} 
}}
\caption{$\xv_i$ and $\bv$ for synthetic data with $\xv\in\mathbb{R}^{10}$.}
\label{tab:xbet}\vspace{-3.5mm}
\end{table}

\begin{algorithm}[!t]
    \caption{Simulation of the survival data with time-varying covariates.}\label{alg:tvsimulation}
  \begin{algorithmic}[1]
    \INPUT Number of subjects $n$, number of competing events $J$, right censoring time $T_{r.c.}$, covariate distribution $P_{\xv}$, maximum number of covariate updates $L\in\mathbb{Z}_+$, potential covariate update times $\tau^{(0)}, \tau^{(1)},\ldots, \tau^{(L)}$ with $\tau^{(L)}<T_{r.c.}$, Weibull parameters $a$ and $\{\lambda_j(\xv)\}_{j=1}^J$
    \OUTPUT $\{t_i, y_i, \xv_i^{(0)},\ldots,\xv_i^{(L_i)}, \tau_i^{(0)},\ldots,\tau_i^{(L_i)}\}_{i=1}^n$
    \FOR{$i=1,\ldots,n$} 
    \STATE $l\leftarrow -1$
    \WHILE{$l <L$}
      \STATE $l \leftarrow l+1$
      \STATE Draw $\xv_i^{(l)}\sim P_{\xv}$  
      \STATE $t_{ij}\sim\mbox{Weibull}_{\tau^{(l)}}(a, \lambda_j(\xv_i^{(l)}))$ or $ \mbox{logNormal}_{\tau^{(l)}}(\mu_j(\xv_i^{(l)})), \sigma^2),~j=1,\ldots,J$
      \IF{$\min_j t_{ij}<\tau^{(l+1)}$}
        \STATE $t_i\leftarrow\min_j t_{ij}$, $y_i\leftarrow\mathop{\mathrm{argmin}}_j t_{ij}$
        \STATE {\bf break}
      \ENDIF    
    \ENDWHILE
    \IF{$t_i>T_{r.c.}$}
      \STATE $t_i \leftarrow  T_{r.c.}$, $y_i\leftarrow 0$ \texttt{\# 0 indicates right censoring}
    \ENDIF
    \STATE  $L_i\leftarrow l$
    \ENDFOR
  \end{algorithmic}
\end{algorithm}

We describe the experiment settings as follows. For the kernel Fine-Gray (KFG) model, we use the radial basis function kernel and select the kernel width from $2^{-5}, 2^{-4}, \ldots, 2^{5}$ by maximizing the partial likelihood of validation data, which are randomly sampled from and accounts for 20\% of the training data. For the random survival forests (RF), we set the number of trees equal to 1000 and
the number of splits equal to 2 if $\xv_i\in\mathbb{R}^{3}$ and equal to 4 if  $\xv_i\in\mathbb{R}^{10}$, which are roughly equal to the square root of the covariate dimensions. For the DeepHit and the piecewise constant hazards (PCH) models, we discretize the continuous time into 20 intervals of an equal length,  in each of which the survival or hazard function is constant, and use a feedforward neural network with the ReLU activation functions and two hidden layers, each of which has 20 nodes. Early stopping is implemented by incorporating a validation set, which is the same as those for the KFG model.
We use \texttt{R} for the MCMC algorithm of WDR and the package \texttt{riskRegression} \citep{riskRegression} for FG, the package \texttt{CoxBoost} \citep{coxboost} for the gradient boosting method in KFG, and the package \texttt{randomForestSRC} \citep{RSF} for RF. For DeepHit and PCH, we use the \texttt{Python} package \texttt{pycox}\footnote{\url{https://github.com/havakv/pycox}. Last access in February 2023.}.

\section{Supplementary Results}\label{app:additional_result}
\subsection{Brier Scores on Synthetic Data with Constant Covariates and No Left Truncation}\label{sec:brier_synthetic}
We take 20 random training-testing partitions of each synthetic data in Section~\ref{sec:synthetic:constant}, and report in Tables~\ref{tab:event1:constant} and~\ref{tab:event2:constant} the Brier scores (mean $\,\pm\,$ standard deviation) on the testing data. A smaller Brier score indicates a better model fit. On data~1 with linear covariate effects, WDR, FG, KFG, and RF are comparable. On other data with nonlinear covariate effects, FG does not perform well, and WDR, KFG, and RF are comparable. Note that DeepHit and PCH have a large variation in prediction accuracy over different time and data, possibly because their piecewise constant survival/hazard functions can be sensitive to time discretization.    

\begin{table}[!t]
\centering
\vspace{-3.5mm}
\begin{tabular}{lccccc}
  \hline
Data 1 & $t=0.1$ & $t=0.3$ & $t=0.5$ & $t=0.7$ & $t=0.9$ \\ 
  \hline
WDR & \textbf{0.123}$\pm$0.006 & \textbf{0.19}$\pm$0.004 & \textbf{0.207}$\pm$0.004 & \textbf{0.213}$\pm$0.004 & \textbf{0.216}$\pm$0.004 \\ 
  FG & 0.124$\pm$0.007 & 0.192$\pm$0.004 & 0.209$\pm$0.004 & 0.216$\pm$0.004 & 0.217$\pm$0.004 \\ 
  KFG & 0.124$\pm$0.007 & 0.193$\pm$0.004 & 0.21$\pm$0.004 & 0.216$\pm$0.004 & 0.218$\pm$0.004 \\ 
  RF & 0.125$\pm$0.007 & 0.198$\pm$0.003 & 0.22$\pm$0.003 & 0.226$\pm$0.003 & 0.231$\pm$0.002 \\ 
  DeepHit & 0.148$\pm$0.009 & 0.288$\pm$0.009 & 0.224$\pm$0.003 & 0.234$\pm$0.003 & 0.243$\pm$0.004 \\ 
  PCH & 0.128$\pm$0.007 & 0.205$\pm$0.005 & 0.225$\pm$0.006 & 0.232$\pm$0.004 & 0.236$\pm$0.005 \\ 
   \hline
Data 2 & $t=0.4$ & $t=0.55$ & $t=0.7$ & $t=0.85$ & $t=1$ \\ 
  \hline
WDR & \textbf{0.052}$\pm$0.004 & \textbf{0.097}$\pm$0.005 & \textbf{0.142}$\pm$0.005 & \textbf{0.164}$\pm$0.005 & \textbf{0.18}$\pm$0.005 \\ 
  FG & 0.054$\pm$0.005 & 0.113$\pm$0.006 & 0.169$\pm$0.005 & 0.199$\pm$0.005 & 0.216$\pm$0.004 \\ 
  KFG & 0.054$\pm$0.005 & 0.112$\pm$0.006 & 0.146$\pm$0.005 & 0.175$\pm$0.005 & 0.192$\pm$0.004 \\ 
  RF & 0.054$\pm$0.005 & 0.109$\pm$0.006 & 0.144$\pm$0.005 & 0.173$\pm$0.005 & 0.19$\pm$0.004 \\ 
  DeepHit & 0.054$\pm$0.005 & 0.108$\pm$0.005 & 0.161$\pm$0.005 & 0.188$\pm$0.004 & 0.206$\pm$0.004 \\ 
  PCH & 0.053$\pm$0.005 & 0.108$\pm$0.006 & 0.161$\pm$0.006 & 0.198$\pm$0.007 & 0.221$\pm$0.007 \\ 
   \hline
Data 3 & $t=0.1$ & $t=0.3$ & $t=0.5$ & $t=0.7$ & $t=0.9$ \\ 
  \hline
WDR & \textbf{0.095}$\pm$0.004 & \textbf{0.175}$\pm$0.005 & \textbf{0.186}$\pm$0.004 & \textbf{0.194}$\pm$0.005 & \textbf{0.199}$\pm$0.004 \\ 
  FG & 0.107$\pm$0.004 & 0.208$\pm$0.004 & 0.235$\pm$0.003 & 0.245$\pm$0.002 & 0.249$\pm$0.002 \\ 
  KFG & 0.105$\pm$0.004 & 0.18$\pm$0.004 & 0.201$\pm$0.002 & 0.209$\pm$0.002 & 0.213$\pm$0.002 \\ 
  RF & 0.103$\pm$0.004 & 0.179$\pm$0.004 & 0.203$\pm$0.002 & 0.214$\pm$0.002 & 0.219$\pm$0.002 \\ 
  DeepHit & 0.121$\pm$0.006 & 0.289$\pm$0.008 & 0.364$\pm$0.009 & 0.225$\pm$0.004 & 0.235$\pm$0.004 \\ 
  PCH & 0.101$\pm$0.004 & 0.188$\pm$0.006 & 0.2$\pm$0.006 & 0.231$\pm$0.006 & 0.24$\pm$0.006 \\ 
   \hline
Data 4 & $t=0.3$ & $t=0.45$ & $t=0.6$ & $t=0.75$ & $t=0.9$ \\ 
  \hline
WDR & 0.02$\pm$0.002 & 0.046$\pm$0.003 & 0.076$\pm$0.003 & \textbf{0.107}$\pm$0.004 & \textbf{0.115}$\pm$0.005 \\ 
  FG & 0.026$\pm$0.004 & 0.075$\pm$0.005 & 0.125$\pm$0.005 & 0.17$\pm$0.004 & 0.189$\pm$0.005 \\ 
  KFG & 0.024$\pm$0.003 & \textbf{0.045}$\pm$0.005 & 0.086$\pm$0.004 & 0.122$\pm$0.003 & 0.134$\pm$0.003 \\ 
  RF & 0.026$\pm$0.003 & 0.055$\pm$0.005 & 0.107$\pm$0.004 & 0.151$\pm$0.003 & 0.17$\pm$0.002 \\ 
  DeepHit & 0.022$\pm$0.003 & 0.055$\pm$0.004 & 0.095$\pm$0.003 & 0.129$\pm$0.004 & 0.142$\pm$0.003 \\ 
  PCH & \textbf{0.019}$\pm$0.002 & 0.046$\pm$0.003 & \textbf{0.072}$\pm$0.004 & 0.113$\pm$0.005 & 0.129$\pm$0.005 \\ 
   \hline
Data 5 & $t=0.9$ & $t=1$ & $t=1.1$ & $t=1.2$ & $t=1.3$ \\ 
  \hline
WDR & \textbf{0.097}$\pm$0.004 & 0.145$\pm$0.004 & \textbf{0.147}$\pm$0.003 & \textbf{0.139}$\pm$0.003 & \textbf{0.133}$\pm$0.003 \\ 
  FG & 0.105$\pm$0.005 & 0.192$\pm$0.005 & 0.227$\pm$0.003 & 0.241$\pm$0.002 & 0.246$\pm$0.001 \\ 
  KFG & 0.1$\pm$0.005 & 0.163$\pm$0.005 & 0.187$\pm$0.003 & 0.194$\pm$0.001 & 0.197$\pm$0.001 \\ 
  RF & 0.098$\pm$0.005 & 0.15$\pm$0.004 & 0.168$\pm$0.003 & 0.179$\pm$0.002 & 0.181$\pm$0.002 \\ 
  DeepHit & 0.099$\pm$0.006 & 0.167$\pm$0.006 & 0.173$\pm$0.006 & 0.168$\pm$0.006 & 0.163$\pm$0.008 \\ 
  PCH & \textbf{0.097}$\pm$0.006 & \textbf{0.144}$\pm$0.005 & 0.148$\pm$0.004 & 0.154$\pm$0.005 & 0.19$\pm$0.007 \\ 
   \hline
Data 6 & $t=1$ & $t=1.1$ & $t=1.2$ & $t=1.3$ & $t=1.4$ \\ 
  \hline
WDR & 0.077$\pm$0.004 & 0.098$\pm$0.005 & 0.098$\pm$0.005 & 0.092$\pm$0.003 & 0.098$\pm$0.003 \\ 
  FG & 0.103$\pm$0.004 & 0.19$\pm$0.004 & 0.213$\pm$0.004 & 0.223$\pm$0.004 & 0.234$\pm$0.003 \\ 
  KFG & 0.094$\pm$0.004 & 0.14$\pm$0.004 & 0.158$\pm$0.004 & 0.164$\pm$0.003 & 0.173$\pm$0.003 \\ 
  RF & 0.094$\pm$0.004 & 0.138$\pm$0.004 & 0.158$\pm$0.004 & 0.164$\pm$0.003 & 0.175$\pm$0.003 \\ 
  DeepHit & 0.099$\pm$0.005 & 0.099$\pm$0.004 & 0.068$\pm$0.003 & \textbf{0.067}$\pm$0.003 & \textbf{0.072}$\pm$0.003 \\ 
  PCH & \textbf{0.074}$\pm$0.004 & \textbf{0.061}$\pm$0.004 & \textbf{0.067}$\pm$0.004 & 0.096$\pm$0.006 & 0.129$\pm$0.006 \\ 
   \hline
\end{tabular}
\caption{BS for event 1 of synthetic data with constant covariates and no left truncation.}\label{tab:event1:constant}
\end{table}

\begin{table}[!t]
\centering
\begin{tabular}{lccccc}
  \hline
 & $t=0.1$ & $t=0.3$ & $t=0.5$ & $t=0.7$ & $t=0.9$ \\ 
  \hline
WDR &\textbf{0.117}$\pm$0.006 &\textbf{0.177}$\pm$0.004 & \textbf{0.203}$\pm$0.004 & \textbf{0.21}$\pm$0.004 & \textbf{0.213}$\pm$0.004 \\ 
  FG & 0.118$\pm$0.006 & 0.179$\pm$0.004 & 0.204$\pm$0.004 & 0.211$\pm$0.004 & 0.215$\pm$0.004 \\ 
  KFG & 0.118$\pm$0.006 & 0.179$\pm$0.004 & 0.205$\pm$0.004 & 0.212$\pm$0.004 & 0.216$\pm$0.004 \\ 
  RF & 0.121$\pm$0.006 & 0.184$\pm$0.005 & 0.213$\pm$0.004 & 0.221$\pm$0.003 & 0.225$\pm$0.003 \\ 
  DeepHit & 0.143$\pm$0.009 & 0.26$\pm$0.01 & 0.22$\pm$0.004 & 0.229$\pm$0.004 & 0.236$\pm$0.004 \\ 
  PCH & 0.125$\pm$0.007 & 0.191$\pm$0.006 & 0.22$\pm$0.007 & 0.23$\pm$0.007 & 0.235$\pm$0.006 \\ 
   \hline
Data 2 & $t=0.4$ & $t=0.55$ & $t=0.7$ & $t=0.85$ & $t=1$ \\ 
  \hline
WDR & \textbf{0.078}$\pm$0.005 & \textbf{0.15}$\pm$0.006 & {0.213}$\pm$0.003 & 0.235$\pm$0.003 & \textbf{0.229}$\pm$0.004 \\ 
  FG & 0.081$\pm$0.006 & 0.155$\pm$0.006 & 0.226$\pm$0.003 & 0.251$\pm$0.001 & 0.25$\pm$0.002 \\ 
  KFG & 0.081$\pm$0.006 & 0.154$\pm$0.006 & 0.214$\pm$0.003 & 0.238$\pm$0.001 & 0.238$\pm$0.001 \\ 
  RF & 0.08$\pm$0.006 & 0.154$\pm$0.006 & \textbf{0.209}$\pm$0.003 & \textbf{0.234}$\pm$0.001 & 0.234$\pm$0.002 \\ 
  DeepHit & 0.081$\pm$0.006 & 0.156$\pm$0.007 & 0.231$\pm$0.005 & 0.255$\pm$0.003 & 0.253$\pm$0.002 \\ 
  PCH & 0.082$\pm$0.006 & 0.158$\pm$0.008 & 0.229$\pm$0.005 & 0.248$\pm$0.002 & 0.249$\pm$0.002 \\ 
   \hline
Data 3 & $t=0.1$ & $t=0.3$ & $t=0.5$ & $t=0.7$ & $t=0.9$ \\ 
  \hline
WDR & \textbf{0.104}$\pm$0.005 & \textbf{0.166}$\pm$0.005 & \textbf{0.184}$\pm$0.005 & \textbf{0.19}$\pm$0.004 & \textbf{0.195}$\pm$0.004 \\ 
  FG & 0.119$\pm$0.006 & 0.212$\pm$0.004 & 0.237$\pm$0.002 & 0.246$\pm$0.002 & 0.25$\pm$0.002 \\ 
  KFG & 0.115$\pm$0.006 & 0.178$\pm$0.004 & 0.202$\pm$0.003 & 0.209$\pm$0.002 & 0.214$\pm$0.002 \\ 
  RF & 0.114$\pm$0.006 & 0.183$\pm$0.004 & 0.208$\pm$0.002 & 0.217$\pm$0.001 & 0.221$\pm$0.001 \\ 
  DeepHit & 0.137$\pm$0.008 & 0.3$\pm$0.009 & 0.371$\pm$0.009 & 0.224$\pm$0.003 & 0.234$\pm$0.004 \\ 
  PCH & 0.11$\pm$0.007 & 0.169$\pm$0.005 & 0.192$\pm$0.005 & 0.211$\pm$0.005 & 0.219$\pm$0.006 \\ 
   \hline
Data 4 & $t=0.3$ & $t=0.45$ & $t=0.6$ & $t=0.75$ & $t=0.9$ \\ 
  \hline
WDR & 0.024$\pm$0.002 & 0.042$\pm$0.003 & 0.074$\pm$0.003 & 0.102$\pm$0.004 & \textbf{0.102}$\pm$0.004 \\ 
  FG & 0.041$\pm$0.003 & 0.084$\pm$0.003 & 0.138$\pm$0.005 & 0.175$\pm$0.005 & 0.182$\pm$0.005 \\ 
  KFG & 0.037$\pm$0.002 & 0.052$\pm$0.003 & 0.096$\pm$0.004 & 0.119$\pm$0.003 & 0.126$\pm$0.003 \\ 
  RF & 0.037$\pm$0.003 & 0.056$\pm$0.003 & 0.111$\pm$0.004 & 0.15$\pm$0.003 & 0.165$\pm$0.002 \\ 
  DeepHit & 0.028$\pm$0.002 & 0.053$\pm$0.003 & 0.093$\pm$0.004 & 0.114$\pm$0.003 & 0.124$\pm$0.003 \\ 
  PCH & \textbf{0.023}$\pm$0.002 & \textbf{0.037}$\pm$0.002 & \textbf{0.068}$\pm$0.004 & \textbf{0.095}$\pm$0.004 & 0.106$\pm$0.006 \\ 
   \hline
Data 5 & $t=0.9$ & $t=1$ & $t=1.1$ & $t=1.2$ & $t=1.3$ \\ 
  \hline
WDR & 0.091$\pm$0.005 & 0.147$\pm$0.005 & 0.151$\pm$0.005 & \textbf{0.135}$\pm$0.003 & \textbf{0.129}$\pm$0.002 \\ 
  FG & 0.092$\pm$0.006 & 0.188$\pm$0.006 & 0.234$\pm$0.004 & 0.246$\pm$0.003 & 0.251$\pm$0.002 \\ 
  KFG & 0.089$\pm$0.006 & 0.164$\pm$0.006 & 0.195$\pm$0.004 & 0.2$\pm$0.003 & 0.201$\pm$0.002 \\ 
  RF & 0.089$\pm$0.006 & 0.156$\pm$0.006 & 0.181$\pm$0.003 & 0.185$\pm$0.002 & 0.187$\pm$0.001 \\ 
  DeepHit & 0.089$\pm$0.006 & 0.165$\pm$0.007 & 0.178$\pm$0.008 & 0.164$\pm$0.008 & 0.16$\pm$0.009 \\ 
  PCH & \textbf{0.084}$\pm$0.006 & \textbf{0.144}$\pm$0.007 & \textbf{0.149}$\pm$0.007 & 0.142$\pm$0.008 & 0.183$\pm$0.01 \\ 
   \hline
Data 6 & $t=1$ & $t=1.1$ & $t=1.2$ & $t=1.3$ & $t=1.4$ \\ 
  \hline
WDR & 0.09$\pm$0.004 & 0.118$\pm$0.004 & 0.103$\pm$0.003 & 0.094$\pm$0.003 & 0.1$\pm$0.003 \\ 
  FG & 0.106$\pm$0.006 & 0.218$\pm$0.005 & 0.232$\pm$0.004 & 0.238$\pm$0.004 & 0.247$\pm$0.003 \\ 
  KFG & 0.1$\pm$0.005 & 0.156$\pm$0.004 & 0.164$\pm$0.004 & 0.168$\pm$0.003 & 0.174$\pm$0.002 \\ 
  RF & 0.098$\pm$0.005 & 0.154$\pm$0.004 & 0.166$\pm$0.004 & 0.174$\pm$0.003 & 0.186$\pm$0.003 \\ 
  DeepHit & 0.104$\pm$0.006 & 0.126$\pm$0.004 & 0.07$\pm$0.003 & \textbf{0.067}$\pm$0.002 & \textbf{0.075}$\pm$0.003 \\ 
  PCH & \textbf{0.082}$\pm$0.004 & \textbf{0.061}$\pm$0.004 & \textbf{0.055}$\pm$0.004 & 0.068$\pm$0.005 & 0.094$\pm$0.005 \\ 
   \hline
\end{tabular}
\caption{BS for event 2 of synthetic data with constant covariates and no left truncation.} 
\label{tab:event2:constant}\vspace{-3.5mm}
\end{table}

\subsection{WDR Estimation on DLBCL and Model Comparison}
\begin{table}[!t]
  \centering
\renewcommand{\arraystretch}{1} % Default value: 1
  \begin{tabular}{lcccc}
    \hline
 $\betav_{jk}$  &  ABC ($j=1$) & GCB ($j=2$) &\multicolumn{2}{c} {T3 ($j=3$)}\\
\cmidrule(lr){4-5} 
 Gene \#     & $k=1$ & $k=1$  & $k=1$  & $k=2$  \\
    \hline
    17482 & 1.943
    %$\pm$0.807 
    & 0.060
    % $\pm$0.381 
    & -0.256
    % $\pm$1.455 
    & -0.497
    % $\pm$1.354 
    \\ 
  24432 & 0.715
%   $\pm$0.549 
  & 0.025
%   $\pm$0.286 
  & 0.248
%   $\pm$0.689 
  & 0.414
%   $\pm$0.904 
  \\ 
  17833 & 0.795
%   $\pm$0.599 
  & 0.128
%   $\pm$0.269 
  & -0.229
%   $\pm$0.811 
  & -0.258
%   $\pm$0.778 
  \\ 
  28193 & 0.096
%   $\pm$0.289 
  & 0.090
%   $\pm$0.236 
  & -0.200
%   $\pm$0.816 
  & -0.077
%   $\pm$0.782 
  \\ 
  28197 & 0.031
%   $\pm$0.168 
  & -0.719
%   $\pm$0.508 
  & -0.021
%   $\pm$0.432 
  & 0.109
%   $\pm$0.489 
  \\ 
  27731 & -0.015
%   $\pm$0.234 
  & -0.081
%   $\pm$0.384 
  & -1.481
%   $\pm$1.822 
  & 1.059
%   $\pm$2.23 
  \\ 
  31456 & 0.246
%   $\pm$0.394 
  & 0.654
%   $\pm$0.556 
  & -2.871
%   $\pm$2.517 
  & -5.568
%   $\pm$2.504 
  \\ 
  \hline
  $r_{jk}$ & 0.122  & 0.131  & 0.030  & 0.031  \\ 
   \hline
  \end{tabular}
  \caption{$\betav_{jk}$ and $r_{jk}$ of WDR for the analysis of  DLBCL.}\label{tab:dlbcl}\vspace{-3.5mm}
\end{table}

We analyze the DLBCL data by WDR and estimate the model by MCMC. The posterior mean of the Weibull shape parameter $a$ is $1.379$ with the posterior standard deviation $0.151$. Provided in Table~\ref{tab:dlbcl} are the posterior means of $\betav_{jk}$'s and $r_{jk}$'s. We see two sub-events under T3 ($j=3$), indicating two potential disease subtypes of T3, and each is linearly accelerated by the expression of the seven genes. 
We compare the performance of WDR, FG, KFG, RF, DeepHit, and PCH on the DLBCL data. To reduce the impact on the training-testing partition, we randomly take 20 partitions, in each of which 200 subjects are used for training and the other 40 for testing. We report the Brier scores in Table~\ref{tab:dlbcl:BS}. As a result, no model consistently outperforms the others, but WDR has a good performance that is comparable to RF. Notably, DeepHit and PCH do not work well and seem to overfit this smaller data, though validation set and early stopping are used.

\begin{table}[ht]
\centering
\renewcommand{\arraystretch}{1} % Default value: 1
\makebox[\linewidth]{
\resizebox{\linewidth}{!}{
\begin{tabular}{llcccccc}
  \hline
& & t=0.5 & t=1 & t=1.5 & t=2 & t=2.5 & t=3 \\ 
  \hline
\multirow{ 6}{*}{ABC} &WDR & 0.11$\pm$0.009 & 0.172$\pm$0.01 & 0.195$\pm$0.011 & 0.202$\pm$0.01 & 0.223$\pm$0.01 & \textbf{0.223}$\pm$0.009 \\ 
&  FG & \textbf{0.108}$\pm$0.009 &  0.172$\pm$0.01 & 0.195$\pm$0.011 & 0.203$\pm$0.01 & 0.224$\pm$0.01 & 0.226$\pm$0.009 \\ 
&  KFG & 0.126$\pm$0.01 & 0.187$\pm$0.013 & 0.201$\pm$0.014 & 0.215$\pm$0.013 & 0.225$\pm$0.014 & 0.257$\pm$0.012 \\ 
&  RF & 0.111$\pm$0.009 & \textbf{0.168}$\pm$0.01 & \textbf{0.191}$\pm$0.011 & \textbf{0.199}$\pm$0.01 & \textbf{0.217}$\pm$0.01 & \textbf{0.223}$\pm$0.008 \\ 
&  DeepHit & 0.122$\pm$0.01 & 0.211$\pm$0.012 & 0.221$\pm$0.013 & 0.231$\pm$0.012 & 0.25$\pm$0.011 & 0.268$\pm$0.01 \\ 
&  PCH & 0.119$\pm$0.01 & 0.197$\pm$0.012 & 0.233$\pm$0.013 & 0.242$\pm$0.013 & 0.265$\pm$0.013 & 0.277$\pm$0.011 \\ 
  \hline
\multirow{ 6}{*}{GCB} &  WDR  & \textbf{0.071}$\pm$0.007 & \textbf{0.134}$\pm$0.008 & 0.196$\pm$0.012 & 0.205$\pm$0.011 & 0.221$\pm$0.011 & 0.247$\pm$0.01 \\ 
&  FG  & 0.072$\pm$0.007 & 0.138$\pm$0.008 & 0.2$\pm$0.011 & 0.213$\pm$0.012 & 0.231$\pm$0.011 & 0.255$\pm$0.01 \\ 
&  KFG  & 0.073$\pm$0.006 & 0.135$\pm$0.006 & \textbf{0.19}$\pm$0.009 & \textbf{0.196}$\pm$0.009 & \textbf{0.212}$\pm$0.008 & \textbf{0.232}$\pm$0.007 \\ 
&  RF  & \textbf{0.071}$\pm$0.007 & 0.135$\pm$0.008 & 0.196$\pm$0.012 & 0.207$\pm$0.012 & 0.224$\pm$0.012 & 0.244$\pm$0.01 \\ 
&  DeepHit  & 0.074$\pm$0.008 & 0.148$\pm$0.009 & 0.214$\pm$0.013 & 0.23$\pm$0.013 & 0.243$\pm$0.013 & 0.276$\pm$0.013 \\ 
&  PCH  & 0.075$\pm$0.008 & 0.146$\pm$0.008 & 0.219$\pm$0.014 & 0.232$\pm$0.014 & 0.255$\pm$0.014 & 0.285$\pm$0.014 \\ 
  \hline
\multirow{ 6}{*}{T3}&  WDR  & \textbf{0.081}$\pm$0.008 & \textbf{0.11}$\pm$0.009 & 0.143$\pm$0.012 & \textbf{0.145}$\pm$0.012 & \textbf{0.158}$\pm$0.011 & \textbf{0.157}$\pm$0.011 \\ 
&  FG  & 0.087$\pm$0.008 & 0.129$\pm$0.009 & 0.163$\pm$0.012 & 0.165$\pm$0.011 & 0.168$\pm$0.011 & 0.168$\pm$0.011 \\ 
&  KFG  & 0.083$\pm$0.007 & 0.118$\pm$0.006 & 0.163$\pm$0.008 & 0.177$\pm$0.008 & 0.189$\pm$0.007 & 0.205$\pm$0.008 \\ 
&  RF  & 0.085$\pm$0.008 & 0.113$\pm$0.009 & \textbf{0.141}$\pm$0.012 & 0.151$\pm$0.012 & 0.161$\pm$0.011 & 0.161$\pm$0.011 \\ 
&  DeepHit  & 0.087$\pm$0.009 & 0.119$\pm$0.01 & 0.153$\pm$0.013 & 0.172$\pm$0.013 & 0.172$\pm$0.011 & 0.172$\pm$0.011 \\ 
&  PCH  & 0.088$\pm$0.009 & 0.122$\pm$0.009 & 0.157$\pm$0.013 & 0.174$\pm$0.012 & 0.185$\pm$0.012 & 0.185$\pm$0.012 \\ 
   \hline
\end{tabular}
}}
\caption{Model comparison on DLBCL in Brier scores (mean$\,\pm\,$standard error).} 
\label{tab:dlbcl:BS}\vspace{-3.5mm}
\end{table}

\subsection{Model Comparison on the AD Data}
\begin{table}[!t]
  \centering
\renewcommand{\arraystretch}{1} % Default value: 1
  \begin{tabular}{lcccccc}
    \hline
   &  \multicolumn{3}{c} {MCI} & \multicolumn{3}{c} {Death}\\
    \cmidrule(lr){2-4}\cmidrule(lr){5-7} 
    & {Age=80} & {Age=85} & {Age=90} & {Age=80} & {Age=85} & {Age=90} \\
    \hline
    WDR & 0.157 & 0.175 & 0.175 & 0.020 & 0.021 & 0.022\\  
    FG & 0.179 & 0.157 & 0.151 & 0.017 & 0.020 & 0.028\\ 
    KFG & 0.242 & 0.265 & 0.264 & 0.020 & 0.021 & 0.025\\
    \hline
  \end{tabular}
   \caption{Brier scores for MCI and death of other causes.}\label{tab:AD:BS}\vspace{-3.5mm}
\end{table}

% \begin{table}[!t]
%   \centering
%   \caption{C-indices for MCI and death of other causes}\label{tab:AD_cidx}\vspace{-3.5mm}
%    \renewcommand{\arraystretch}{1} % Default value: 1
%   \begin{tabular}{lcccccc}
%     \hline
%    &  \multicolumn{3}{c} {MCI} & \multicolumn{3}{c} {Death}\\
%     \cmidrule(lr){2-4}\cmidrule(lr){5-7} 
%     & {Age=80} & {Age=85} & {Age=90} & {Age=80} & {Age=85} & {Age=90} \\
%     \hline
%     WDR & 0.977 & 0.977 & 0.977 & 1.000 & 1.000 & 1.000\\  
%     FG & 0.947 & 0.953 & 0.953 & 0.833 & 0.778 & 0.778\\ \hline
%   \end{tabular}
% \end{table}

We compare the performance of WDR, FG, and KFG on the testing set of the Alzheimer's disease data. Since RF, DeepHit, and PCH are not designed for left truncation and time-varying covariates, we do not run these models.
We report the Brier scores in Table~\ref{tab:AD:BS}, 
%and the C-indices in Table~\ref{tab:AD_cidx}, 
which are evaluated at ages 80, 85, and 90. As a result, WDR is comparable to FG and outperforms KFG in predicting MCI.

\section{WDR for Classification and Additional Experiments}\label{app:discrete_choice}
We show WDR as a nonlinear discrete choice model for classification problems. Considering the existence of many sophisticated classification models, our goal is not the state-of-the-art prediction accuracy, but to provide an alternative approach for interpretable classification as a supplement to the linear ones, like probit and logistic regressions.
Weibull (delegate) racing can be regarded as a discrete choice model where the decision of categorization is made to minimize the waiting time for the arrival of the first candidate choice, which is~$y=\mathop{\mathrm{argmin}}_j t_j$. Therefore, WDR classification inherits all the advantages of the WDR survival model such as data-adaptive nonlinearity and interpretability as shown in Section \ref{sec:experiments}. 
In practice, we adopt finite truncation of the gamma processes of WDR classification by allowing each category $j\in 1,\ldots,J$ to consist of up to $K$ subtypes. Consequently, the probability of $y$ given $\{\lambda_{jk}\}_{j,k}$ is 
\begin{align}
   \mbox{Pr}(y=j\given \{\lambda_{jk}\}_{j,k}) = \frac{\sum_{k=1}^{K}\lambda_{jk}}{ \sum_{j'=1}^J\sum_{k'=1}^{K}\lambda_{j'k'}}  \label{eq:class_prob}
\end{align}
where $\lambda_{jk}\sim \mbox{Gamma}(r_{jk}, \exp({\xv^\T\betav_{jk}}))$. This probability can be estimated by Monte Carlo methods if we have point or posterior estimates of $r_{jk}$'s and $\betav_{jk}$'s. Note that \eqref{eq:class_prob} does not depend on the Weibull shape parameter $a$. So we fix $a$ at an arbitrary constant value without sacrificing modeling capacity (we set $a=1$ for all the experiments of WDR classification). The inference by MCMC or MAP follows the same algorithm as for the WDR survival model as if all the event times are missing, except that the step of estimating $a$ is skipped. In this way, the MCMC for WDR classification turns out to be a Gibbs sampler. For identifiability and good mixing of MCMC, we should fix one of the $\betav_{jk}$'s, say $\betav_{11}$, equal to a zero vector. 
\subsection{WDR Classification on Toy Data}
\begin{figure}[!t]
\centering
%\begin{minipage}{.485\textwidth}
\includegraphics[width=1\columnwidth]{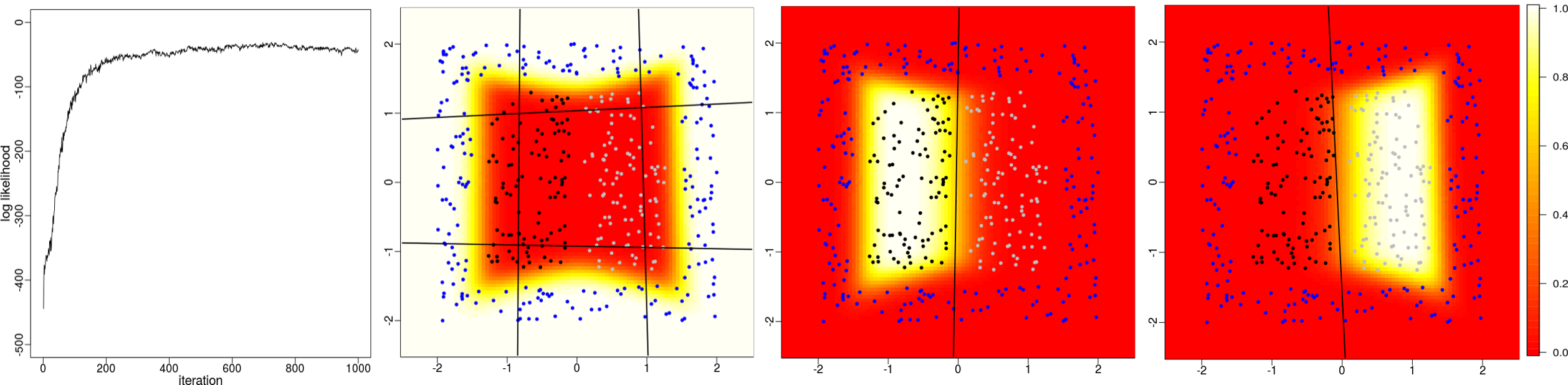}\vspace{-3mm}
\caption{\small Trajectory plots of log likelihood (column 1) and predictive probability heat maps and hyperplanes for category 1 to 3 (column 2 to 4) of \textit{square} data. Blue points are labeled as category 1, black 2 and gray 3.}\label{square}
\end{figure}

We first illustrate the data-adaptive nonlinearity of WDR for the classification of \textit{square}, which is synthesized with two-dimensional covariates and $J=3$ categories. Figure \ref{square} shows the classification of the three categories, two rectangles (denoted by black and grey dots) in one square frame (blue dots), by MCMC with $K$ set to be $10$. The panel on the left gives the trajectory plot of the log likelihood by MCMC iterations, and the other three panels show the heat maps of predictive probabilities of category 1 (blue dots), 2 (black dots), and 3 (gray dots), respectively. The solid lines on these three panels denote the hyperplanes $\bm x^\T\hat{\betav}_{jk}=0$ where $\hat{\betav}_{jk}$ is the estimated posterior means of ${\betav}_{jk}$. We see the three categories have been perfectly separated within 1,000 MCMC iterations and WDR have found four subtypes of category 1 and one subtype for category 2 or 3, respectively. For illustration purpose, we do not care for identifiability and have not fixed any $\betav_{jk}$ equal to zero in Figure~\ref{square}.

\begin{figure}[!t]
%\begin{wrapfigure}{R}{0.66\textwidth}
 \centering
 \includegraphics[width=0.4\linewidth]{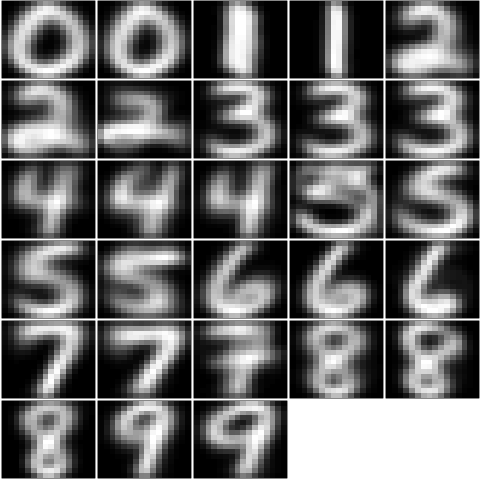}\vspace{-2mm}
 \caption{Subtypes of \textit{usps} handwritten numbers by WDR.  }\vspace{-3mm}\label{expert_visual}
%\end{wrapfigure}
\end{figure}

We further visualize WDR's capability of finding subtypes on data \textit{usps} that have handwritten numbers zero to nine. Figure \ref{expert_visual} shows the subtypes of each number found by WDR. Specifically, we visualize $\textstyle{\sum_i \frac{\hat\lambda_{ijk}}{\hat\lambda_{ijk}+\sum_{j'\neq j}\sum_{k'} \hat\lambda_{ij'k'}}\xv_i}$, a weighted average of all the training samples $\xv_i$'s, for subtype $k$ of category $j$ where $\hat\lambda_{ijk}$ is the estimated posterior mean of $\lambda_{ijk}$. We see that the number five has four subtypes while the number zero, one, or nine has only two subtypes, indicating the different amount of nonlinear capacity is required when depicting the classification boundaries.

\subsection{Improving Nonlinear Capacity by Data Transformation}
If used in classification settings, WDR is not primarily focused on highly accurate prediction for big and complex data. Instead, it is advantageous in interpretation and data-adaptive nonlinearity; WDR  can be ideal if one wants to discover subtypes of a category or evaluate how much nonlinearity is relatively needed for each category. In case WDR is applied to reasonably complex classification problems, we propose a data transformation scheme to enhance the nonlinear capacity and make its classification accuracy comparable to kernel support vector machines on moderate-sized data sets.  

Unlike kernel methods such as support vector machines \citep{boser1992training,SVM} and the relevance vector machine \citep{tipping2001sparse} which make categories more linearly separable in a  higher-dimensional transformed covariate space, or neural networks using complex data transformations, WDR uses interactions of linear hyperplanes to construct nonlinear decision boundaries, and hence may have insufficient capacity if the class boundaries are highly complex. To tackle such a problem we can stack another WDR on a previously trained one to enhance its capacity, which is a similar strategy of \citet{zhang2017permuted}. Concretely, we first run a WDR to obtain a finite set of hyperplanes denoted as $\tilde{\betav}_{jk}$, and then augment the original covariates $\xv_i$ into
\begin{align}
\tilde{\xv}_i:=\left[ \xv_i^\T, \log\big(1+\exp{(\xv_i^\T  \tilde{\betav}_{11})}\big),
 \log\big(1+\exp{(\xv_i^\T  \tilde{\betav}_{12})}\big),\cdots, 
 \log\big(1+\exp{(\xv_i^\T \tilde{\betav}_{SK})}\big)\right]^\T\label{eq:data_trans},
\end{align}
and then run another WDR with the transformed covariate $\tilde{\xv}_i$.

\begin{figure}[!t]
\centering
%\begin{minipage}{.485\textwidth}
 \includegraphics[width=0.675\columnwidth]{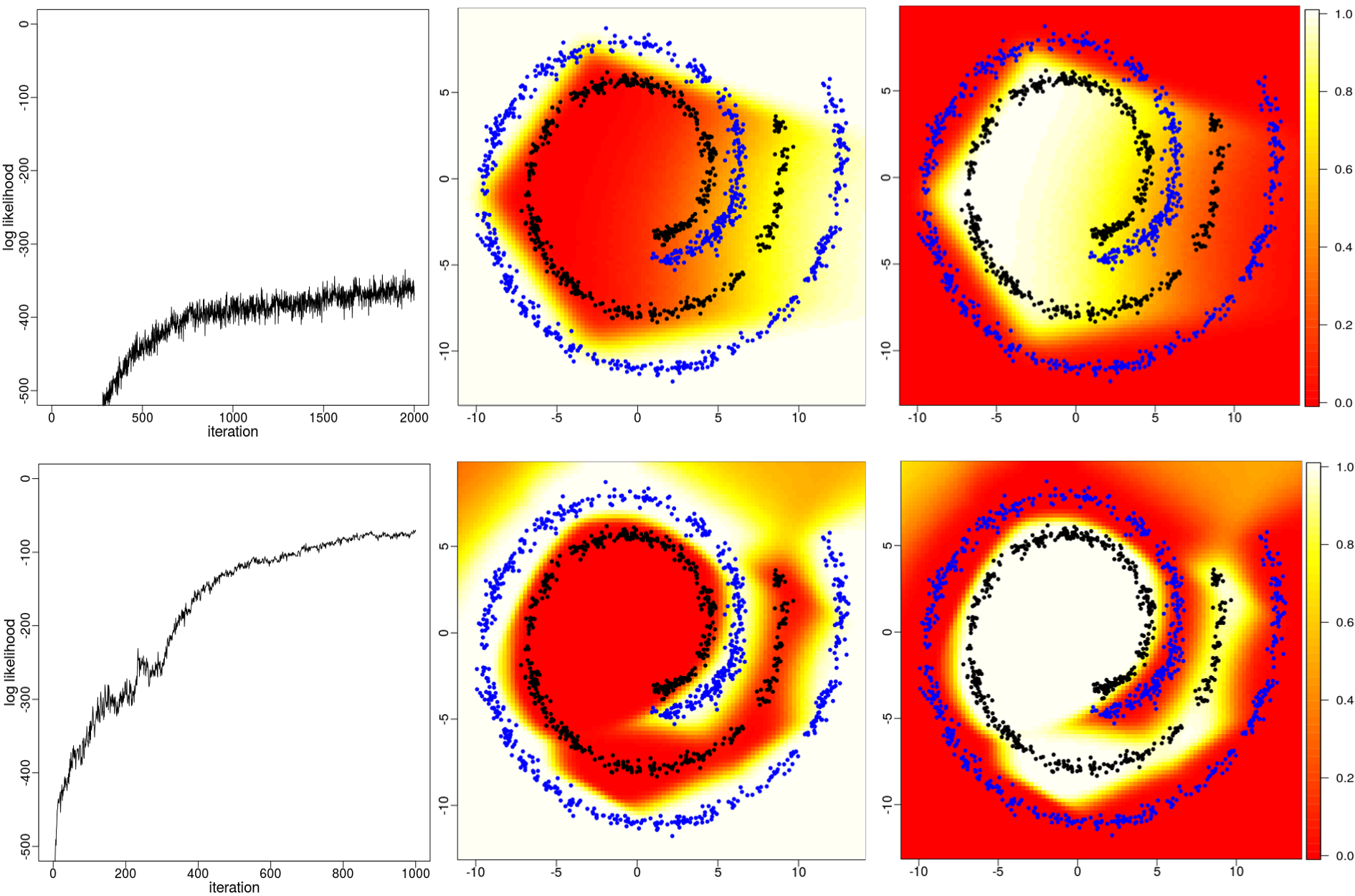}\vspace{-3mm}
 \caption{Trajectory plot of log likelihood and predictive probability heat map for \textit{swiss roll} data. Blue points are labeled as category 1 and black 2. Row~1 results from WDR with the original covariates and row~2 with data transformation as in \eqref{eq:data_trans}.}\label{swiss}
\end{figure}

We show the increased capacity of WDR using the data transformation on a synthetic 2-D \textit{swiss roll} data in Figure \ref{swiss}. Row~1 shows the results using the data with original covariates and row~2 using the transformed data by \eqref{eq:data_trans} where $\tilde{\betav}_{jk}$ are from the estimation in row~1. The improvement is remarkable in not only in-sample fits reflected by log likelihood trajectory plots but also in the out-of-sample prediction shown in the heat maps. 

\subsection{Model Comparison}
We compare the classification performance of WDR on real data sets. Table \ref{tab:class_data_des} summarizes the data sizes and the numbers of covariates and categories. The training and testing sets are predefined for \textit{vehicle}, %vowel, 
\textit{dna}, \textit{satimage}, \textit{usps} and \textit{mnist} where the validation sets are merged into training. We divide the other data sets into training and testing as follows. For \textit{iris}, \textit{wine}, and \textit{glass}, five random partitions are taken such that for each partition the training set accounts for 80\% of the whole data while the testing set accounts for 20\%. The classification error rate is calculated by averaging the error rates of all the five partitions. For %\textit{square}, 
\textit{waveform} and \textit{segment}, one random partition is taken and
%70\% observations of \textit{square} are used as training and the remaining 30\% as testing, and 
10\% of the data points %of \textit{waveform} and \textit{segment} 
are used for training and the other 90\% for testing.
\begin{table}[!t]
\centering
%\captionsetup{labelformat=empty}
\makebox[\linewidth]{
\resizebox{\linewidth}{!}{
%\scriptsize
%\begin{tabular}{l P{0.3cm}P{0.3cm}P{0.4cm}P{0.6cm}P{1cm}P{0.9cm}P{0.4cm}P{0.9cm}P{0.5cm}P{0.6cm}P{0.6cm}}
\begin{tabular}{lccccccccccc}
 \hline
 & iris & wine & glass & vehicle & waveform & segment & % vowel & 
 dna & satimage & usps & mnist \\% & square \\ 
 \hline
Train size  & 120 & 142 & 171 & 592 & 500 & 231 & %528 & 
2000 & 4435 & 7291 & 60000 \\% & 294\\ 
 Test size  & 30 & 36 & 43 & 254 & 4500 & 2079 & %462 & 
 1186 & 2000  & 2007 & 10000 \\% & 126\\ 
 Covariate number & 4 & 13 & 9 & 18 & 21 & 19 & %10 & 
 180 & 36 & 256 & 784 & \\% 2\\ 
 Category number  & 3 & 3 & 6 & 4 & 3 & 7 & %11 & 
 3 & 6 & 10 & 10 & \\% 3\\ 
 \hline
\end{tabular}
%}
}
}
\caption{Multiclass datasets used in experiments for model comparison.}\label{tab:class_data_des}
\end{table}

\begin{table}[!t]
\centering
%\makebox[\linewidth]{
%\resizebox{\linewidth}{!}{
\begin{tabular}{lccccccc}
%\begin{tabular}{l P{1.6cm}P{1.7cm}P{1.6cm}P{1.7cm}P{1.4cm}P{1.4cm}P{1.4cm}P{1.4cm}}
  \hline
 & WDR  & WDR w. data trnsf  & $L_2$-MLR & SVM\\% & AMM \\ 
  \hline
  iris &4.00$\pm$2.79  & \textbf{2.67}$\pm$2.79 & 3.33$\pm$3.33  & 4.00$\pm$1.83 \\% & 4.67$\pm$3.80 \\ 
  wine &3.89$\pm$3.17 & 3.33$\pm$3.04 & 3.89$\pm$3.17 & \textbf{2.78}$\pm$1.47 \\%& 3.89$\pm$3.17 \\ 
  glass &29.77$\pm$8.76 & \textbf{26.98}$\pm$8.32 & 33.02$\pm$3.82 & 28.84$\pm$5.74 \\%& 37.67$\pm$14.75 \\ 
  vehicle &  17.24$\pm$1.16 & \textbf{14.44}$\pm$0.23 & 22.83 & 18.50 \\%& 21.89 \\ 
  waveform & \textbf{14.39}$\pm$0.29 & 14.95$\pm$0.26 & 15.60 & 15.22 \\%& 18.54 \\ 
  segment & 7.09$\pm$0.65 & 6.62$\pm$0.16 & 8.56 & \textbf{6.20} \\%& 12.47 \\ 
  dna & \textbf{4.06}$\pm$0.28 & 4.69$\pm$0.22 & 5.98 & 4.97 \\%& 5.43 \\ 
  satimage & 11.42$\pm$0.20 & 9.37$\pm$0.26 & 17.80 & \textbf{8.50} \\%& 15.31 \\ 
  usps & 5.98$\pm$0.17 & 5.64$\pm$0.14 & 8.47 & \textbf{4.78} \\%& 7.87 \\ 
 mnist & 2.71$\pm$0.17 & 1.78$\pm$0.22 & 7.40	& \textbf{1.48}	\\%& 6.89\\
  %square & \textbf{0.00}$\pm$0.00 & 0.22$\pm$0.33 & 62.29 & 4.76 \\%& 16.67 \\ 
   \hline
\end{tabular}
%}}
\caption{Comparison of prediction error rate (\%).}\label{tab:classificationerror}
\end{table}

% 3 other algorithms
We compare the WDR classification model with an $L_2$-regularized multinomial logistic regression ($L_2$-MLR) and support vector machines (SVMs) with radial basis function (RBF) kernels. An observation in testing sets is classified into the category associated with the largest predictive probability if provided by the model. For WDR models we use the Monte-Carlo average for predictive probabilities and report the mean and standard deviation. For the WDR with data transformation, we first run WDR with $K=10$, using the original training data to obtain $\tilde{\betav}_{jk}$'s, and then we run another WDR with $K=10$ with the transformation of \eqref{eq:data_trans}.

We use the \texttt{R} package \texttt{LiblineaR} \citep{LiblineaR} for $L_2$-MLR where a bias term is included and the regularization parameter %$cost$
 is selected by a five-fold cross-validation on the training set from $(2^{-10}, 2^{-9},\ldots, 2^{15})$. 
 For SVMs, we use the LIBSVM \citep{LIBSVM} provided by the \texttt{R} package \texttt{e1071} \citep{e1071}.  %and paSB-MSVM, 
A Gaussian RBF kernel is used and a three-fold cross validation is adopted to tune both the regularization parameter %$cost$ 
and kernel width from $(2^{-10}, 2^{-9},\ldots, 2^{10})$  
on the training set. 
We choose the default settings for all the other parameters. 

We report the classification error rates on the testing sets in Table \ref{tab:classificationerror} together with standard errors by all the models on \textit{iris}, \textit{wine}, and \textit{glass} (recall that we have five training-testing partitions ), as well as the standard errors by WDR on other data sets.  We can see the classification accuracy by WDR is comparable to fine-tuned SVMs with RBF kernels.

%\vskip 0.2in
%\bibliography{sample}
\bibliography{wdr.bib,References052016.bib,References112016.bib}
\end{document}

%% file: revisit/Command.tex
\newcommand{\beq}{\vspace{0mm}\begin{equation}}
\newcommand{\eeq}{\vspace{0mm}\end{equation}}
\newcommand{\beqs}{\vspace{0mm}\begin{eqnarray}}
\newcommand{\eeqs}{\vspace{0mm}\end{eqnarray}}
\newcommand{\barr}{\begin{array}}
\newcommand{\earr}{\end{array}}

\def\T{{ \mathrm{\scriptscriptstyle T} }}

\newcommand{\bv}[0]{{\boldsymbol{b}}}

\newcommand{\xv}{\boldsymbol{x}}

\newcommand{\cdotv}{\boldsymbol{\cdot}}

\newcommand{\given}{\,|\,}

\newcommand{\betav}[0]{{\boldsymbol{\beta}}}

\newcommand{\E}{\mathbb{E}}

\newtheorem{thm}{Theorem} %[section]
\newtheorem{cor}[thm]{Corollary}

\usepackage{booktabs}
\usepackage{array}
\newcolumntype{L}[1]{>{\raggedright\let\newline\\\arraybackslash\hspace{0pt}}m{#1}}
\newcolumntype{C}[1]{>{\centering\let\newline\\\arraybackslash\hspace{0pt}}m{#1}}
\newcolumntype{R}[1]{>{\raggedleft\let\newline\\\arraybackslash\hspace{0pt}}m{#1}}